\documentclass[12pt]{article}

\usepackage{amstext,amsmath,amssymb,amsfonts}
\usepackage[latin1]{inputenc}
\usepackage{epsfig}
\usepackage{hyperref}
\usepackage{amsthm}
\usepackage{subfigure}
\usepackage{color}
\usepackage{multirow}
\usepackage[english]{babel}

\newtheorem{corollary}{Corollary}

\newtheorem{theorem}{Theorem}
\newtheorem{proposition}{Proposition}

\newcommand{\R}{\mathbb R}
\newcommand{\C}{\mathbb C}
\newcommand{\N}{\mathbb N}
\newcommand{\I}{\mathbb I}

\newcommand{\beq}{\begin{eqnarray}}
\newcommand{\eeq}{\end{eqnarray}}

\newcommand{\bea}{\begin{equation}}
\newcommand{\eea}{\end{equation}}

\newcommand{\iche}{\check{\imath}}
\newcommand{\iiche}{\check{\iche}}

\newcommand{\fun}{ \;  _{1}F_{1} }

\makeatletter

\begin{document}

\begin{titlepage}
\begin{flushright}
ICMPA-MPA/010/2011\\
pi-other-237
\end{flushright}
\begin{center}

\vspace{20pt}

{\Large\bf {Deepening the vector coherent state analysis:\\\vspace{10pt}
Revisiting the harmonic oscillator}}

\vspace{20pt}

 I. Aremua$^{a,b}$, J. Ben Geloun$^{a,c}$
and M. N. Hounkonnou$^{a,}$\footnote{Correspondence author,
norbert.hounkonnou@cipma.uac.bj  with copy to hounkonnou@yahoo.fr.}

\vspace{20pt}

{\em $^a$ International Chair in Mathematical Physics
and Applications} \\
{\em ICMPA-UNESCO Chair}\\
{\em University of Abomey-Calavi}\\
{\em  072 B.P. 50 Cotonou, Republic of Benin}\\
\vspace{5pt}
{\em $^b$ Institut de Math\'ematiques et de Sciences Physiques}\\
{\em University of Abomey-Calavi}\\
{\em 01 B.P. 613 Porto-Novo, Republic of Benin}\\
\vspace{5pt}
{\em  $^c$ Perimeter Institute for Theoretical Physics} \\
{\em 31 Caroline St. N., ON, N2L 2Y5, Waterloo, Canada} \\
\vspace{10pt}

\vspace{5pt}
\today

\begin{abstract}
Vector coherent states (VCS) viewed as a generalization of ordinary
coherent states for higher rank tensor Hilbert spaces are investigated.
We consider a systematic way of generating classes of VCS which
are solvable (i.e., in the present context, normalizable states
satisfying a resolution of the identity) on the Hilbert space of  $2D$ and $3D$  harmonic oscillators.
Thanks to the type of  construction,   these VCS are classified
according to  specific criteria. Furthermore,
in many cases,  the found classes of VCS
are continuously deformable one onto another,  still
remaining solvable.
\end{abstract}

\end{center}

\end{titlepage}

\setcounter{footnote}{0}

\section{Introduction}
 \label{intro}

Vector coherent states (VCS) are well-known objects in mathematical physics when they are particularly defined as orbits of vectors under  operators of unitary representations of groups and  used in a variety of symmetry problems  in quantum mechanics \cite{ali-antoine-gazeau}.
In some earlier works, a fairly systematic method has been introduced for constructing VCS over various types of matrix domains
\cite{thirulogasanthar-ali, ali-englis-gazeau} in analogy with the
canonical coherent states (CS),
under the additional assumption of the existence of a resolution of the identity. Besides, VCS are also formulated for quantum optical models
with spin-orbit interactions among which the Jaynes-Cummings model \cite{thirulo,ab1,hus1}
and its deformed versions \cite{joben1, joben2,joben3,joben4}.
Furthermore, in \cite{ab1},  the study of the Landau levels
has been achieved and different classes of VCS  have been rigorously defined by taking into account the degeneracy.
Among precursor works on some closely related topics,
one also quotes the multidimensional generalization of CS introduced in \cite{gazeau-novaes} defined for Hamiltonians
with non-degenerate discrete spectrum \cite{gazeau-klauder}.
These multidimensional CS serve for the determination
of the thermodynamic potential of a $2D$ electron gas
in a perpendicular magnetic field.
At the theoretical level, the latter work extends the results given in \cite{gazeau-klauder} to a system with several degrees of freedom.
 An analogous procedure was  used in \cite{thirulo-saad} in
order to obtain the CS for a free magnetic Schr\"{o}dinger operator, and in \cite{thirulo} by introducing a class of VCS  derived with  matrices viewed as simple vectors in an enlarged Hilbert space.
The present work deals with an extension of
these three contributions by Gazeau and Novaes \cite{gazeau-novaes} and Thirulogasanthar {\it et al} \cite{thirulo,thirulo-saad}.

Let us come back for the moment on basic facts on CS.
For any given multidimensional system with associated quantum
Hilbert space spanned by some basis $\{|[n]\rangle\}$,
 $[n]$ being some multi-valued index labeling the
eigenvalues of some commuting observables,
there is a straightforward way to generate a CS
 for the $r$-th degree of freedom \cite{gazeau-novaes}:
\bea
|J_r,\gamma_r,[n]\rangle =
[\mathcal{N}_r(J_r)]^{-\frac12} \sum_{n_r} \frac{J_r^{\frac{n_r}{2}}}{\sqrt{\rho_r}}
e^{-i \gamma_r e_r([n])} |[n]\rangle,
\label{multics}
\eea
where $(J_r,\gamma_r)$ are Gazeau-Klauder
action angle variables, $\mathcal{N}_r(J_r)$ is a normalization
factor\footnote{The normalization factor may depend on
remaining indices and on the form of
the quantity $\rho_{r}$.},
$e_r([n])$ some eigenvalue of $r$-th observable,
$\rho_r$ a free quantity at the moment which may depend on the remaining indices in $[n]$ and so on the label of the energy level.
In fact, even for the simple harmonic oscillator,
there is always a freedom in the definition of CS.
Indeed, in obvious notations, we have:
 $ |z\rangle =\mathcal{N}(|z|,\rho)^{-1/2} \sum_{n=0}^\infty
z^n/\sqrt{\rho(n)} |n \rangle$, where $\rho(n)$
is still a free function of the energy level $n$.
In the specific instance of canonical CS,
$\rho(n)=n!$. We will
call these functions $\rho_r$ {\it generalized factorials}.
Dealing with a generalized version of CS (\ref{multics}),
the function $\rho_r([n])$ may have different forms
and, moreover,
its dependency on the indices of $[n]$ may entail
drastic consequences on the solvability  of the CS
with respect to some set of axioms.
Hence an issue  worthwhile to be investigated is
the definition of particular classes of functions $\rho$ making the
CS solvable.

In the particular instance of \cite{thirulo-saad}, the authors treat
some CS in a rank two Hilbert space, i.e.  a Hilbert
space of the tensored form $\mathcal H \otimes \mathcal H'$. Peculiar classes of CS were defined by a procedure which
takes into account some particular generalized factorials
in such a way that the consequent states fulfill Gazeau-Klauder axioms \cite{gazeau-klauder}.
By scrutinizing that procedure, one realizes
that the set of CS that the authors consider is not complete
and therefore can be enlarged and, more to the point,
even systematized. This is the bottom line
of our investigations.

It is then valuable to investigate how the above scheme can be extended and systematized to more involved Hilbert spaces
provided one could make a sense of it for some basic example
at first. This is what we propose to investigate here.
Any systematic approach foreseeing technicalities,
we will consider simple harmonic oscillators
as toy models for which the same above questions
could be naturally asked and indeed find nontrivial issues
as we will see.

In the present study, based on the prime scheme developed
in \cite{gazeau-novaes,thirulo,thirulo-saad}, we perform a systematic analysis of VCS associated with
the harmonic oscillator in $2D$ and then in $3D$.
The VCS are built using different generalized factorials,
are normalizable and have a resolution of unity.
We implement a way to classify these VCS which is given by,
roughly speaking, their increasing number of complex parameters
(called degrees of freedom)
and increasing complexity of their generalized factorials.
It turns out that the VCS classes can be also
understood, from another point of view,
as continuous deformed classes of one into another
by different frequency limits, providing a possible
second type of classification.
All the VCS highlighted in this work can be extended
without ambiguity to VCS of more complex systems having
at least two tensor copies of
the harmonic oscillator as an underlying system
(Landau problem with harmonic potential,
$n$-level system with at least a two-bosonic
modes such as the $n$-mode Jaynes-Cummings model)
or matrix VCS \cite{ali-englis-gazeau}.

In addition, we emphasize that

(a) we  perform the investigations not at the CS
but at the VCS level.
The reason for that is the following: the resolution
of the identity of VCS is, in a sense, weaker than
the one of ordinary CS.
Hence, we expect to solve more classes.

(b) as far as we are concerned with our current
analysis, we  restrict the sense of
{\it solvable} VCS to {\it normalizable} VCS,
namely with $\mathcal{N}(J)<\infty$,
and satisfying a \emph{partial} resolution of the identity
on the Hilbert space. The continuity in label
will be obvious. However, if the procedure  only ensures
that the VCS  satisfy these basic requirements,
it is not excluded at all that their properties
could be improved with respect to Gazeau-Klauder
physical axioms \cite{gazeau-klauder}.
For instance, both temporal stability
could be implemented with extra parameters,
taken case by case, and action angle constraints
investigated afterwards.

 The outline of paper is the following.
Section \ref{sect:revis} is devoted to a pedagogical
review of the main aspects of solvable classes of CS and VCS
associated with the harmonic oscillator. These aspects
give farther motivations for this work. Next, we briefly recall the
work performed in \cite{thirulo,thirulo-saad} fitting it with our specific
notations and main objectives.
Then Section \ref{sect:ho2D} initiates the analysis
of VCS with one degree and two degrees of freedom
for the $2D$ harmonic oscillator.
Starting first with simple cases of one degree of freedom,
we go into the analysis in depth, revealing
some internal deformation structure and symmetries between
 the classes of VCS. The connection with the Landau
problem is discussed also therein.
Section \ref{sect:ho3D}
focuses on the VCS of the $3D$ harmonic oscillator.
The discussion is directly settled on two degrees
of freedom while the case of three degrees of freedom
is slightly mentioned, for the sake of brevity.
Section \ref{concl} provides
a summary of our results and an outlook of this
work. Finally,  an appendix collects complementary
proofs and identities used in the text.

\section{Revisiting harmonic oscillators:
the VCS method}
\label{sect:revis}

One of the  notorious  forms of
coherent states of the quantum harmonic oscillator
in $1D$ can be written
\bea
|z\rangle = \mathcal{N}^{-\frac12}(z) \sum_{n=0}^\infty
\frac{z^n}{\sqrt{\rho(n)}} |n \rangle,
\label{coh}
\eea
where $z\in \C$, $\mathcal{N}(z)$ is a normalization factor,
$\{ |n\rangle, n\in \N \}$
forms an eigenstate basis of
the number operator $N|n\rangle =n|n\rangle$,
$N=a^\dag a$ associated with the Heisenberg operators
$a$ and $a^\dag$ obeying $[a,a^\dag]=\I$.
The function
$
\rho(n) = \prod_{i=0}^{n} x_{i}$,
$\rho(0):=1$,
goes under the name of generalized factorial, with arguments
$x_i= x(i)$ which are energy-built quantities related to the spectrum
of the system.

Gazeau and Klauder \cite{gazeau-klauder} proposed a set
of axioms that can be implemented on the states (\ref{coh})
before calling these states CS. We will focus on two of
them: (a) a normalization condition, meaning that the
normalization factor
\bea
\mathcal{N}(z) = \mathcal{N}(|z|)= \sum_{n=0}^\infty
\frac{|z|^{2n}}{\rho(n)}
\eea
should be finite and (b) CS should satisfy a
resolution of the identity, i.e. there exists
a measure $d\mu(z)$ such that, on a complex domain
$\mathcal{D}\subset \C$,
\bea
\int_{z\in \mathcal{D}} d\mu(z) |z\rangle \langle z| = \sum_{n=0}^\infty |n\rangle \langle n| .
\eea
These two axioms are clearly mathematical statements
that one can roughly summarize as to be a proof of existence (a)
and the fact that this set is an overcomplete basis (b).

It is then striking that to satisfy both axioms intimately depends
 on the content of $\rho(n)$.  Indeed,
requiring (a) is equivalent to have $\overline\lim_{n\to \infty} \,
^{n} \sqrt{\rho(n)}=R\neq 0$ meanwhile,
using for instance polar
coordinates to parameterize the complex plane $z=re^{i\theta}$,
and a measure factorized as $d\mu(z) = (1/\pi) \mathcal{N}(r) rdrd\theta \varrho(r)$,
with $\varrho(r)$ a positive density function of unit weight,
 what boils down in (b) is simply a Stieljes like moment problem
\bea
\int_{r\in [0,R)} 2rdr\varrho(r) r^{2n} = \varrho(n).
\eea
Dealing with the well known canonical CS, we set
$x_n = n$ representing  directly the energy level itself,
the generalized factorial $\rho(n)=n!$,
the radius of convergence of the norm series is infinite
and the resolution of the identity turns out to
be solved by $\varrho(r)=\exp[-r^2]$.
Note that $x_n=n$ is intimately rooted in
group theoretical considerations since this choice
appears to be the one associated with the expansion of
CS using a displacement type operator
$|z\rangle =e^{za^\dag - \bar z a}|0\rangle$.
A natural question is then:
What can one put in $\rho(n)$ without breaking the
normalizability and integrability of the CS ?

Remarkably, in the search of solvable classes of CS even for simple harmonic oscillator or related systems invoking only\footnote{We exclude
here any general group theoretical
consideration {\it{\`a la}} Perelomov \cite{perel}.}
Heisenberg algebras
and their deformations as an underlying group theoretical framework, only
a few number of classes was achieved.
In a broad view, models with solvable sets of CS include
the model of a particle in a plane subject  to a magnetic field
also called the Landau problem \cite{landau2},
atomic $n$-level systems and quantum optics models
such as the Jaynes-Cummings model \cite{jcum},   to mention but a few. Let us emphasize some other instructive systems. The $f$- \cite{jannus},
 $q$- \cite{ar,bie} and $(q,p)$- \cite{cha}
deformed harmonic oscillators have been solved and their CS studied. Besides, CS of the
Landau problem \cite{fak},
Jaynes-Cummings model \cite{hus1,hus2}
and its deformed versions
were also exactly solved (\cite{chan,previous,bal2} and
see more references therein).
In a quantum deformed framework,
$\rho(n)$ can be related to the eigenvalue  $\{n\}$
of the deformed number operator $\{N\}$.
Hence, deforming the algebra in a well
controlled way is an  acknowledged efficient
way to map a set of solvable non deformed CS
to a set of solvable  deformed CS.

The inception of VCS \cite{ali-antoine-gazeau}
for a higher rank tensor-like Hilbert space has enable
to extend the notion of CS and thereby to achieve more
in the quest of solvable classes of CS of physical model.
Indeed, it has been highlighted a significant number
of physical appearances of VCS
\cite{ab1}. Moreover, the VCS formalism has been put forward
for quantum systems with many degrees of freedom
including two-level systems with possible degeneracy
\cite{ab1,hus2}. The resulting states
remain integrable when fully deformed and prove
to preserve regular properties of CS \cite{joben1,joben2,joben3,joben4}.
More theoretically, they have
opened the door to a wide range of applications
by extending the notion of CS defined with
a unique complex variable $z$ to the notion of CS defined over complex matrix $\mathcal Z$ (shortly called matrix VCS or MVCS) \cite{ali-englis-gazeau}, quaternions and complex tensor domains \cite{thirulo}. By extending the Barut-Girardello eigenvalue problem $a|z\rangle=z|z\rangle$, to a matrix eigenvalue problem of the type
$a|\mathcal Z \rangle= \mathcal Z|\mathcal Z \rangle$,
the associated with MVCS have meaningful consequences
at the group representation level.
In a nutshell, the VCS formalism gives
a new point of view of the CS definition:
it considers each set of CS as  \emph{embedded} in a Hilbert
subspace $\mathcal H_k$ of a larger Hilbert space $\mathcal H=\otimes_k \mathcal H_k$, reaping the benefit of the higher rank structure of the latter.

More closely related to our present concern,
it has been unraveled in \cite{thirulo,thirulo-saad}
new classes of VCS. Let us give a digest of these
results which will be at the basis of our ensuing construction.
Consider the operator
(in units such that $c=1$)
\bea
H= \frac{1}{2m} (\vec p - \text{e} \vec A)^2, \qquad \vec p = -i \hbar \vec \nabla,
\label{landau}
\eea
 describing the motion in an infinite layer of width $d$,
namely $\Sigma = \R^2 \times [0,d]$,
of a particle of mass $m$ and charge $\text{e}$ subject to a
magnetic field of vector potential taken in the symmetric gauge form  $\vec A = (1/2) \vec B \times \vec r$, where $\vec B =(0,0,B)$ is the magnetic field, $\vec r=(x,y,z)$
 the coordinate position.
Moreover, the state function $\psi(x,y,z)\in L^2(\Sigma)$
satisfy Dirichlet boundary conditions $\psi(x,y,z=d)=0=\psi(x,y,z=0)$,
$(x,y)\in \R^2$, $z\in [0,d]$ being the height.
The Hamiltonian (\ref{landau}) proves to be  diagonalizable
with energies given by (considering only null or negative angular
momentum modes  $l\leq 0$,
the spectrum becoming infinitely degenerate)
\bea
 E_{k,(l\leq 0), n} =: E_{k,n}=\omega_1( 2k +1) +\omega_2\left(\frac{\pi (n+1)}{d} \right)^2
\eea
where $\text{e}|B|/(2m)=\omega_1$ is the cyclotron frequency,
$k$ labels the Landau levels, $ \hbar^2/2m=\omega_2$ and
$n$ are simply the frequency and quantum number
associated with the motion of a particle on a segment under the same boundary conditions. We will omit to report the orthonormal eigenfunctions keeping only the formal expression $\psi_{k,n}:=\psi_{k,(l\leq 0),n}$ for sake of simplicity.

The next stage was to define classes of CS.
Keeping $n_2$ fixed and writing
\beq
&&
E_{n_1,n_2} = (2\omega_1) \left[ n_1 + \frac12 +
\frac{\omega_2}{2\omega_1} \left(\frac{\pi (n_2+1)}{d} \right)^2\right] , \crcr
&&
\rho(n_1,n_2) = (2\omega_1)^n_1 (\gamma)_{n_1},
\quad
\gamma =
1+
\frac{ \omega_1 d^2 + \omega_2\pi (n_2+1)^2}{2\omega_1 d^2},
\quad (a)_{n} = \frac{\Gamma[a+n]}{\Gamma[a]},
\eeq
where $\Gamma$ is the Euler gamma function
and $(a)_n$ the Pochhammer symbol, and
using just one variable $z$,\footnote{
Note that the authors used action-angle variables $(J,\alpha)$
defining their CS. We reformulate all their results using our notations
and our considerations pertaining to the only
two mentioned axioms. }
the first kind of state can be defined with one summation in $n_1$
\bea
|z,n_2\rangle = \mathcal N^{-\frac12}(z,n_2)
\sum_{n_1=0}^\infty \frac{z_1^{n_1}}{\sqrt{\rho_1(n_1,n_2)}} |\psi_{n_1,n_2}\rangle.
\label{thirsta}
\eea
Due to the presence of the \emph{vector} index $n_2$,
one can actually call $|z,n_2\rangle$ as VCS provided
it is normalizable and satisfies a resolution of the identity.
The normalization factor is reduced to a hypergeometric
function $\fun\big(1,\gamma; |z|^2/(2\omega_1)\big)$
converging everywhere in $[0,\infty)$.
The state (\ref{thirsta}) also obeys a resolution of
identity in the sense of VCS, i.e. a partial
resolution of the identity of the entire Hilbert
space:
\beq
&&
\int d\mu(z,n_2) |z,n_2 \rangle \langle z,n_2| = \sum_{n_1=0}^\infty
|\psi_{n_1,n_2}\rangle \langle \psi_{n_1,n_2} |, \\
&&
d\mu(z,n_2) =\mathcal N(r,n_2) rdrd\theta \varrho(r,n_2),
\qquad \varrho(r,n_2) = \frac{r^{2(\gamma-1)}}{(2\omega_1)^\gamma\Gamma[\gamma]} \exp\left\{-\frac{r^2}{2\omega_1} \right\},
\eeq
where in the last equation we have introduced a polar
parametrization of $z=re^{i\theta}$.

A second kind of VCS was introduced by switching the role
of $n_1$ and $n_2$, {\it viz} fixing $n_1$ and then
summing $n_2$, and defining
\beq
&&
E_{n_1,n_2} =\omega_2\left(\frac{\pi}{d}\right)^{2}
\left[\frac{id}{\pi} \sqrt{\frac{\omega_1}{\omega_2}}
\sqrt{2n_1+ 1} + n_2+1\right]\left[\frac{-id}{\pi} \sqrt{\frac{\omega_1}{\omega_2}}
\sqrt{2n_1+ 1} + n_2+1\right]  , \crcr
&&
\rho(n_1,n_2) = \left(\frac{\pi}{d}\right)^{2n_2} (\beta)_{n_2}
(\bar\beta)_{n_2} \qquad
\beta = 2 + \frac{i\, d}{\pi} \sqrt{\omega_1(2n_1 + 1)}.
\eeq
The resulting state which
can be naturally written  as $|z,n_1\rangle$
is normalizable and, if integrated, gives another
partial identity in the second sector.
These CS were the first types of CS, VCS in fact,
defined with one degree of freedom issued from the system.
Indeed, one can proceed farther and introduces
the second species of CS (these states cannot be called VCS since
they do not possess a vector dependence), by
considering two degrees of freedom
associated with each sector $n_1$ and $n_2$, respectively.
To make matters worse, one can couple the sectors each to other: the sums performed on
a unique label $n_i$ of one sector become dependent on the label $n_j$ of another
sector  through the generalized factorial
$\rho(n_1,n_2)$. It becomes a non trivial
issue to prove the CS axioms in this case.
The following cases prove to be solvable: for  \emph{independent} sums, (i.e. the sum over the label of one sector does
not depend on another sector label),
\beq
&&
\rho_1(n_1) = (2\omega_1)^{n_1} \left(\frac32\right)_{n_1} ,\quad
\text{by factorizing the partial energy} \quad
e_{n_1} = 2\omega_1\left( n_1 + \frac12\right) ,\crcr
&&
\rho_2(n_2) = \left(\frac{\pi}{d}\right)^{2n_2} (2)_{n_2} \, (2)_{n_2},
\quad
\text{by factorizing the partial energy} \quad
e_{n_2} =\left(\frac{\pi}{d}\right)^{2}  (n_2 + 1), \crcr
&&
\eeq
whereas for \emph{dependent} sums, the following
quantities lead to well defined CS:
\beq
&&
\rho_1(n_1,n_2) = \left(\frac{\pi}{d}\right)^{2n_2} (\beta)_{n_2}
(\bar\beta)_{n_2}  ,\quad
\text{by factorizing the total energy} ,\\
&&
\rho_2(n_1) = (2\omega_1)^{n_1} \left(\frac32\right)_{n_1} ,\quad
\text{by factorizing  the partial energy} \quad
e_{n_1} = 2\omega_1\left( n_1 + \frac12\right) .
\nonumber
\eeq
Using these states, the authors then discussed the axioms of temporal stability and action-identity. The same ideas
can be found in \cite{thirulo} dealing with  another kind
of physical model: the two-mode Jaynes-Cummings model.

Clearly, by  simple combinatorics,
the picture is far to be complete:
there are many cases which remain to be studied.
Furthermore, for a more simple situation
and not even for the above mentioned Landau-like problem,
 similar ideas could be applied and might
lead to results not yet investigated to the best of our
knowledge. Indeed,
we can simplify the analysis by considering
 the simple harmonic oscillator in
$2D$ and write its dimensionless energy spectrum as
\bea
E_{n_1,n_2} = e_{n_1,n_2} -\text{const.} = \omega_1(n_1 +\frac{ \omega_2}{ \omega_1} n_2)
 =  \omega_2(n_2 +\frac{ \omega_1}{ \omega_2} n_1).
\eea
According to the formalism so far, we can built four
generalized factorials associated with the two harmonic
subsystems
\bea
\rho_{1,2}(n_{1,2}) = (\omega_{1,2})^{n_{1,2}} n_{1,2} !,
\qquad
\rho_{1,2}(n_{1,2},n_{2,1}) = (\omega_{1,2})^{n_{1,2}} (\gamma_{1,2})_{n_{1,2}},
\qquad
\gamma_{1,2} = 1 + \frac{\omega_{2,1}}{\omega_{1,2}} n_{2,1}.
\eea
Unexpectedly, the number of solvable
VCS classes which can be built from
these quantities is really significant.
Then arises a question: Is there a definite way
to understand these classes of CS and to give them
a substantive structure?

The specific purpose of this paper is the following:
Using a rigorous combinatorics,
provide the largest possible set of VCS
of the harmonic oscillator in $2D$ and $3D$
fulfilling the normalizability and resolution of the
identity requirements. We also investigate a way to classify the
VCS,
to explore possible  links between them and thereby
giving them a sense on their own. Our formulation completes
in a more precise way the aforementioned
study and furthermore improve the formulation
of both \cite{gazeau-novaes} and \cite{thirulo,thirulo-saad}
when restricted to the harmonic oscillator.

\section{2D Harmonic oscillator}
\label{sect:ho2D}

Consider the Hilbert space of the quantum harmonic oscillator
in $2D$:
\bea
\mathcal{H}_{2D} = \text{span}\left\{ |n_1,n_2\rangle, \;
 n_i \in \mathbb{N} \right\},
\eea
where $ |n_1,n_2\rangle$ is the two-mode eigenstates of the
bosonic  number operators $N_i = a_i^\dag a_i$ of two decoupled
Heisenberg algebras obeying $[a_i,a_i^\dag] = \mathbb{I}_i$.
The dimensionless Hamiltonian $H_{2D}$ associated with
this system, including different frequencies $\omega_i$ for each
sector, and its eigenvalues in this basis can be written as
\beq
H_{2D} = \frac{1}{\hbar}H'_{2D} =  \sum_{i=1,2}\omega_i (a_i^\dag a_i  + \frac12) \;,\qquad
e_{n_1,n_2} = \omega_1 n_1 + \omega_2 n_2 +\frac12 (\omega_1 + \omega_2).
\eeq
One can shift the Hamiltonian $H_{2D}$ by the
constant $-(1/2)(\omega_1 + \omega_2)$ giving
the operator $\widetilde{H}_{2D}$ with
eigenvalues $E_{n_1,n_2} = \omega_1 n_1 + \omega_2 n_2$.
At the end, we will come back on the consequences
of having an unshifted spectrum.
We will construct various solvable
classes of VCS spanning $\mathcal{H}_{2D}$ by scrutinizing the
\emph{two-tower}
structure of the eigenstates regarding the energies $E_{n_1,n_2}$.

In the sequel, Subsection \ref{sub:ho1d1c}
is quite well-known but for completeness purpose
it is convenient to include it as a starting point.
Subsection \ref{sub:ho1d2c} follows ideas of
\cite{thirulo} and \cite{gazeau-novaes} studying multidimensional
CS (i.e. CS with many parameters) on the same type of Hilbert spaces. We  also introduce therein some taxonomy.
Subsection \ref{sub:1ddual}
starts our analysis: we improve the formulation of the
above works by noting some useful facts
shaded in these prime studies
which will enable us to systematize the determination
and classification of the VCS with two
degrees of freedom in Subsection \ref{sub:ho2d}.

\subsection{VCS with one degree of freedom}
\label{sub:1d}

We recall some terminology:
 a VCS \emph{ degree of freedom} is a variable $\mathcal{Z}$ belonging to some continuous domain in terms of which the VCS is expanded.
Dealing with the harmonic oscillator in this section,
VCS will be defined with one degree of freedom $\mathcal{Z}=z$ which simply stands for a complex variable.
Two distinct classes of VCS are introduced below.
They are built on the Hilbert subspace
spanned by one tower $i=1$ or $2$,
 the other sector being maintained fixed.
The ensuing calculations are performed by selecting the tower with label
$n_1$, and, obviously, to each highlighted class  corresponds another
set of VCS obtained by choosing instead the tower coined by $n_2$
and doing the calculation.
We will review, in a pedagogical spirit, the elementary constructions in order to prepare the reader to more combinatorial developments induced by an increasing number of degrees of freedom.

\subsubsection{First class: Canonical CS}
\label{sub:ho1d1c}

This class (and the similar one in the case of higher number
of degrees of freedom) corresponds to a straightforward
extension of ordinary canonical CS associated with annihilation
operator eigenvalue problem for the $1D$ harmonic oscillator.
The set of VCS is therefore merely built with pure factorials
and, in particular, for one degree of freedom, with a unique
factorial.

We define
\beq
\rho(n_{1})
= (\omega_{1})^{n_1}\, n_{1} !\,,
\label{hoqua01}
\eeq
and consider the set of states
\beq
|z, n_{2}\rangle
= \mathcal N(z)^{-\frac12}
\sum_{ n_{1} = 0}^{\infty}  \rho(n_{1})^{-\frac12}
z^{n_{1}}|n_{1}, n_{2} \rangle,
\label{ho1d2c}
\eeq
 $z$ being a complex variable.
The next stage is to normalize these states and to find a resolution of
the identity that they should satisfy.

The normalization to unity of the states (\ref{ho1d2c}) is fulfilled under the condition
\beq
\langle z, n_{2}|z,  n_{2}\rangle = 1
\Leftrightarrow
\mathcal N(|z|)
= \exp\left\{ \frac{|z|^{2}}{\omega_{1}}\right\}.
\eeq
The states (\ref{ho1d2c}) must form also an overcomplete basis of states
and so we seek for a measure $d\mu(z)$ such that a \emph{partial} resolution
of the identity\footnote{In order to obtain the resolution of the identity
on the entire Hilbert space $\mathcal{H}$ one should sum over the
index $n_2$ in (\ref{horesol2}).} should be satisfied:
\beq
 \int_{\mathcal D}
|z,  n_2 \rangle \langle  z, n_2|
 \, d\mu(z)
= I_{n_2},
\label{horesol2}
\eeq
where $I_{n_2}$ is the projector onto the subspace of $\mathcal H$
obtained by keeping $n_2$ fixed,
namely
\beq
I_{n_2}
 = \sum_{n_1=0}^\infty
|n_1,n_2\rangle \langle n_1,n_2|.
\label{hoprojec}
\eeq
 Using polar coordinates for the variable $z =r e^{i \theta}\in \C$, the measure is of the form
\beq
d\mu(z) = \frac{1}{\pi}
{\mathcal N}(r)
\varrho(r) r\; drd\theta.
\eeq
The integration domain is $\mathcal D=\C$ since the norm converges everywhere.
The relation (\ref{horesol2}) translates into the Stieljes moment problem
\beq
\label{homom05}
2\int_{0}^{\infty} \, r^{2n_1 +1}\varrho(r)dr
= (\omega_{1})^{n_1}n_1 !,
\eeq
which is solved by the density
\beq
\label{hodens01}
\varrho(r) = \frac{1}{\omega_{1}}
\exp\left\{-\frac{r^{2}}{\omega_1}\right\}.
\eeq
The above scenario is straightforward from what
one could expect for getting canonical CS. As was claimed
at the very beginning, the class of VCS (\ref{ho1d2c}) can be
simply viewed as CS of a harmonic oscillator in $1D$
\emph{attached} (i.e. tensored) to some fixed vector of an
abstract Hilbert space: $|z,n_2\rangle = |z\rangle \otimes |n_2\rangle$.

\subsubsection{Second class: $\gamma$-deformed CS}
\label{sub:ho1d2c}
The second class VCS cannot be simply associated with
ordinary canonical CS as it was the case for the first class VCS.
To be properly defined,
here the system of VCS requires at least a harmonic oscillator in $2D$ with unbalanced frequencies in each direction. On the computational side,
the states do not involve a simple factorial but a \emph{generalized}
factorial: the Pochhammer symbol.
Dealing with one degree of freedom, we just have one
such a symbol.

One starts by observing that, setting $\omega_i\neq 0$, the eigenenergies can be factorized
as
\beq
E_{n_1,n_2} =
\omega_{1}\left[n_1  + \frac{\omega_{2}}{\omega_{1}}n_2 \right],
\label{var01}
\eeq
and fixing again the tower labeled by $n_2$, the following
quantities can be defined
\beq
&&
\rho(n_1)
= \prod_{k=1}^{n_1} \omega_{1}\left[k + \frac{\omega_{2} }{\omega_1 }\,n_2\right]
= (\omega_{1})^{n_1} (\gamma)_{n_1}, \label{hoqua00}\\
&&
\gamma = 1 + \frac{\omega_{2}}{\omega_{1}}n_{2}, \qquad
(\gamma)_{n_{1}} = \frac{\Gamma(n_{1} + \gamma)}{\Gamma(\gamma)},
\label{hogamma}
\eeq
where $(\gamma)_{n}$ stands for the Pochhammer symbol. Note that, implicitly, $\gamma$ depends on $n_2$.

We introduce the set of vectors\footnote{We will always use the same
notation for different VCS made with
the same dependencies. For instance the VCS (\ref{ho1d2c}) and (\ref{ho1d1c}) are both denoted $|z, n_2 \rangle$. This
is to avoid useless proliferation of notations and,
as noted, the sense they will refer to remains unambiguous.}
\begin{equation}
|z, n_2 \rangle = \mathcal N(z,n_2)^{-\frac12}
\sum_{ n_1 = 0}^{\infty}  \rho(n_1 )^{-\frac12}
z^{ n_1 } |n_{1}, n_2 \rangle
\label{ho1d1c}
\end{equation}
which can be normalized to unity according to
\begin{equation}
\label{honorm03}
\langle z, n_2 |z, n_2 \rangle = 1,
\qquad
\mathcal N(|z|, n_2 ) =\sum_{n_1=0}^\infty
\frac{1}{(\gamma_1)_n} \frac{|z|^{n}}{\omega_1^n}
=  \fun\left(1;\gamma;\frac{|z|^{2}}{\omega_{1}}\right),
\end{equation}
where $\fun (\cdot)$ denotes the
ordinary confluent hypergeometric function.
The convergence radius of the series $\fun$,
as for any other hypergeometric function, can be determined by
a simple ratio test (or by comparison test since $1/\Gamma[\gamma_1 + n_1]\leq 1/n_1!$ for $\gamma\geq 1$.). It can be checked that (\ref{honorm03}) converges everywhere in the complex plane.
The form for this series can be given in general by
\bea
\fun\left(1;a;z\right) =
e^z z^{-a } ( \Gamma [a
   +1]-a\Gamma [a,z] ),
\eea
where
$ \Gamma [a,z]= \int_z^{\infty} t^{a-1} e^{-t}dt$
is the incomplete Euler-gamma function with ordinary
conditions on the complex number $a$.

On the domain $\mathcal D = \C$, consider the measure
\begin{equation}
d\mu(z, n_2) = \frac{1}{\pi}
{\mathcal N(z, n_2)}
\varrho(r, n_2)\, r drd\theta.
\end{equation}
The set of states (\ref{ho1d1c}) ought to satisfy the partial resolution of the identity
\begin{equation}
\int_{\mathcal D}
|z, n_2 \rangle \langle z, n_2 |
\; d\mu(z,n_2)
= I_{n_2},
\label{horesol1}
\end{equation}
where $I_{n_2}$ is again the projector (\ref{hoprojec}).
From (\ref{horesol1}), one infers the moment problems
\begin{equation}
\label{homom00}
2\int_{0}^{\infty}\,r^{2n_1+1}\varrho(r, n_2) dr_{j} = \rho(n_{1}),
\end{equation}
solved by
\begin{equation}
\label{hodens00}
\varrho(r, n_2) = \frac{1}{\Gamma(\gamma)(\omega_{1})^{\gamma}}
r^{2(\gamma - 1)}\exp\left\{-\frac{r^{2}}{\omega_{1}}\right\}.
\end{equation}
Hence the states (\ref{ho1d1c}) define a different class of VCS.

\subsubsection{Lessons from the construction with
one degree of freedom}
\label{sub:1ddual}

Let us start our deepening analysis of these results
that will enable us to improve the above procedure
of building the VCS and, from that, extending them for more
degrees of freedom.

There is a \emph{dual} picture to the above construction of VCS,
as performed in Subsections \ref{sub:ho1d1c} and \ref{sub:ho1d2c},
that is interesting to point out and to investigate for its properties.
Indeed, the role of $n_1$ and $n_2$ being interchangeable,
this implies that the other generalized factorial
\beq
\rho (n_2) = (\omega_2)^{n_2} \,n_2!
\label{genfact}
\eeq
could equally serve to construct another set of VCS with one degree of freedom and meeting all requirements, that we call
\emph{dual} class associated with (\ref{ho1d2c}) and that we denote
\beq
|z, n_2\rangle^*  =
|z, n_1\rangle
= \mathcal N(z)^{-\frac12}
\sum_{ n_2 = 0}^{\infty}\rho(n_2)^{-\frac12}
z^{n_2}|n_{1}, n_{2} \rangle.
\label{ho1d2cdual}
\eeq
One notes that the initial class and its dual  are both of first class.
The first class therefore contains two canonical sets of VCS in addition
with the similar one  when we will be dealing with
higher number of degrees  of freedom.

In analogy with the dual VCS (\ref{ho1d2cdual}),
a second dual class of the set of VCS (\ref{ho1d1c}) can be built.
Again by switching $(1\leftrightarrow 2)$, the number $\gamma$ (\ref{hogamma}) has a dual counterpart
\beq
\gamma_1 = 1 + \frac{\omega_2}{\omega_1} n_2, \qquad
\gamma_2 = 1 + \frac{\omega_1}{\omega_2} n_1 .
\eeq
Within this framework, since all derivations remain the same,
under $(1 \leftrightarrow 2)$,
a set of VCS dually associated with (\ref{ho1d1c}) can be generated.
We have:
\beq
|z, n_2 \rangle^*_{\gamma} =
|z, n_1\rangle_{\gamma} ,\qquad
|z, n_1 \rangle_{\gamma}=
\mathcal N(z, n_1 )^{-\frac12}
\sum_{n_2 = 0}^{\infty}  \rho(n_2)^{-\frac12}
z^{ n_2 } |n_{1}, n_{2}\rangle,
\label{ho1d1cdual}
\eeq
where we add an index $\gamma$ in order to distinguish
 the above VCS second class from the first class one.

One notices the following interesting fact which has been
never discussed so far in the literature, to the best
of our knowledge.
At the limit
$\gamma_{1,2} \to 1$ (limit when one of the two frequencies
becomes much greater than the other,
namely $\omega_{1,2}>> \omega_{2,1}$), the first class set of VCS (\ref{ho1d1c})
and its dual (\ref{ho1d1cdual})
smoothly tend  to the set of VCS (\ref{ho1d2c}) and its dual (\ref{ho1d2cdual}), respectively,
with smooth measure deformations:
\beq
\lim_{\gamma_{1,2} \to 1} \frac{1}{\omega_{1,2}^{\gamma_{1,2}}\Gamma(\gamma_{1,2})}r^{2(\gamma_{1,2}-1)}
\exp\left\{-\frac{r^{2}}{\omega_{1,2}}
\right\} = \frac{1}{\omega_{1,2}}\exp\left\{-\frac{r^{2}}{\omega_{1,2}} \right\}.
\eeq
Hence the name of $\gamma$-deformed VCS.
For one degree of freedom, each of the two classes of VCS may
define a unique set of VCS in that particular limit.

Two VCS classes are said to be of the same \emph{type}
if there exists a continuous limit under which one of them
can be mapped onto the other. The first on which the
limit is performed will be called \emph{ancestor}
and the second resulting state \emph{descendant}.
Thus, an ancestor and a descendant are of the same type.
For example, above, the second classes are ancestors
while the first class limits are descendants.

\noindent{\bf Remark 1 -} The construction of some  VCS classes starting from
building at first the second class VCS (ancestor) might be more efficient
since, by a large frequency limit, one could deduce
the corresponding first class (descendant) of the same type.

Another relevant remark on the above construction
is the following: consider the moment problem for the
second class that is given by
\beq
\label{homoment}
\int_{0}^{\infty}\,r^{2n_{1}+1}\varrho(r, n_{2}) dr
= (\omega_1)^{n_{1}}(\gamma_1)_{n_1},
\eeq
with solution
\beq
\label{hodensity}
\varrho(r, n_{2}) = \frac{1}{\Gamma(\gamma_1)(\omega_{1})^{\gamma_1}}
r^{2(\gamma_1 - 1)}\exp\left\{-\frac{r^{2}}{\omega_{1}}\right\}.
\eeq
A closer look on this expression (\ref{hodensity}) shows that $\gamma_1 -1 = \frac{\omega_2}{\omega_1} n_2$.
Then, we would like to trade an extra factor in
the measure, say $r^{2(\gamma_1 - 1)}$,
for an extra variable in the VCS. In order to do so,
let us introduce the new and modified generalized factorial (to be compared with (\ref{hoqua00}))
\beq
\rho(n_1,n_2)
= \omega_1^{n_1 + \frac{\omega_2}{\omega_1}  n_2} \;\Gamma(\gamma_1) (\gamma_1)_{n_1}
= \omega_1^{n_1 + \frac{\omega_2}{\omega_1}  n_2}
 \Gamma(\gamma_1+ n_{1})
\label{hofactosmart}
\eeq
for which a second class of VCS can be defined as well as
\beq
&&
|z, n_2\rangle' = \mathcal N(z,n_{2})^{-\frac12}
\sum_{ n_{1} = 0}^{\infty}   \rho(n_{1},n_{2})^{-\frac12}
z^{ n_{1}+\frac{\omega_2}{\omega_1} n_2}\, |n_{1}, n_{2} \rangle,
\label{ho1d1csmart} \\
&&
\mathcal N(|z|, n_{2}) = \frac{1}{\Gamma(\gamma_1)}
\left[\frac{|z|^{2}}{\omega_1}\right]^{\frac{\omega_2}{\omega_1}n_2}
 \fun\left(1;\gamma_1;\frac{|z|^{2}}{\omega_{1}}\right) .
\label{honormsmart}
\eeq
The measure density integrating to unity these states at fixed $n_2$
has to be solution of
\beq
\label{homomsmart}
2\int_{0}^{\infty}\,r^{2(n_{1} + \frac{\omega_2}{\omega_1}n_2)+1}\varrho(r,n_2) dr =
\omega_1^{n_1 + \frac{\omega_2}{\omega_1}  n_2} \; \Gamma(\gamma_1+ n_{1})
\eeq
yielding a simpler formula
\beq
\label{hodenssmart}
\varrho(r) = \frac{1}{\omega_{1}}\exp\left\{-\frac{r^{2}}{\omega_{1}}\right\}.
\eeq
Note that the two VCS, (\ref{ho1d1csmart}) and (\ref{ho1d1c}),
are simply connected by a factor
\bea
|z,n_2\rangle' = \left(\frac{z}{\sqrt{\omega_1}}\right)^{\frac{\omega_1}{\omega_2}n_2}
|z,n_2\rangle.
\eea
Finally, the choice (\ref{ho1d1csmart}) for defining the second class of VCS  does have the advantage to display how explicitly, in the limit $\gamma_{1,2} \to 1$ generated by $\kappa_{1,2}\to 0$,
the class of VCS (\ref{ho1d1csmart}) converges
to the class (\ref{ho1d2c}).
This is our

\noindent{\bf Remark 2 -} Defining the second class of VCS, use the
$\gamma$-modified generalized factorial (\ref{hofactosmart}).

Finally, there can be variant forms of the previous VCS
remaining still integrable to unity
that, for completeness purpose, one should also address
and list.
So far, we emphasized
the generalized factorials of the form
(\ref{hoqua01}) or (\ref{hofactosmart}) for building the VCS.
However, for one degree of freedom and
still summing only on $n_1$, there is some freedom
in the choice of the exponents of the prefactor
$\omega_1$ and the complex variable $z$.
Indeed, comparing (\ref{hoqua01}), (\ref{hoqua00})
and (\ref{hofactosmart}), we see that the exponent of $\omega_1$
changes of form.The same observation holds for
the exponent of the variable $z$ on which the different classes of VCS are based.
We can  think of them as new classes, called below \emph{sub-classes}, of VCS defined as (still summing on the tower $n_1$)
\bea
|z,n_2\rangle =
\mathcal N(z,n_{2})^{-\frac12}
\sum_{n_{1} = 0}^{\infty}   a(n_{1},n_{2},z)\, |n_{1}, n_{2} \rangle,
\label{subgener1d}
\eea
with general term $a(n_{1},n_{2},z)=\omega_1^{-\frac12(n_1 + \bullet)}
R^{-\frac12}(n_1,n_2) z^{n_1 + \bullet'}$ where $\bullet,\bullet' \in\{0, \kappa_1 n_2\}$ and
$R(n_1,n_2)$ is a (generalized) factorial.
 One may wonder if having introduced these sub-classes
is not in contradiction with the improvement procedure
of the previous remark. Clearly, doing so will have again
the effect to modify the exponents which could have led to a simple density solution of the moment problem for these states.
Hence, at this point, the answer is yes. However, in general,
we will see that proceeding in the same manner, when one has
 more degrees of freedom,
will have an effect during the integration
and will lead to new classes.

The measure
$d\mu(z,n_2)= (1/\pi)\mathcal{N}(|z|,n_2)$ $rdrd\theta\varrho(r,n_2)$
integrating to unity these variant states ($4$ for each class) can be determined by solving the following generalized moment
problem with parameters $\alpha,\beta,\alpha',\beta'$,
(to be fixed later),
\bea
2\int rdr\; \varrho(r,n_2)\;
\frac{r^{\alpha n_1 + \beta\kappa_1 n_2}}{
\omega_1^{\alpha' n_1 + \beta'\kappa_1 n_2}}
 = R(n_1,n_2) \;.
\eea
Its solutions  are given by:
\beq
&&
R(n_1,n_2) = n_1! \,,\qquad \varrho_1(n_1,n_2)   =
\frac{ \alpha }{ \omega^{\alpha'}_1 } \,
\left( \omega^{\beta'}_1r^{-2\beta} \right)^{\kappa_1 n_2}
\frac{1}{r^{2(1-\alpha)}} e^{- \frac{r^{2\alpha}}{\omega^{\alpha'}} } \;,
\crcr
&&
R(n_1,n_2) = \Gamma[n_1 + \gamma_1] \,,\qquad \varrho_1(n_1,n_2)   =
\frac{\alpha}{\omega^{\alpha'}_1}\,
\left( \omega^{\beta'-\alpha'}_1r^{2(\alpha-\beta)} \right)^{\kappa_1 n_2}
\frac{1}{r^{2(1-\alpha)}} e^{-\frac{r^{2\alpha}}{\omega^{\alpha'}}}
\eeq
so that (\ref{subgener1d}) defined with free parameters
\bea
a(n_{1},n_{2},z)=\omega_1^{-\frac12(\alpha' n_1 + \beta' \kappa_1 n_2)}
R(n_1,n_2) z^{\alpha n_1 + \beta \kappa_1 n_2}
\eea
 determines the most extended class of VCS
generating all sub-classes characterized  by
$\bullet,\bullet' \in\{0, \kappa_1 n_2\}$.
The parameters are to be fixed as
$\alpha,\beta,\alpha',\beta' \in \{0,1\}$.
Given $f(r,\omega)= (1/\omega) e^{-r^2/\omega}$,
the following tables yield the different sub-classes
generated by this restriction:

{\small
\begin{center}
\begin{tabular}{cc}
 &First class (1):  $R(n_1)=n_1!$ \\
\hline\hline
&\\
A &
$a(n_1,z)=[n_1!]^{-1/2} (z/\omega_1^{1/2})^{n_1}$\;;
\qquad   $\varrho(r)= f(r,\omega_1)$  \quad(\ref{ho1d2c}) \\
B& { $a(n_1,n_2,z)= [(\omega_1)^{\kappa_1 n_2 } n_1!]^{-1/2}  (z/\omega_1^{1/2})^{n_1}$ \;;
\qquad
$\varrho(r,n_2) = \omega_1^{\kappa_1n_2} f(r,\omega_1)$   }  \\
C& {  $a(n_1,n_2,z) =  [n_1!]^{-1/2}
(z/\omega_1^{1/2})^{n_1} z^{ \kappa_1 n_2}$ \;;
\qquad
$\varrho(r,n_2) = (r^{2\kappa_1 n_2})^{-1} f(r,\omega_1)  $ }  \\
D& {  $ a(n_1,n_2,z) = [n_1!]^{-1/2}  (z/\omega_1^{1/2})^{n_1+ \kappa_1 n_2}$ \;;
\qquad
$\varrho(r,n_2) =   [\omega_1/r^2]^{\kappa_1n_2}  f(r,\omega_1)$ }   \\
&\\
\hline\hline
\end{tabular}
\end{center}
}
{\small
\begin{center}
\begin{tabular}{cc}
 &Second class ($\gamma_1$-deformed): $R(n_1,n_2)=\Gamma[\gamma_1 + n_1])$\\
\hline\hline
&
\\
A &
 $a(n_1,n_2,z) = [\Gamma[\gamma_1+ n_1]]^{-1/2} (z/\omega^{1/2}_1)^{n_1+ \kappa_1 n_2}$ \;;
\qquad
$\varrho(r) =f(r,\omega_1)$ \quad (\ref{ho1d1csmart}) \\
B &
 $a(n_1,n_2,z)= [(\omega_1)^{ \kappa_1 n_2}\Gamma[\gamma_1+ n_1]]^{-1/2}  (z/\omega_1^{1/2})^{n_1}$ \;;
\qquad
$\varrho(r,n_2) = r^{2\kappa_1 n_2}f(r,\omega_1)$     \\
C &
   $a(n_1,n_2,z) = [\Gamma[\gamma_1+ n_1]]^{-1/2}
(z/\omega_1^{1/2})^{n_1} z^{ \kappa_1 n_2}$ \;;
\qquad
$\varrho(r,n_2) =\omega_1^{-\kappa_1n_2}
f(r,\omega_1) $
 \\
D &
   $a(n_1,n_2,z)= [ \Gamma[\gamma_1+ n_1] ]^{-1/2}
 (z/\omega_1^{1/2})^{n_1}$ \;;
\qquad
$\varrho(r,n_2)=
[r^{2}/\omega_1]^{\kappa_1 n_2}f(r,\omega_1)$ \quad
$\propto$ (\ref{ho1d1c})  \\
&\\
\hline\hline
\end{tabular}
\end{center}
}
As expected, the first and second sub-classes B, C and D
(denoted by $(1)$B, $(1)$C, $(1)$D and
$(\gamma_1)$B, $(\gamma_1)$C, $(\gamma_1)$D, respectively)
are not so enlightening: they simply appear as
factors of the VCS (first and second) sub-class of the kind A
(denoted by $(1)$A and $(\gamma_1)$A, respectively),
the latter being at the basis of the previous analysis.
Indeed, to get the sub-class $(1)$B, $(1)$C and $(1)$D,
one has just to multiply the VCS sub-class $(1)$A by the factor $\omega_1^{\kappa_1 n_2/2}$,
 $z^{\kappa_1n_2}$ and $(z/\omega^{1/2}_1)^{\kappa_1n_2}$,
respectively. Similar relations hold for the second
sub-classes $(\gamma_1)$A, $(\gamma_1)$B, $(\gamma_1)$C
and $(\gamma_1)$D, as it can be easily checked.
Note also that the second class VCS (\ref{ho1d1c})
coincides with the second sub-class $(\gamma_1)$D
up to a $\Gamma(\gamma_1)$ factor.
Hence defining (\ref{ho1d1csmart}) or (\ref{ho1d1c})
as a second class does not have any importance: they only differ
by a factor.

In the present instance, we introduce the following
definition: Given a number of degrees of freedom
and a VCS class, a sub-class of states is called \emph{irrelevant}
or \emph{a factor} if any of its VCS representative can be explicitly written
as a product involving a previous VCS of some different sub-class.
Otherwise, a sub-class is called \emph{relevant}
and will matter in our classification.
Note that it becomes a matter of choice to identify
a prime set of states from which one
determines if other sub-classes are factors of this set or not.

Dual classes can be introduced again by
$(1\leftrightarrow2 )$ and will share similar properties.
Last, all sub-classes introduced so far are of the same type
as sub-class $(1)$A (\ref{ho1d2c}) by observing the limit $\kappa_1 \to 0$.

\noindent{\bf Remark 3 -} A special combinatorics has to be taken
on the exponents of the frequencies $\omega_i$ and degree
of freedom $z_i$ as these could generate relevant VCS sub-classes.

\begin{figure}[htb]
  \centering
\begin{minipage}[t]{0.65\textwidth}
      \centering
\centering{
\includegraphics[angle=0, width=55mm, height= 20mm]{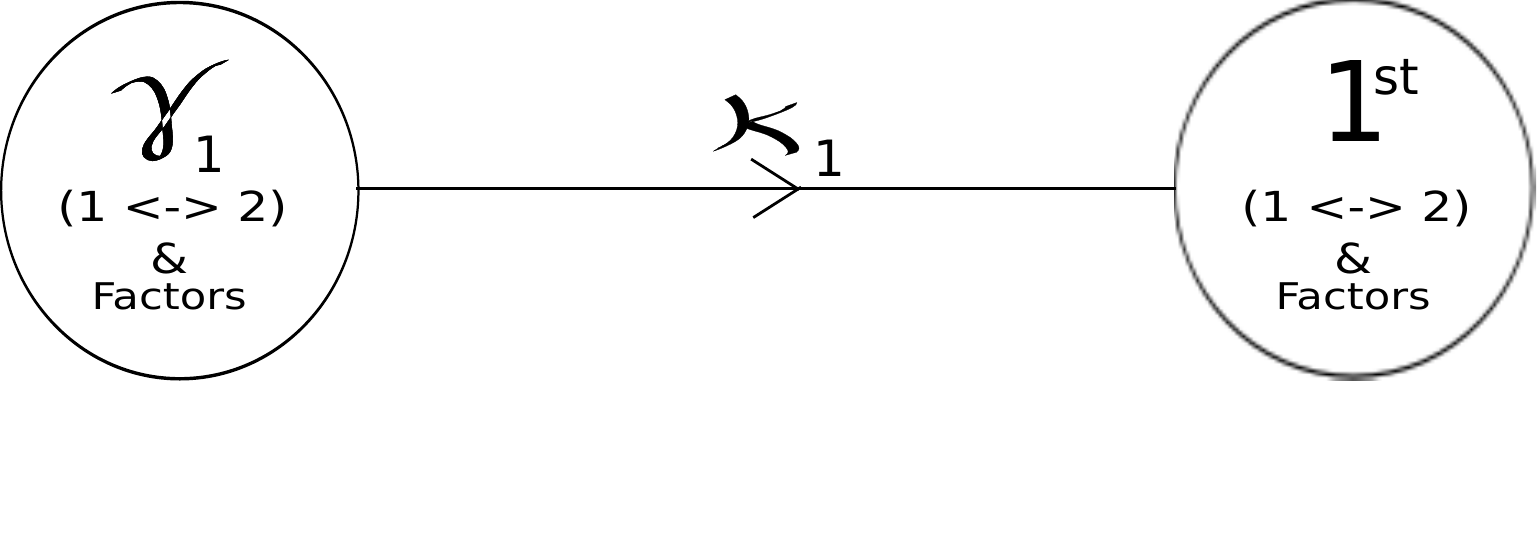}}
\caption{{\small The ancestor $\gamma_{1,2}$-deformed VCS class and its
descendant $1^{\text{st}}$-class limit as $\kappa_{1,2}\to \infty$ for one degree of freedom.}}
\label{fig:1d1c2c}
 \end{minipage}
\end{figure}

Figure \ref{fig:1d1c2c} gives a diagrammatical
summary of the VCS classes studied  so far as well as their dependence.

\subsection{VCS with two degrees of freedom}
\label{sub:ho2d}

In this subsection, the classes of VCS are equipped with
two complex variables,
so according to our definition, two degrees of freedom.
 There is another subtlety here due to
the fact that the sectors $1$ and $2$ can be coupled
or not in the definition of the generalized factorial.

\subsubsection{Second and first classes: Generators of bi-CS}
\label{sub:ho2d2c}

These classes are direct generalization of VCS first and second classes
 as defined in Subsections \ref{sub:ho1d1c} and \ref{sub:ho1d2c}.
By Remark 1, we notice that the first class can be deduced
from the second one, and so we will start by building the
second class.
Besides, the following construction is made using both the
towers $i=1$ and $2$ but in an asymmetric way.
The dual construction can be easily recovered by
switching the role of $1$ and $2$.
In the present case, computations involve one simple and
one generalized factorials.

\noindent{\bf Second class: $(\gamma,1)$-deformed VCS -}
Consider $n_{1}$ and $n_{2}$ fixed by the energy spectrum,
$\rho_{1}(n_{1},n_2)$ given by (\ref{hofactosmart}) that
is, in new notations,
\begin{equation}
\rho_1(n_1,n_2) =  (\omega_{1})^{n_1 + \kappa_1 n_2}
\Gamma(\gamma_1 + n_1),
\qquad
\gamma_{1} =\gamma_{1} (n_{2}) = 1 + \kappa_{1}\, n_{2}, \qquad
\kappa_{1} = \frac{\omega_{2}}{\omega_{1}},
\end{equation}
 and $\rho_{2}(n_{2})$ given by a simple factorial (\ref{hoqua01}).
Then, the set of states with two degrees of freedom
\beq
|z_{1}, z_{2}, n_2 \rangle =
\mathcal N(z_{1}, z_{2},n_2)^{-\frac12} \sum_{n_{1} = 0}^{\infty}
\left[\rho_{1}( n_{1},n_{2}) \rho_{2}( n_{2})\right]^{-\frac12}
z^{ n_{1} + \kappa_1 n_2}_{1} \,z^{n_{2}}_{2}\, |  n_{1}, n_{2}\rangle ,
\label{ho2d1c}
\eeq
where $z_i \in \C$, $i=1,2$, will draw our attention.

A direct inspection shows us that this state is a factor
of the second class VCS with one degree of freedom
(\ref{ho1d1csmart})
\bea
|z_{1}, z_{2}, n_2 \rangle
 =z_2^{n_2} [(\omega_2 )n_2!]^{-\frac12} |z_1,n_2\rangle.
\eea
However, having more degrees of freedom, it then defines
another relevant class. In fact, due to this extra degree of freedom,
(\ref{ho2d1c}) generates sub-classes which are not factors
of any of the sub-classes previously defined.

The states (\ref{ho2d1c}) satisfy the normalization condition
\begin{equation}
\langle z_{1}, z_{2},n_2  | z_{1}, z_{2},n_2 \rangle = 1,
\quad
\mathcal N(|z_{1}|, |z_{2}|,n_2) =
\frac{1}{\Gamma(\gamma_1)\,n_2!}
\left[\frac{ |(z_{1})^{\kappa_1} z_2 |^{2} }
 { (\omega_1)^{\kappa_1}\,\omega_2 }\right]^{ n_2}
\fun\left(1;\gamma_1;\frac{|z_{1}|^{2}}{\omega_{1}}\right) .
\end{equation}
which converges everywhere in the complex plane.
Defining the measure, in polar coordinate
$z_k = r_k e^{i \theta_k}$,
\begin{equation}
d\mu(z_{1}, z_{2},n_{2}) = \frac{1}{\pi^2}\;
{\mathcal N}(z_{1}, z_{2}, n_{2}) \;
\varrho_1(r_{1}, n_{2}) \, r_{1} dr_{1} d\theta_{1}\;
\varrho_2(r_{2}) \, r_{2} dr_{2} d\theta_{2}
\end{equation}
on $D_{1} \times D_{2}= \C^2$,
the VCS satisfy the partial resolution of the identity:
\begin{equation}
 \int_{D_{1} \times D_{2}}
|z_{1}, z_{2},n_2 \rangle \langle z_{1}, z_{2},n_2  |\;
 d\mu(z_{1}, z_{2},n_{2})
= I_{n_{2}}.
\label{horesolu3}
\end{equation}
The moment problems issued from (\ref{horesolu3})
are of two forms:  one satisfied by $\varrho_1(r_{1},n_2)$
which is of the kind (\ref{homomsmart}),
therefore the corresponding density solution $\varrho_1(r_{1},n_2)$ does not actually depend on $n_2$ and coincides with (\ref{hodenssmart});
another moment problem for $\varrho_2(r_2)$
which is of the kind (\ref{homom05})
and so is solved by (\ref{hodens01}).
Thus the relation (\ref{ho2d1c}) forms a VCS class that we call
$(\gamma_1,1)$-class.

\noindent{\bf First class: (1,1)-generators of bi-CS -} In order to obtain the VCS first class, one performs
the continuous limit $\kappa_1 \to 0$ in (\ref{ho2d1c}) and gets the set
of VCS:
\bea
|z_{1}, z_{2},  n_{2} \rangle =
\mathcal N(z_{1}, z_{2},  n_{2})^{-\frac12} \sum_{n_{1} = 0}^{\infty}
\left[\rho_{1}( n_{1}) \rho_{2}( n_{2})\right]^{-\frac12}
z^{ n_{1} }_{1} z^{ n_{2} }_{2}| n_{1}, n_{2}\rangle .
\label{ho2d2c}
\eea
Summing on the remaining index $n_2$,
the states (\ref{ho2d2c}) generate the so called \emph{bi-CS} as constructed in \cite{ab1}. Hence,
$|z_{1}, z_{2},  n_{2} \rangle \propto z_2^{n_2} |z_1\rangle \otimes |n_2\rangle$, with $|z_1\rangle$ the canonical CS.

The norm series
\bea
\mathcal N(|z_{1}|, |z_{2}|,n_2)  =  \frac{1}{n_2!}\left[ \frac{|z_2|^2}{\omega_2}\right]^{n_2}\frac{1}{\omega_1}
\exp\left\{\frac{|z_{1}|^{2}}{\omega_{1}} \right\}
\eea
 converges everywhere in $\C$.
The following measure
\beq
d\mu(z_{1}, z_{2}) = \frac{1}{\pi^{2}}\,
{\mathcal N}(z_{1}, z_{2})\prod_{k=1}^{2}
\varrho_k(r_{k}) \, r_{k} dr_{k}d\theta_{k},
\eeq
on $D_{1} \times D_{2}= \C^2$, is considered.
The class of VCS (\ref{ho2d2c}) satisfies a partial resolution of the identity
like (\ref{horesolu3}); its moment problems
are identical to (\ref{homom05})
and again its densities $\varrho_k(r_k)$ are given by (\ref{hodens01}).
As a result, the states (\ref{ho2d2c}) consist in a
$(1,1)$-nondeformed VCS class.

\subsubsection{ Solvable sub-classes }
\label{sovsubs}

Let us now discuss on sub-classes which occur in the present study.
As previously performed, we proceed in three phases: (I)
to solve the most general class of VCS with deformation parameters;
(II) to restrict these parameters to be valued in $\{0,1\}$
in order to get the simplest sub-classes and
 to specify which of these classes are relevant
in the sense that we have already defined.
The following discussion will be valid for the remaining
subsections.

\noindent(I) {\bf Solving the generalized moment problem.} Writing the generalized state parametrized by the
real numbers $\alpha_i\neq 0 , \beta_i,\alpha'_i\neq 0 ,\beta'_i$, $i=1,2$,
\beq
&&
|z_{1}, z_{2}, n_2 \rangle =
\mathcal N(z_{1}, z_{2},n_2)^{-\frac12}\times \label{ho2d1csub}\\
&& \sum_{n_{1} = 0}^{\infty}
\frac{1}{[
\omega_{1}^{[\alpha'_1 n_1 + \beta'_1\kappa_1n_2]} R_1( n_{1},n_{2})\, \omega_2^{ [\alpha'_2 n_2 +\beta'_2 \kappa_2n_1]} R_{2}( n_{2})
]^{\frac12}}\,
z^{\alpha_1  n_{1} +\beta_1\kappa_1n_2 }_{1} \,z^{\alpha_2 n_{2}+ \beta_2\kappa_2n_1}_{2}\, |  n_{1}, n_{2}\rangle ,
\nonumber
\eeq
where $R_1(n_1,n_2)\in \{\Gamma[\gamma_1 + n_1],n_1!\}$
and $R_2(n_2) =n_2!,$
one can check that the state (\ref{ho2d1csub})
is normalizable with a norm
series of infinite radius of convergence.

The moment problem associated with the generalized
class (\ref{ho2d1csub}) is given by (we work up to unessential
$\alpha^{(')}_i,\beta^{(')}_i$ constant dependencies
in the measure $d\mu$ obtained after phase integrations)
\bea
2.2\int r_1dr_1 r_2dr_2\; \chi(r_1,r_2,n_2)\;
\frac{r_1^{2(\alpha_1 n_1 + \beta_1 \kappa_1 n_2)}}{
\omega_1^{\alpha'_1 n_1 + \beta'_1 \kappa_1 n_2}}
\frac{r_2^{2( \alpha_2 n_2 + \beta_2 \kappa_2 n_1) }}{
\omega_2^{\alpha'_2 n_2 + \beta'_2 \kappa_2 n_1}}
 = R_1(n_1,n_2) R_2(n_2)
\eea
where $\chi(r_1,r_2,n_2)$ is the generalized measure density that we have to determine.
Passing to square variables $u_i =r^2_i$, $i=1,2$, the moment
problem takes the form:
\beq
\int du_1 du_2\; \chi(u_1,u_2,n_2)\;
\left[\frac{ u_1^{\alpha_1} u_2^{\beta_2 \kappa_2 } }
{ \omega_1^{\alpha'_1} \omega_2^{ \beta'_2 \kappa_2} }
\right]^{n_1}
\left[
\frac{ u_1^{ \beta_1 \kappa_1 } }{
\omega_1^{ \beta'_1 \kappa_1} }
\right]^{n_2}
\left[
\frac{ u_2^{ \alpha_2 } }{ \omega_2^{\alpha'_2 } }
\right]^{n_2}= R_1(n_1,n_2) R_2(n_2).
\label{genmom}
\eeq
Interesting properties now emerge from this
multivariate moment problem that we ought to underline.
Indeed, one realizes that there exist many solutions $\chi(u_1,u_2,n_2)$ of (\ref{genmom}) and therefore
one is led to the non unicity of the measure which
integrates the VCS. The simple reason why this holds
is the freedom afforded by the index $n_2$ present in
the density.
This non unicity could be studied for its own interest
using a weaker version of the multivariate
Carleman-Nussbaum criterion  \cite{nuss}\cite{lassere}.

To find a density solution of (\ref{genmom})
can be tackled in different ways (see Appendix \ref{app:solving}).
The main issue here is to find an efficient and
non singular change of variables pertaining to (\ref{genmom}),
so that this problem is reduced to a simpler one
yielding solutions in such a way that
 our classification may be still achieved.

 The following densities given in radial variables solve the problem (\ref{genmom}) for $R_1(n_1)=n_1!$ and $R_2(n_2)=n_2!$
describing a first class of VCS of the kind (\ref{ho2d2c})
 (Appendix \ref{app:solving} provides details on this result):
\beq
&&
\varrho_1(r_1,r_2,n_2) =
 \alpha_1
\frac{r^{2(\alpha_1-1)}_1r_2^{2 \beta_2 \kappa_2 }
}{\omega^{\alpha'_1}_1  \omega_2^{\beta'_2 \kappa_2 }}
\left[
\frac{ \omega_1^{\beta'_1  \kappa_1} }
{r_1^{ 2\beta_1\kappa_1 }}
\right]^{n_2}
e^{ - \frac{ r^{2\alpha_1}_1r_2^{ 2\beta_2 \kappa_2 }}{ \omega^{\alpha'_1}_1  \omega_2^{\beta'_2 \kappa_2 } }}, \quad
\varrho_2(r_2)=\alpha_2 \frac{1}{\omega^{\alpha'_2}_2}
r_2^{2(\alpha_2-1)} e^{-\frac{r^{2\alpha_2}_2}{\omega^{\alpha'_2}_2}},\crcr
&&
\chi(r_1,r_2,n_2) = \alpha_1 \alpha_2
\frac{r^{2(\alpha_1-1)}_1r_2^{2 (\alpha_2 +\beta_2 \kappa_2 -1)}
}{\omega^{\alpha'_1}_1
\omega_2^{ \alpha'_2 +\beta'_2 \kappa_2}}
\left[
\frac{ \omega_1^{\beta'_1  \kappa_1} }
{r_1^{ 2\beta_1\kappa_1 }}
\right]^{n_2}
e^{ - \frac{ r^{2\alpha_1}_1r_2^{ 2\beta_2 \kappa_2 }}{ \omega^{\alpha'_1}_1  \omega_2^{\beta'_2 \kappa_2 } }
-\frac{r^{2\alpha_2}_2}{\omega^{\alpha'_2}_2}}.
\label{soludefor}
\eeq
Considering on the contrary $R_1(n_1,n_2)= \Gamma[\gamma_1+ n_1]$ and
$R_2(n_2)=n_2!$ being data for second class
VCS (\ref{ho2d1c}), one gets the solutions (see Appendix \ref{app:solving}):
\beq
&&
\varrho_1(r_1,r_2,n_2) = \alpha_1
\frac{r_1^{2(\alpha_1-1)} r_2^{ 2\beta_2 \kappa_2 }
}{\omega^{\alpha'_1}_1  \omega_2^{\beta'_2 \kappa_2 }}
\left[\frac{\omega_1^{(\beta'_1 - \alpha'_1) \kappa_1}
r_2^{ 2 \beta_2  }}
{r_1^{ 2(\beta_1 - \alpha_1 )\kappa_1 }
\omega_2^{ \beta'_2 }}
\right]^{n_2}
e^{ - \frac{ r^{2\alpha_1}_1r_2^{ 2\beta_2 \kappa_2 }}{ \omega^{\alpha'_1}_1  \omega_2^{\beta'_2 \kappa_2 } }},\crcr
&&
\varrho_2(r_2)=\alpha_2 \frac{1}{\omega^{\alpha'_2}_2}
r_2^{2(\alpha_2-1)} e^{-\frac{r^{2\alpha_2}_2}{\omega^{\alpha'_2}_2}}.
\label{soludefor2}
\eeq

\noindent(II) {\bf Extracting the relevant sub-classes.}
We can first analyze these solutions
and consequently organize the sub-classes.
First a quick checking shows that
the VCS second class  (\ref{ho2d1c}) corresponds to
the sub-class called A, (denoted by $(\gamma_1,1)$A), defined by the
8-tuple
\bea
(\alpha_1,\beta_1,\alpha'_1,\beta'_1,\alpha_2,\beta_2,\alpha'_2,\beta'_2)= (1,1,1,1,1,0,1,0)
\eea
which yields the correct solutions of the moment
problem, $\varrho_1$ and $\varrho_2,$ substituting these
parameters in (\ref{soludefor2}).
The first class limit (\ref{ho2d2c}),
also named sub-class A, (denoted by $(1,1)$A), is defined by
\bea
(\alpha_1,\beta_1,\alpha'_1,\beta'_1,\alpha_2,\beta_2,\alpha'_2,\beta'_2)= (1,0,1,0,1,0,1,0)
\eea
and here $\varrho_1$ and
$\varrho_2$ (\ref{soludefor})
resolve the moment problem for this class,
given  these parameters.

Henceforth, we can restrict to the situation
where $\alpha^{(\prime)}_i,\beta^{(\prime)}_i\in \{0,1\}$
and, since $\alpha_i^{(\prime)}$ should be always
fixed to $1$, we have just to analyze different cases
for the quadruple $(\beta_1,\beta'_1,\beta_2,\beta'_2)$.
The sub-classes which are irrelevant are just
factors of the first or second class of this section. These include
$
(1,0,0,0), (0,1,0,0).
$
The relevant sub-classes
are defined by tuples which contain an exponent $n_1$
in the sector $(\omega_2,z_2)$,
since they are getting involved in the series.
The sub-classes are defined
by
\beq
&&
(0,0,0,1),(0,0,1,0),
(0,0,1,1),
\label{relevho2d1c} \\
&&
(1,1,0,1),(1,1,1,0),
(1,1,1,1).
\label{relevho2d2c}
\eeq
Hence it remains six tuples which can be listed as:
\beq
(1,0,0,1),(1,0,1,0),
(1,0,1,1) ,
\label{irrho2d1c}\\
(0,1,0,1),(0,1,1,0),
(0,1,1,1).
\label{irrho2d2c}
\eeq
Each VCS defined by (\ref{irrho2d1c}) is a
factor of a sub-class in (\ref{relevho2d1c})
and each VCS  defined by a tuple in (\ref{irrho2d2c}) becomes
also a factor of some sub-class already listed in (\ref{relevho2d2c}).
To be even more precise, we do not need to compute six
sub-classes for each class: only the quadruples in
(\ref{relevho2d1c}) are relevant with respect to the first class
(\ref{ho2d1c})
whereas the quadruples in (\ref{relevho2d2c}) are
the only relevant ones for the second class (\ref{ho2d2c}).

Given a general form of the VCS with two degrees of
freedom
\bea
|z_{1}, z_{2}, n_2 \rangle =
\mathcal N(z_{1}, z_{2},n_2)^{-\frac12} \sum_{n_{1} = 0}^{\infty}
a( n_{1},n_{2},z_1,z_2) \, |  n_{1}, n_{2}\rangle ,
\eea
the following table gives the values of $a\equiv a( n_{1},n_{2},z_1,z_2)$
and measure densities $\varrho_k$, $k=1,2$, corresponding to the relevant first sub-classes:

{ \footnotesize
\begin{center}
\begin{tabular}{cc}
 &\small{ First class $(1,1)$-generator of bi-CS}: $R_1(n_1)=n_1!$,\;\; $R_2(n_2) =n_2!$\\
\hline\hline
&\\
A &
$a=\frac{\left[\frac{z_1}{\omega_1^{\frac12}}\right]^{n_1}
\left[\frac{z_2}{\omega_2^{\frac12}}\right]^{n_2}}{[n_1!n_2!]^{\frac12}}
$;\quad
   $\varrho_k(r_k)= f(r_k,\omega_k)$,\;
$k=1,2$ \;\;(\ref{ho2d2c}) \\
B&  $a=\frac{\left[\frac{z_1}{\omega_1^{\frac12}}\right]^{n_1}\left[\frac{z_2}{\omega_2^{\frac12}}\right]^{n_2}}{[ n_1! (\omega_2)^{\kappa_2 n_1 }n_2!]^{\frac12} } $;\quad
$\varrho_1(r_1)=f(r_1,\omega_1\omega_2^{\kappa_2})$;\;
$\varrho_2(r_2) = f(r_2,\omega_2)$    \\
C&   $a =  \frac{\left[\frac{z_1}{\omega_1^{\frac12}}\right]^{n_1} \left[\frac{z_2}{\omega_2^{\frac12}}\right]^{n_2}  z_2^{ \kappa_2 n_1}}{[n_1!n_2!]^{\frac12}}
$;\quad
$\varrho_1(r_1,r_2)=r_2^{2\kappa_2}f(r_1r_2^{\kappa_2},\omega_1)$;\;
$\varrho_2(r_2) =  f(r_2,\omega_2)$ \\
D&   $ a = \frac{\left[\frac{z_1}{\omega_1^{\frac12}}\right]^{n_1} \left[\frac{z_2}{\omega_2^{\frac12}}\right]^{n_2+ \kappa_2 n_1}}{[n_1!n_2!]^{\frac12}}
$;\quad
 $\varrho_1(r_1,r_2)=r_2^{2\kappa_2}f(r_1r_2^{\kappa_2},\omega_1\omega_2^{\kappa_2})$;
$\varrho_2(r_2) =    f(r_2,\omega_2)$  \\
&\\
\hline\hline
\end{tabular}
\end{center}
}
\noindent where the function $f$ is given by
 $f(r,\omega) =(1/\omega)e^{-r^2/\omega}$.
As a quick inspection in order to be certain that
the tuples  (\ref{relevho2d2c}),  (\ref{irrho2d1c}) or (\ref{irrho2d2c})
do not define new classes, we can check that
quadruples included in (\ref{relevho2d2c}),
taken in that order,
 the elements of the list (\ref{irrho2d1c}),
in that order, and those of the list (\ref{irrho2d2c}), in that order,
determine  equivalently the classes  $(1,1)$B, $(1,1)$C  and $(1,1)$D, respectively, up to the factors $(z_1/\omega^{1/2}_1)^{\kappa_1 n_2}$,
$z_1^{\kappa_1 n_2}$ and $\omega_1^{\kappa_1 n_2/2}$,
respectively.
These factors can be reabsorbed in the measure
density $\varrho_1(r_1,r_2,n_2)$ without complication.

The following table gives the relevant second sub-classes:
{  \footnotesize
\begin{center}
\begin{tabular}{cc}
 &\small{ Second class $(\gamma_1,1)$-deformed VCS}:  $R_1(n_1,n_2)=\Gamma[\gamma_1 + n_1]$,\;\; $R_2(n_2) =n_2!$\\
\hline\hline
&
\\
A &
 $a =\frac{\left[\frac{z_1}{\omega^{\frac12}_1}\right]^{n_1+ \kappa_1 n_2}\left[\frac{z_2}{\omega^{\frac12}_2}\right]^{n_2}}{ [\Gamma[\gamma_1+ n_1]n_2!]^{\frac12}}
$;\quad
$\varrho_k(r_k)= f(r_k,\omega_k)$,\;
$k=1,2$  (\ref{ho2d1c}) \\
B &
 $a=\frac{\left[\frac{z_1}{\omega^{\frac12}_1}\right]^{n_1+ \kappa_1 n_2}\left[\frac{z_2}{\omega^{\frac12}_2}\right]^{n_2}}{ [\Gamma[\gamma_1+ n_1](\omega_2)^{ \kappa_2 n_1}n_2!]^{\frac12}} $;
\quad
$\varrho_1(r_1,n_2) =\frac{1}{\omega_2^{n_2}}f(r_1,\omega_1\omega_2^{\kappa_2})$;
$\varrho_2(r_2) =f(r_2,\omega_2) $     \\
C &
   $a = \frac{\left[\frac{z_1}{\omega^{\frac12}_1}\right]^{n_1+ \kappa_1 n_2}
\left[\frac{z_2}{\omega^{\frac12}_2}\right]^{n_2} z_2^{\kappa_2 n_1}}{[\Gamma[\gamma_1+ n_1]n_2!]^{\frac12}}
$;\quad
$\varrho_1(r_1,n_2) =r_2^{2(\kappa_2 +n_2)}f(r_1r_2^{\kappa_2},\omega_1)$;
$\varrho_2(r_2) = f(r_2,\omega_2)$
 \\
D &
   $a=\frac{\left[\frac{z_1}{\omega^{\frac12}_1}\right]^{n_1+ \kappa_1 n_2}\left[\frac{z_2}{\omega^{\frac12}_2}\right]^{n_2+ \kappa_2 n_1}}{ [ \Gamma[\gamma_1+ n_1] n_2!]^{\frac12}}
$;\quad
$\varrho_1(r_1,n_2) =\frac{r_2^{2(\kappa_2 +n_2)}}{\omega_2^{n_2}}f(r_1r_2^{\kappa_2},\omega_1\omega_2^{\kappa_2})$;
$\varrho_2(r_2) = f(r_2,\omega_2)$
\\
&\\
\hline\hline
\end{tabular}
\end{center}
}
\medskip
\noindent One notes that all these sets of states are new classes of VCS
that our systematic analysis has allowed to generate.
They consist mainly, on the non singular change of
variables in the VCS (\ref{ho2d1c}) and (\ref{ho2d2c})
without breaking the solvability of the VCS.
Each of them will also span new generalized or
$(\gamma,1)$-deformed bi-CS by summing on
the remaining index $n_2$.

Let us discuss the type of the above VCS.
As $\kappa_2 \to 0$, all sub-classes
denoted by $(1,1)$B, $(1,1)$C and $(1,1)$D
tend to the sub-class $(1,1)$A. The sub-class $(\gamma_1,1)$A
is also of this type but under the other limit $\kappa_1\to 0$.
However, the sub-classes $(\gamma_1,1)$B, $(\gamma_1,1)$C and $(\gamma_1,1)$D
are not well-defined under the limits $\kappa_{1}\to 0$
or $\kappa_2 \to 0$,  and so are not of the type of the sub-class $(1,1)$A.

\subsubsection{Third class: The fake dual}
\label{sub:ho2d3c}

\noindent{\bf Third class: $(1,\gamma)$-deformed VCS -}
In the  above construction, the indices $1$ and $2$
do not play  a symmetric role since, at least,
the index $n_1$ is summed and the index $n_2$ is not.
In the following, we still assume that $n_1$ is summed but
use instead  different (generalized) factorials.
We choose $\rho_{1}(n_1)$ given by a simple factorial (\ref{hoqua01}) and
$\rho_{2}(n_{2},n_{1})$ given by (\ref{hofactosmart}) with
\begin{equation}
\gamma_{2} =\gamma_{2} (n_{1}) = 1 + \kappa_{2}\, n_{1}, \qquad
\kappa_{2} = \frac{\omega_{1}}{\omega_{2}}.
\end{equation}
Then, we get a different set of states
\bea
|z_{1}, z_{2}, n_2 \rangle =
\mathcal N(z_{1}, z_{2},n_2)^{-\frac12} \sum_{n_{1} = 0}^{\infty}
\left[\rho_{1}( n_{1}) \rho_{2}(n_{2}, n_1)\right]^{-\frac12}
z^{ n_{1} }_{1} \,z^{n_{2}+ \kappa_2 n_1}_{2}\, |  n_{1}, n_{2}\rangle ,
\label{ho2d3c}
\eea
normalized provided that the factor
\bea
\mathcal N(|z_{1}|,|z_{2}|,n_2) =
\left[ \frac{|z_2|^2}{\omega_2} \right]^{n_2}
\sum_{n_{1} = 0}^{\infty} \frac{1}{n_1!\, \Gamma(\gamma_2) (\gamma_2)_{n_2}}
\left[ \frac{|z_1 (z_2)^{\kappa_2}|^2}{\omega_1 (\omega_2)^{\kappa_2}}\right]^{n_1} \,
\eea
is converging. We can use the ratio test in order to check that
$\mathcal N(z_{1}, z_{2},n_2)$ is absolutely convergent
everywhere in $\C$. Indeed, since $\gamma_2 =1+ \kappa_2 n_1\geq 1$ and using a comparison test, we can bound each term of this
series by the term of an exponential series,
\bea
\frac{1}{n_1!\, \Gamma(\gamma_2) (\gamma_2)_{n_2}}
\left[ \frac{|z_1 (z_2)^{\kappa_2}|^2}{\omega_1 (\omega_2)^{\kappa_2}}\right]^{n_1} \leq
\frac{1}{n_1!\,n_2 !}
\left[ \frac{|z_1 (z_2)^{\kappa_2}|^2}{\omega_1 (\omega_2)^{\kappa_2}}\right]^{n_1}.
\eea
The latter is a term of a convergent series for all
$|z_1 (z_2)^{\kappa_2}|>0$,
implying $|z_1|$ and $|z_2|>0$.
A resolution of the identity can be found making use
of the measure
\bea
d\mu(z_1, z_2, n_2) = \frac{1}{\pi^2}\mathcal N(z_{1}, z_{2}, n_{2}) \;
\varrho_1(r_{1}) \, r_{1} dr_{1} d\theta_{1}\;
\varrho_2(r_{2}) \, r_{2} dr_{2} d\theta_{2}
\eea
and
\bea
\int_{D_{1} \times D_{2}}
|z_{1}, z_{2}, n_2 \rangle \langle z_{1}, z_{2}, n_2  |\;
 d\mu(z_{1}, z_{2},n_{2})
= I_{n_{2}}
\label{horesolu4}
\eea
is satisfied for the solutions to the moment problems
(\ref{homom05})  and  (\ref{homomsmart}) for
$\varrho_1(r_1)$ and $\rho_2(r_2)$, respectively.

This VCS third class should not be confused with
the dual class  (hence the nickname
of \emph{fake dual}) obtained from the VCS second class (\ref{ho2d1c})
by switching completely the role of $1$ and $2$.
Note also that this third class has its own dual partner.
Finally, from (\ref{ho2d3c}), by taking the limit
$\kappa_2 \to 0$, one generates the VCS first class $(1,1)$A (\ref{ho2d2c}).

\noindent{\bf Solvable sub-classes -}
Let us investigate sub-classes associated with
(\ref{ho2d3c}). As customary,  we introduce
the general state
\beq
&&
|z_{1}, z_{2}, n_2 \rangle =
\mathcal N(z_{1}, z_{2},n_2)^{-\frac12}\times \label{ho2d3csub}\\
&& \sum_{n_{1} = 0}^{\infty}
\frac{1}{[\omega_{1}^{[\alpha'_1 n_1 + \beta'_1\kappa_1n_2]}
R_1(n_{1}) \, \omega_2^{ [\alpha'_2 n_2 +\beta'_2 \kappa_2n_1]} R_{2}(n_{2},n_1)]^{\frac12}}\,
z^{\alpha_1  n_{1} +\beta_1\kappa_1n_2 }_{1} \,z^{\alpha_2 n_{2}+ \beta_2\kappa_2n_1}_{2}\, |  n_{1}, n_{2}\rangle .
\nonumber
\eeq
The formalism for solving the associated generalized moment problem
 has been already introduced.
Given
 $\chi(r_1,r_2,n_2)  = \varrho_1(r_1,r_2,n_2) \varrho_{2}(r_2)$,
one solves the problem for (see Appendix \ref{app:solving}):
\beq
&&
\varrho_1(r_1,r_2,n_2) = \alpha_1
\frac{r_1^{2(\alpha_1-1)} r_2^{ 2(\beta_2-\alpha_2)\kappa_2 }
}{\omega^{\alpha'_1}_1  \omega_2^{(\beta'_2-\alpha'_2) \kappa_2 }}
\left[\frac{\omega_1^{\beta'_1}}
{r_1^{ 2\beta_1 }}
\right]^{ \kappa_1n_2}
e^{ - \frac{ r^{2\alpha_1}_1r_2^{ 2(\beta_2-\alpha_2) \kappa_2 }}{ \omega^{\alpha'_1}_1  \omega_2^{(\beta'_2-\alpha'_2) \kappa_2 } }},\crcr
&&
\varrho_2(r_2)=\alpha_2 \frac{1}{\omega^{\alpha'_2}_2}
r_2^{2(\alpha_2-1)} e^{-\frac{r^{2\alpha_2}_2}{\omega^{\alpha'_2}_2}}.
\label{soludefor3}
\eeq
Replacing $(\alpha_1,\beta_1,\alpha'_1,\beta'_1,\alpha_2,\beta_2,\alpha'_2,\beta'_2)=(1,0,1,0,1,1,1,1)$, one gets
the correct densities  associated with (\ref{ho2d3c}).

Discussing the relevant classes
by considering the quadruple  $(\beta_1,\beta'_1,\beta_2,\beta'_2)$
and keeping fixed $\alpha_i^{(')}=1$,
note first that $(0,0,1,1)$ defines
(\ref{ho2d3c}). We need to consider as a basic sub-class
the one defined by $(0,0,0,0)$.
Focusing on states involving at least an exponent $n_1$
in the tower $2$, only the following quadruples are relevant
\bea
(0,0,1, 0), (0,0,0,1) .
\label{ho2d3csubirr}
\eea
The left over tuples determine nothing but factors
of the sub-classes defined by the above set of quadruples.
In the following table are collected the relevant third sub-classes
(using previous notations):
{  \footnotesize
\begin{center}
\begin{tabular}{cc}
 &\small{ Third class $(1,\gamma_2)$-deformed VCS}:  $R_1(n_1)= n_1! ,\;\; R_2(n_2,n_1)=\Gamma[\gamma_2 + n_2]$\\
\hline\hline
&
\\
A &
 $a =\frac{\left[\frac{z_1}{\omega^{\frac12}_1}\right]^{n_1}\left[\frac{z_2}{\omega^{\frac12}_2}\right]^{n_2+ \kappa_2 n_1}}{ [n_1!\Gamma[\gamma_2 + n_2]]^{\frac12}}
$;\quad
$\varrho_k(r_k)= f(r_k,\omega_k)$,\;
$k=1,2$  (\ref{ho2d3c}) \\
B &
 $a=\frac{\left[\frac{z_1}{\omega^{\frac12}_1}\right]^{n_1}
\left[\frac{z_2}{\omega^{\frac12}_2}\right]^{n_2}}{
[n_1!(\omega_2)^{ \kappa_2 n_1}\Gamma[\gamma_2+ n_2]]^{\frac12}} $;
\quad
$\varrho_1(r_1) =f(r_1,\omega_1\omega_2^{-\kappa_2})$;
$\varrho_2(r_2) = f(r_2,\omega_2)$
   \\
C &
   $a = \frac{\left[\frac{z_1}{\omega^{\frac12}_1}\right]^{n_1}
\left[\frac{z_2}{\omega^{\frac12}_2}\right]^{n_2} z_2^{\kappa_2 n_1}}{[n_1! \Gamma[\gamma_2+ n_2]]^{\frac12}}
$;\quad$\varrho_1(r_1,r_2) =r_2^{-2\kappa_2}
f(r_1r_2^{-\kappa_2},\omega_1)$;
$\varrho_2(r_2) =f(r_2,\omega_2) $
 \\
D &
   $a=\frac{\left[\frac{z_1}{\omega^{\frac12}_1}\right]^{n_1}\left[\frac{z_2}{\omega^{\frac12}_2}\right]^{n_2}}{ [ n_1!\Gamma[\gamma_2+ n_2] ]^{\frac12}}
$;\quad
$\varrho_1(r_1,r_2) =r_2^{-2\kappa}
f(r_1r_2^{-\kappa_2},\omega_1\omega_2^{-\kappa_2})$;
$\varrho_2(r_2) = f(r_2,\omega_2)$
\\
&\\
\hline\hline
\end{tabular}
\end{center}
}
\medskip
\noindent
As $\kappa_2 \to 0$, these states are all of the type of $(1,1)$A. Meanwhile, all the factors
associated with that third class involving an additional
factor $\kappa_{1}$ will be not of that type.

\subsubsection{Fourth class: $(\gamma_1,\gamma_2)$- or doubly-deformed CS}
\label{sub:ho2d4c}
We pursue the analysis by introducing doubly dependent generalized factorials
$\rho_{1,2}(n_{1,2},n_{2,1})$ still given  by (\ref{hofactosmart}) with
\begin{equation}
\gamma_{1,2} =\gamma_{1,2} (n_{1,2}) = 1 + \kappa_{1,2}\, n_{2,1}, \qquad
\kappa_{1,2} = \frac{\omega_{2,1}}{\omega_{1,2}}.
\end{equation}
The corresponding set of states can be built as
\bea
|z_{1}, z_{2}, n_2 \rangle =
\mathcal N(z_{1}, z_{2},n_2)^{-\frac12} \sum_{n_{1} = 0}^{\infty}
\left[\rho_{1}( n_{1},n_{2}) \rho_{2}( n_{2},n_{1})\right]^{-\frac12}
z^{ n_{1} + \kappa_1 n_2}_{1} \,z^{n_{2} +  \kappa_2 n_1}_{2}\, |  n_{1}, n_{2}\rangle ,
\label{ho2d4c}
\eea
with the normalization factor
\bea
N(|z_{1}|, |z_{2}|,n_2)^{-\frac12}
 = \sum_{n_1 =0}^{\infty}
\frac{1}{\Gamma[\gamma_1 + n_1]\Gamma[\gamma_2 + n_2]}
\left(\frac{|z_{1}|^2}{\omega_1}\right)^{n_1+\kappa_1 n_2}
\left(\frac{|z_{2}|^2}{\omega_2}\right)^{n_2+\kappa_2 n_1}
\eea
converging everywhere in $\C$ since
the following inequality holds
\bea
\frac{1}{\Gamma[\gamma_1 + n_1]\Gamma[\gamma_2 + n_2]}
\left(\frac{r_1^2}{\omega_1}\right)^{n_1+\kappa_1 n_2}
\left(\frac{r_2^2}{\omega_2}\right)^{n_2+\kappa_2 n_1}
\leq
\frac{1}{n_1! n_2!}
\left(\frac{r_1^2r_2^{2\kappa_2}}{\omega_1\omega_2^{\kappa_2}}\right)^{n_1}
\left(\frac{r_2^2r_1^{2\kappa_1}}{\omega_2\omega_1^{\kappa_1}}\right)^{n_2}.
\eea
The resolution of the identity of these states
uses the measure
\bea
d\mu(z_1,z_2,n_2) =
\frac{1}{\pi^2}\mathcal N(z_{1}, z_{2}, n_{2}) \;
\varrho_1(r_{1}) \, r_{1} dr_{1} d\theta_{1}\;
\varrho_2(r_{2}) \, r_{2} dr_{2} d\theta_{2}
\eea
so that
\bea
\int d\mu(z_1,z_2) |z_1,z_2 ,n_2 \rangle
\langle z_1,z_2,n_2| = I_{n_2}
\eea
holds for $\rho_1(r_1)$ and $\rho_{2}(r_2)$
satisfying the moment problems of the kind (\ref{homomsmart})
and hence with the solutions given by (\ref{hodenssmart}).

\noindent{\bf Solvable sub-classes -}
Determining sub-classes of (\ref{ho2d4c})
can be discussed by
introducing the parameters $\alpha_i^{(')}$ and $\beta_i^{(')}$
and the generalized state
\beq
&&\hspace*{-0.7cm}
|z_{1}, z_{2}, n_2 \rangle =
\mathcal N(z_{1}, z_{2},n_2)^{-\frac12}\times
\crcr
&&\hspace*{-0.7cm}
 \sum_{n_{1} = 0}^{\infty}
\frac{1}{
[\omega_{1}^{[\alpha'_1 n_1 + \beta'_1\kappa_1n_2]}
R_1(n_{1},n_2)
\omega_2^{[\alpha'_2 n_2 +\beta'_2 \kappa_2n_1]}
R_{2}(n_{2},n_1) ]^{\frac12} }
z^{\alpha_1  n_{1} +\beta_1\kappa_1n_2 }_{1} \,z^{\alpha_2 n_{2}+ \beta_2\kappa_2n_1}_{2}\, |  n_{1}, n_{2}\rangle .
\crcr
&&\label{ho2d4csub}
\eeq
The associated moment problem is written, with $u_i = r_i^2,$
\beq
\int du_1 du_2\; \chi_1(u_1,u_2,n_2)
\left[
\frac{ u_2^{\alpha_2 }  }
{ \omega_2^{\alpha'_2 }  }
\right]^{n_2 + \kappa_2 n_1}
\left[\frac{ u_1^{ \alpha_1 }u_2^{ (\beta_2-\alpha_2)\kappa_2} }
{\omega_1^{\alpha'_1 }\omega_2^{ (\beta'_2-\alpha'_2)\kappa_2 } }
\right]^{n_1+\kappa_1n_2 }
\left[\frac{ u_1^{\beta_1 -\alpha_1 }u_2^{- (\beta_2-\alpha_2)\kappa_2} }
{\omega_1^{\beta'_1-\alpha'_1 }\omega_2^{- (\beta'_2-\alpha'_2)\kappa_2 } }
\right]^{\kappa_1n_2 }
\label{genmom5}
\eeq
and is solved by the following densities (see Appendix \ref{app:solving})
\beq
&&
\varrho_1(r_1,r_2,n_2) = \alpha_1
\frac{r_1^{2(\alpha_1-1)} r_2^{ 2(\beta_2-\alpha_2)\kappa_2 }
}{\omega^{\alpha'_1}_1  \omega_2^{(\beta'_2-\alpha'_2) \kappa_2 }}
\left[\frac{ u_1^{\alpha_1 - \beta_1}\omega_2^{ (\alpha'_2-\beta'_2)\kappa_2 }}
{\omega_1^{\alpha'_1-\beta'_1 } u_2^{ (\alpha_2-\beta_2)\kappa_2} }
\right]^{\kappa_1n_2 }
e^{ - \frac{ r^{2\alpha_1}_1r_2^{ 2(\beta_2-\alpha_2) \kappa_2 }}{ \omega^{\alpha'_1}_1  \omega_2^{(\beta'_2-\alpha'_2) \kappa_2 } }},\crcr
&&
\varrho_2(r_2)=\alpha_2 \frac{1}{\omega^{\alpha'_2}_2}
r_2^{2(\alpha_2-1)} e^{-\frac{r^{2\alpha_2}_2}{\omega^{\alpha'_2}_2}}.
\label{soludefor4}
\eeq
For the present study,
the quadruple $(\beta_1,\beta'_1,\beta_2,\beta'_2)$ such that
$(1,1,1,1)$ defines (\ref{ho2d4c}).
We choose another sub-class defined by $(1,1,0,0)$.
Proceeding as previously,
the following quadruples
\bea
(1,1,0, 1),  (1,1,1, 0)
\eea
are relevant.
The ingredients defining the relevant classes, above listed,
are given by  the following table (in anterior notations):
{  \footnotesize
\begin{center}
\begin{tabular}{cc}
 &\small{ Fourth class $(\gamma_1,\gamma_2)$-deformed VCS}:  $R_1(n_1,n_2)= [\gamma_1+ n_1], \;\; R_2(n_2,n_1)=\Gamma[\gamma_2 + n_2]$\\
\hline\hline
&
\\
A &
 $a =\frac{\left[\frac{z_1}{\omega^{\frac12}_1}\right]^{n_1+ \kappa_1 n_2}\left[\frac{z_2}{\omega^{\frac12}_2}\right]^{n_2+ \kappa_2 n_1}}{ [\Gamma[\gamma_1 + n_1]\Gamma[\gamma_2 + n_2]]^{\frac12}}
$;\quad
$\varrho_k(r_k)= f(r_k,\omega_k)$,\;
$k=1,2$  (\ref{ho2d4c}) \\
B &
 $a=\frac{\left[\frac{z_1}{\omega^{\frac12}_1}\right]^{n_1+ \kappa_1 n_2}
\left[\frac{z_2}{\omega^{\frac12}_2}\right]^{n_2}}{
[\Gamma[\gamma_1 + n_1](\omega_2)^{ \kappa_2 n_1}\Gamma[\gamma_2+ n_2]]^{\frac12}} $;
\quad$\varrho_1(r_1) =\frac{1}{r_2^{2(\kappa_2+n_2)}} f(r_1r_2^{-\kappa_2},\omega_1)$;
$\varrho_2(r_2) = f(r_2,\omega_2)$
     \\
C &
   $a = \frac{\left[\frac{z_1}{\omega^{\frac12}_1}\right]^{n_1+ \kappa_1 n_2}
\left[\frac{z_2}{\omega^{\frac12}_2}\right]^{n_2} z_2^{\kappa_2 n_1}}{[\Gamma[\gamma_1 + n_1] \Gamma[\gamma_2+ n_2]]^{\frac12}}
$;\quad
$\varrho_1(r_1,r_2) =\omega_2^{n_2}
f(r_1,\omega_1\omega_2^{-\kappa_2})$;
$\varrho_2(r_2) =f(r_2,\omega_2) $
 \\
D &
   $a=\frac{\left[\frac{z_1}{\omega^{\frac12}_1}\right]^{n_1+ \kappa_1 n_2}\left[\frac{z_2}{\omega^{\frac12}_2}\right]^{n_2}}{ [ \Gamma[\gamma_1 + n_1]\Gamma[\gamma_2+ n_2] ]^{\frac12}}
$;\quad
$\varrho_1(r_1,r_2) =\frac{\omega_2^{n_2}}{r_2^{2(n_2+\kappa_2)}}
f(r_1r_2^{-\kappa_2},\omega_1\omega_2^{-\kappa_2})$;
$\varrho_2(r_2) = f(r_2,\omega_2)$
\\
&\\
\hline\hline
\end{tabular}
\end{center}
}
\medskip
\noindent
These states are not of the type of any previous classes.

\begin{figure}[t]
\centering{
\includegraphics[width=105mm]{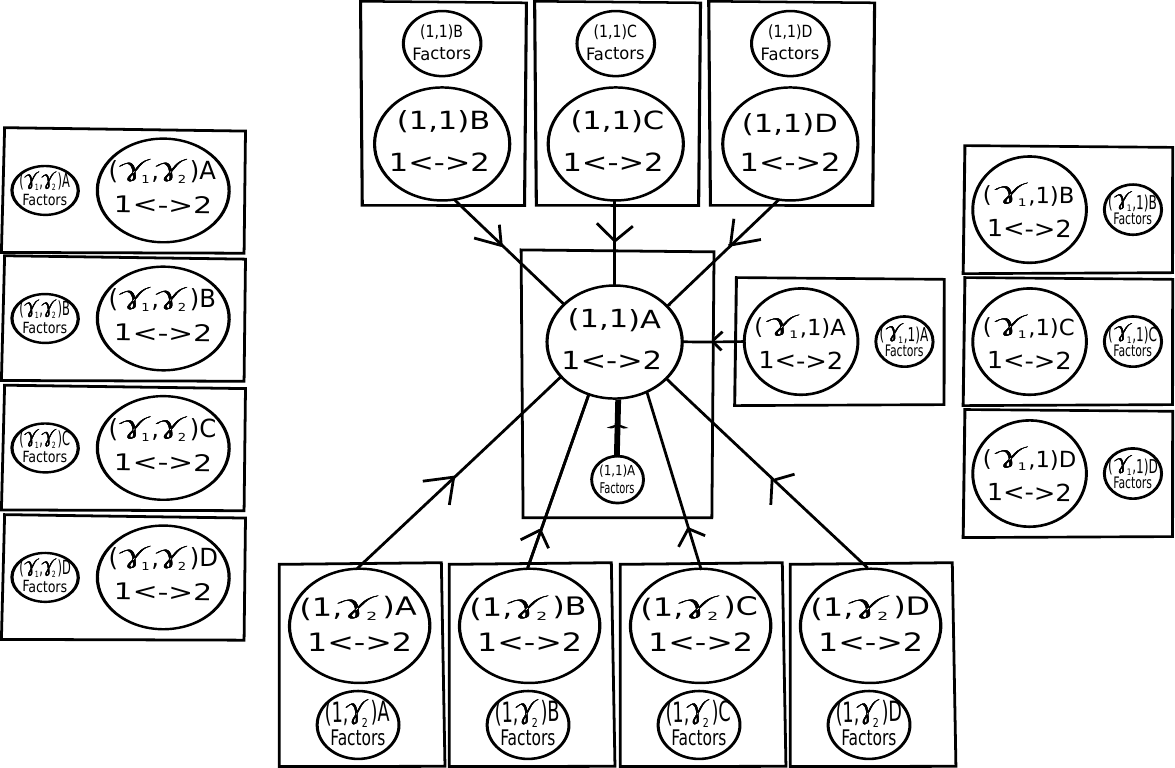}}
\caption{The four deformed classes $(\gamma_1,1)$,
$(1,\gamma_2)$ and $(\gamma_1,\gamma_2)$
and their relation with the  $(1,1)$-class for two degrees of freedom.}
\label{fig:2d1c-4c}
\end{figure}

We recapitulate these results on two degrees of freedom
by a diagram given by Figure \ref{fig:2d1c-4c}.

\subsection{VCS without energy shift and
 Landau level VCS}

There is another way to generalize the above VCS
to classes built out of energy levels with or without
energy shift by a fixed amount:
\beq
\label{var07}
\mathcal E_{n_{1}, n_{2}} =  \omega_{1}\left(n_{1} + \alpha_1 \right)
 +  \omega_{2}\left(n_{2} +\alpha_2 \right).
\eeq
Typically, of course,  $\alpha_i$ is simply the ground
state energy of the harmonic oscillator,
namely $\alpha_i = 1/2$ and the states built out
of  these energy levels can be called \emph{unshifted}. The VCS of this system
are still solvable in full generality for any value of $\alpha_i \geq 0$,
giving another full set of $\alpha_i$-deformed classes of VCS
such that at the limit $\alpha_i \to 0$, one continuously recovers
 earlier  computed classes.

We sketch how the main quantities involved in the
construction of the VCS get modified by these shifts.
We introduce the modified set of generalized factorials as
\beq{\label{var10}}
\tilde{\rho}_{i}(n_{i}) = (\omega_{i})^{n_i}\prod_{k=1}^{n_{i}}\left(k + \alpha_i \right)
= (\omega_{i})^{n_{i}}\left(1+\alpha_i\right)_{n^{i}_{i}}, \quad i=1,2,
\eeq
which at the limit $\alpha_i \to 0$ converge indeed
to plain factorial $(\omega_{i})^{n_{i}}n_i!$.
Meanwhile, the previous Pochhammer symbols find the generalization:
\beq
&&
{\mathcal E}_{n_1,n_2}=
\omega_{1,2} \left( n_{1,2} + \alpha_{1,2} + \frac{\omega_{2,1}}{\omega_{1,2}}(n_{2,1} + \alpha_{2,1})\right), \\
&&
\tilde{\rho}_{1,2}(n_{1,2})
= (\omega_{1,2})^{n_{1,2}}(\gamma_{1,2})_{n_{1,2}}, \quad
\gamma_{1,2} = 1 + \alpha_{1,2}
+ \frac{\omega_{2,1}}{\omega_{1,2}}\left(n_{2,1}  +\alpha_{2,1} \right),
\eeq
so that all solvable VCS classes can be directly extended using these quantities.

Some comments on another particular
physical system  called the Landau problem
which can be exactly implemented
with the above construction are in order.

The Landau problem is the quantum mechanical system
describing the motion of a charged particle in a plane subjected
to a magnetic field perpendicular to that plane \cite{landau2,goerbig,lederer}.
The Hamiltonian of this model which can be written as (\ref{landau}) has been successfully applied to various
condensed matter system and gave rise to interesting
results among which one quotes
the (simple, fractional, spin) quantum Hall effect \cite{lederer}.
By adding further a harmonic potential to its Hamiltonian,
the Landau model is still solvable and the previous model
with its infinite degeneracy for each energy level
becomes lifted.

In symbol, after diagonalization, the dimensionless
Landau Hamiltonian including a harmonic potential with frequency $\omega$ is given by
\beq
\frac{1}{\hbar}H_{L}= \Omega_+ (N_+ +\frac12) + \Omega_- (N_- +\frac12 ) +\text{const.} ,\qquad
N_\pm = a_\pm^\dag a_\pm, \qquad [a_\pm,a_\pm^\dag]=\mathbb{I}_\pm,
\eeq
where $\Omega_\pm = \pm \varpi + \sqrt{\varpi^2 + \omega^2};$,
$\varpi= \text{e}|B|/m$ is the cyclotron frequency, $a_\pm,a_\pm^\dag$ are two
bosonic Heisenberg modes (generating $N_\pm$
as number operators) describing the particle
helicity or the possible winding orientations
of the particle. Note that turning off the harmonic
potential ($\omega \to 0$), the spectrum of operator becomes
infinitely  degenerate with $(+)$ as a remaining helicity sector.

The VCS formalism can be obviously applied here.
The above model is indeed nothing but a harmonic oscillator in $2D$
with only a particular feature to be degenerate at a certain
limit. So all the previous sets of VCS are valid for the
Landau problem with shifted or unshifted spectrum.
Here, the VCS possess a vector index given by
one or  other  helicity sector either corresponding to
$1$ or $2$ in the anterior study. Hence, besides of the
VCS as computed by Ali and Bagarello \cite{ab1}
and then improved in \cite{thirulo}, we have shown
that other VCS classes can be defined from the same
model.

\section{Harmonic oscillator in 3D}
\label{sect:ho3D}

In this section, we consider the harmonic
oscillator in $3D$,
\beq
H_{3D} = \frac{1}{\hbar}H'_{3D} =  \sum_{i=1}^3\omega_i (a_i^\dag a_i  + \frac12) ,\qquad
e_{n_1,n_2,n_3} = \sum_{i=1}^3 \omega_i (n_i +\frac12).
\eeq
Our aim is to find an extension of the above
VCS classes in the higher rank Hilbert space
of the $3D$ harmonic oscillator system, namely
$\mathcal{H}_{3D} = \text{span}\{ |n_1,n_2,n_3\rangle, \;
 n_i \in \mathbb{N} \}$. Here $ |n_1,n_2,n_3\rangle$
are again the eigenstates of the number operators of three
Heisenberg algebras $[a_i,a_i^\dag] = \mathbb{I}_i$.

Clearly, it can easily establish that all VCS classes with one
degree and two degrees of freedom as worked out
in Section \ref{sect:ho2D} can be reported here.
Concerning the case of one degree of freedom,
assuming again the sum is performed on $n_1$,
the VCS become tensor coherent states labeled by $n_2,n_3$.
Dealing with two degrees of freedom and one sum on $n_1$,
the whole discussion so far is again valid in
the present situation.
We will not deal with these states being already well listed in
the anterior study since they can be simply obtained by
tensoring the ket $|n_3\rangle$ to all these classes
and consequently modifying  the resolution of the identity.
 However, it should be emphasized that the classes
of states contains more than the above and include
any symmetric one obtained by interchanging
the role of 1,2 and 3.

The interesting goal here is to investigate other classes of VCS labeled by only one index, for example $n_3$ in the following,  whereas the two other indices, say $n_1$ and $n_2$, are summed.
We are led to new classes of VCS made with
two degrees of freedom. Indeed, many of the techniques
worked out before do no  longer be applied. Specifically,
when summing on both indices,
by the likely dependence of generalized factorials
on many indices, this generates what  is usually referred
to as VCS  with \emph{dependent or independent} sums.
It might become difficult to find measure densities solving the VCS since
they should not be dependent on two indices $n_1$ and
$n_2$. The analysis becomes more involved but still
a systematic approach for listing these states
remains possible.

The case of three degrees of freedom
will be not treated in length here. As we will show, the number of cases
which may lead to solvable VCS is really significant due
to the freedom in the remaining index $n_3$.
Only two simplified situations will be exhibited.

\subsection{Preliminaries: Class counting, strategy
and normalization}
\label{sub:3Dprel}

Due to the increasing number tensor indices,
the number of classes  rapidly proliferate.
We have to perform some combinatorics in order to
identify the meaningful cases which have to be studied.

\noindent{\bf Class counting -}
The definition of the VCS rests on the
generalized factorials. Combinatorially, the energy
$E_{n_1,n_2,n_3} = \omega_1n_1 + \omega_{2} n_2 +
\omega_3n_3$ can be dissected in different ways
each of which generating a different generalized
factorial associated with a degree of freedom.
Seeking for solvable VCS with two and three degrees of
freedom, the following energy-built quantities will
define the possible generalized factorials
\beq
&&
\omega_i n_i,\quad \;\; i=1,2,3,\crcr
&&
\omega_i(n_i + \kappa_{i\iche} n_{\iche}),
\quad \kappa_{i \iiche} = \frac{\omega_{\iche}}{\omega_i},\quad \;\;  i,\iche =1,2,3,\;\;  i\neq \iche, \crcr
&&
 \omega_i(n_i + \kappa_{i\iche} n_{\iche} + \kappa_{i \iiche} n_{\iiche}),
\quad
\;\;  i,\iche,\iiche =1,2,3,\;\;  \iiche \neq i\neq \iche
< \iiche,
\eeq
so that each generalized factorial $\rho_i(n_i,n_{\iche},n_{\iiche})$
can be of $4$ different forms:
\beq
\rho_i(n_i,n_{\iche},n_{\iiche}) &\in& \Big\{ \omega_i^{n_i} n_{i}!\;,\;\;
\omega_i^{n_i + \kappa_{i\iche} n_{\iche}} \Gamma[\gamma_{i\iche} + n_{\iche}]\;,\crcr
&&
 \omega_i^{n_i + \kappa_{i\iiche} n_{\iiche}} \Gamma[\gamma_{i\iiche} + n_{\iiche}]\;,\;\;
\omega_{i}^{n_i+ \kappa_{i\iche} n_{\iche} + \kappa_{i\iiche} n_{\iiche}}\Gamma[\gamma_{i} + n_i] \Big\}, \crcr
&&
\gamma_{i\iche} = 1 + \kappa_{i\iche} n_{\iche}, \qquad
\gamma_{i} = 1 + \kappa_{i\iche} n_{\iche} + \kappa_{i\iiche} n_{\iiche},
\eeq
with the same convention for the triple $(i,\iche,\iiche)$ as above.

For definiteness, we will assume that
$n_1$ and $n_2$ are summed while $n_3$ is kept fixed as
mentioned earlier.
Any over set of VCS induced by another choice
taking two over the three indices in $\{n_1,n_2,n_3\}$, will be
considered as another representative of a class
which will be listed in the sequel.
The notion of class takes now a broader but justified
sense.

Coming back to our particular choice,
$n_1$ and $n_2$ play now a completely symmetric role\footnote{
This was not the case in Section \ref{sect:ho2D},
a summation being performed only on $n_1$. }
and, discussing a class, we implicitly include  in that class
any symmetric of any representative.
Using two degrees of freedom $z_i$ and $z_j$
associated with two different towers of states,
one has two possibilities:
either the VCS is expanded versus $z_1$ and $z_2$
(tower 1 and 2) or versus $z_{1,2}$ and $z_3$ (tower $i=1$ or $2$
and $3$).
Focusing on the $(z_1,z_2)$ case with a total of $4^2$ classes,
 there are $10$ non symmetric
(i.e. any of these classes cannot be recovered from another
class in this list by just renaming $1\leftrightarrow 2$, hence
are unequivalent) classes according
to the distribution of $4$ different generalized factorials
to each variable. Besides, for the series expansion in
$(z_i,z_3)$, one has again a total of $4^2$ classes corresponding
to the number of all pairs $(\rho_i, \rho_3)$.
However, not all these pairs lead to a sensible notion of VCS class,
because, as the sum is performed on towers labeled by
$n_1$ and $n_2$, only a VCS series expansion dependent on
both $n_1$ and $n_2$
might lead to a normalizable VCS. Fixing the pair $(\rho_1, \rho_3)$ and having a closer look on which kind of generalized factorials $(\rho_1, \rho_3)$
could lead {\it a priori} to that condition, one agrees with that

- $\rho_1\equiv  \omega_1^{n_1} n_1!$ could only have
2 partners involving in their definition the index $n_{2},$
which are $\rho_3\in \{\omega_3^{n_3 + \kappa_{23} n_2}
\Gamma[\gamma_{23} + n_3], \omega_3^{n_3 + \gamma_3 - 1} \Gamma[\gamma_{3} + n_3]\}$;

- $\rho_1\in \{\omega_1^{n_1 + \kappa_{12} n_2}
\Gamma[\gamma_{12} + n_1], \omega_1^{n_1 + \gamma_1 - 1} \Gamma[\gamma_{1} + n_1]\} $ can be associated with any
$\rho_3$, since it already contains a $n_2$: we have here
(4+4) cases;

- $\rho_1\equiv \omega_1^{n_1 + \kappa_{13} n_3}
\Gamma[\gamma_{13} + n_1]$ has the same partners
has $\rho_1(n_1)= \omega_1^{n_1} n_1!$ and so
2 cases should be studied here.

Thus, the number of classes to be studied for two degrees
of freedom is $10+12=22$.

The number of unequivalent classes with three
degrees of freedom is much greater than the latter
and the total number of possible classes is $4^3$.
The two integers $n_1$ and $n_2$ playing a symmetric role, we expect a number of unequivalent
classes less than this total number. We have
seen that the number of unequivalent cases
expanding the VCS in terms of $z_1,z_2$ is $10$.
Then it remains to connect these cases to the $4$
provided by the last variable $z_3$. A rapid checking show
that these cases are all unequivalent.
Hence, for three degrees of freedom, we obtain {\it a priori}
$40$ classes.

\noindent{\bf Strategy and omissions -}
Needless to emphasize that an optimal way to study these states
is necessary. The above combinatorics have already introduced
some simplifications.
 The exchange $1\leftrightarrow 2$ becomes now a symmetry and therefore
should be extensively used in order to enlarge  the notion of class.
Furthermore,  the idea of defining some most general class and then continuously tuning the $\kappa$'s parameters in order to get simpler
cases will be used.  Hence, a key ingredient is the notion of type.
Nevertheless, at each limit,  the basic axioms should be checked
 (we will only reveal the intermediate
steps when  they have not  been  treated earlier).
Henceforth, we will not organize the remaining part of the text
centered on the notion of class, but more using the
notion of type. Practically, we will start by a general deformation
class, the \emph{ancestor},
prove that it is solvable and then derive by
continuous limit its \emph{descendant} states
(checking implicitly that they remain solvable).

The notion of sub-classes could be also introduced along
ideas of Section \ref{sect:ho2D}
but will be omitted for the sake of simplicity.
VCS with three degrees of freedom will be not treated
in detail also because of their important number.
The resulting VCS may be understood as extensions
of the VCS treated with two degrees of freedom
for the $2D$ harmonic oscillator. In that study,
three complex variables will be introduced
and only a slight discussion in the furthest situations
will be done: when the state is maximally deformed and
when it is not at all.

\noindent{\bf On the normalization condition -}
Most importantly, the axioms of VCS, namely the normalization
condition and resolution of the identity,
are  to be checked here.
Due to our particular choice
of VCS construction, i.e. by fitting the correct
exponent of the complex variables in order to match with
the generalized factorials, the resolution of the identity
will be trivialized showing thereby the robustness of our
formalism using a unique kind of moment problem.
Nevertheless, and interestingly, we discover that
the complications have migrated:
the convergence of the norm series, the latter becoming
a double power series in $\C^2$,
is  far from a trivial problem whenever one replaces ordinary
factorials by generalized factorials.
Indeed, the issue of convergence of double series is a whole
subject of investigation on its own \cite{mariaz,Bel70}. The following
theorems give some criteria for convergence of a
double series that will be extensively used in the
remaining part of the text.
To start with,  let us introduce some basic definitions.

Let $(a_{k,\ell}) \equiv (a_{k,\ell})_{k,\ell\in \mathbb{N}}$ be a double
sequence of nonzero real numbers. The double series
$\sum_{k,\ell=0}^{\infty} a_{k,\ell}$ defines two
partial series: one called \emph{row} series defined
by $\sum_{k=0}^{\infty} a_{k,\ell}$  and the other \emph{column}
series defined by $\sum_{\ell=0}^{\infty} a_{k,\ell}$.
We recall a statement that will be considered
as a definition:

\begin{proposition}[Lemma 2.1 in \cite{mariaz}]
\label{prop}
 A double series $(a_{k,\ell})$ is absolutely convergent if and only if the following conditions hold:

(i) There are $(k_0,\ell_0)\in \N$ and $\alpha_0 > 0$ such that
\bea
\sum_{k=k_0}^{m}\sum_{\ell=\ell_0}^{n} |a_{k,\ell}| \leq \alpha_0 ,\quad
\forall \, (m,n)\geq (k_0,\ell_0).
\eea
(ii) Each row series as well as each column series is absolutely convergent.
\end{proposition}
We use the shorthand notation $(k,l)\leq (m,n)$ for the
partial order in $\N^2$, $k\leq m$ and $l \leq n$.
A well known consequence of this statement is the following:
 \begin{corollary}[Comparison test]
\label{coro}
Let $(a_{k,\ell})$ and $(b_{n,k})$ be double sequences
of nonzero numbers. Assume that $\exists K_0,L_0 \in \N$ such that
$\forall (k,\ell)\geq (K_0,L_0)$, $|a_{k,\ell}|\leq |b_{n,k}|$.
If $\sum_{k,\ell=0}^{\infty} b_{k,\ell}$ is absolutely convergent
then so is $\sum_{k,\ell=0}^{\infty} a_{k,\ell}$.
\end{corollary}
\noindent{\bf Proof.} Let us consider $k_0,\ell_0 \in \N$ be such
that $\exists\alpha_0>0$ and $\sum_{k=k_0}^{m}\sum_{\ell=\ell_0}^{n} |b_{k,\ell}|\leq \alpha_0 $.
Then, four cases may occur:
\begin{enumerate}
\item[(i)] $(k_0,\ell_0)\geq (K_0,L_0)$, then for
$\forall (k,\ell)\geq (k_0,\ell_0)$, we have $|a_{k,\ell}|\leq |b_{n,k}|$ and therefore
\bea
\forall (m,n)\geq (k_0,\ell_0)\geq (K_0,L_0)  \;\;
\Rightarrow  \;\;
\sum_{k=k_0}^{m}\sum_{\ell=\ell_0}^{n} |a_{k,\ell}|\leq
\sum_{k=k_0}^{m}\sum_{\ell=\ell_0}^{n} |b_{k,\ell}|\leq
\alpha_0
\eea
\item[(ii)] $k_0\geq K_0$ and $\ell_0 < L_0$
(resp. $\ell_0\geq L_0$ and $k_0 < K_0$),
and so
\beq
&&
\forall m \geq k_0 \geq K_0 , \;\;\forall n \geq  L_0 >\ell_0
 \;\;
\Rightarrow  \;\;
\sum_{k=k_0}^{m}\sum_{\ell=L_0}^{n} |a_{k,\ell}|\leq
\sum_{k=k_0}^{m}\sum_{\ell=L_0}^{n} |b_{k,\ell}|\leq
\alpha_0 \\
&&\Big(\text{resp.}\;\;
\forall m \geq K_0 > k_0 , \;\;\forall n \geq  L_0 \geq \ell_0
 \;\;
\Rightarrow  \;\;
\sum_{k=K_0}^{m}\sum_{\ell=\ell_0}^{n} |a_{k,\ell}|\leq
\sum_{k=K_0}^{m}\sum_{\ell=\ell_0}^{n} |b_{k,\ell}|\leq
\alpha_0 \Big)
\nonumber
\eeq
\item[(iii)] $(k_0,\ell_0) < (K_0,L_0)$, and then
\bea
\forall (m,n)\geq (K_0,L_0)\geq (k_0,\ell_0)  \;\;
\Rightarrow  \;\;
\sum_{k=K_0}^{m}\sum_{\ell=L_0}^{n} |a_{k,\ell}|\leq
\sum_{k=K_0}^{m}\sum_{\ell=L_0}^{n} |b_{k,\ell}|\leq
\alpha_0 .
\eea
\end{enumerate}
Finally, the ordinary theorem of comparison
for simple series proves that row and
column series of $\sum_{k,\ell}|a_{k,\ell}|$
are convergent if row and column series
of $\sum_{k,\ell} |b_{k,\ell}|$
are, respectively. One concludes by Proposition \ref{prop}.
\qed

More involved ratio tests are also useful.

\begin{theorem}[Ratio test, Theorem 2.7 in \cite{mariaz}]
\label{ratio}
Let $(a_{k,\ell})$ be a double sequence of nonzero numbers
such that either $|a_{k,\ell+1}|/|a_{k,\ell}| \to a$ or
$|a_{k+1,\ell}|/|a_{k,\ell}| \to \tilde{a}$ as both $k\to \infty$ and $\ell\to \infty$, where
$a,\tilde{a} \in \mathbb{R}\cup \{\infty\}$.

(i) Suppose each row-series as well as each column-series corresponding to $\sum_{k,\ell=0}^{\infty} a_{k,\ell}$ is absolutely
convergent.
If $a < 1$ or $\tilde a < 1$, then $\sum_{k,\ell=0}^{\infty} a_{k,\ell}$
 is absolutely convergent.

(ii) If $a > 1$ or  $\tilde a > 1$, then
$\sum_{k,\ell=0}^{\infty} a_{k,\ell}$ is divergent.
\end{theorem}
Furthermore the following statement holds
\begin{theorem}[Ratio-Comparison test, Theorem 2.9 in \cite{mariaz}]
\label{compa}
Let $(a_{k,\ell})$ and $(b_{k,\ell})$ be double
sequences with $b_{k,\ell} > 0$ for all $(k,\ell) \in \N$.

(i) Suppose each row-series as well as each column-series corresponding to $\sum_{k,\ell=0}^{\infty} a_{k,\ell}$ is convergent.
If
$|a_{k,\ell+1}|b_{k,\ell}\leq |a_{k,\ell}|b_{k,\ell+1}$
and $|a_{k+1,\ell}|b_{k,\ell}\leq |a_{k,\ell}|b_{k+1,\ell}$
whenever $k$ and $\ell$ are large, and if
$\sum_{k,\ell=0}^{\infty} b_{k,\ell}$ is
convergent, then so is $\sum_{k,\ell=0}^{\infty} a_{k,\ell}$.

(ii) If $|a_{k,\ell+1}|b_{k,\ell}\geq |a_{k,\ell}|b_{k,\ell+1}>0$
whenever $\ell$ is large and $k\in \N$,
and $|a_{k+1,\ell}|b_{k,\ell}\geq |a_{k,\ell}|b_{k+1,\ell}>0$
whenever $k$ is large and $\ell\in \N$, and if
$\sum_{k,\ell=0}^{\infty} b_{k,\ell}$ is divergent,
then so is $\sum_{k,\ell=0}^{\infty} |a_{k,\ell}|$.
\end{theorem}

\subsection{VCS with two degrees of freedom}
\label{sub:ho3D2d}

The VCS classes studied here possess two degrees
of freedom and they are generated by a
double series coined by two towers, $n_1$ and $n_2$,
(over the three) of the Hilbert space $\mathcal{H}_{3D}$.

There are, in conformity with the discussion of Subsection
\ref{sub:3Dprel},  two kinds of classes
either  introduced  by the couple $(z_1,z_2)$
or by the couple  $(z_1,z_3)$
(the case $(z_2,z_3)$ will fall into the same classes
of this latter).

{\bf Case $(12)$ -}
There is two main situations:  first, the generalized
factorials $\rho_1$ or $\rho_2$ do not depend
 on both $n_1$ and $n_2$, then the sums in $n_1$
and $n_2$ become independent  of one another;
this case occurs when
\bea
\begin{array}{ccc}
\hspace*{-1.8cm}
\rho_1(n_1) =\omega_1^{n_1} n_1!
&\text{and}&\!
\hspace*{-1.5cm}
\rho_2(n_2) =\omega_2^{n_2} n_2! \\
\hspace*{-1.8cm}
\rho_1(n_1) =\omega_1^{n_1} n_1!
&&\quad\!\!
\rho_2(n_2,n_3) =\omega_2^{n_2+\kappa_{23}n_3} \Gamma[\gamma_{23} +n_2] \\
\rho_1(n_1,n_3) =\omega_1^{n_1+\kappa_{13}n_3 } \Gamma[\gamma_{13} +n_1]
&&\quad
\rho_2(n_2,n_3) =\omega_2^{n_2+\kappa_{23}n_3}
\Gamma[\gamma_{23} +n_2];
\end{array}
\label{indeptsum}
\eea
second, one of $\rho_1$ or $\rho_2$ does involve
both variables $n_1$ and $n_2,$ then the sums
in these indices become dependent;  this situation
concerns the following factorials:
\bea
\begin{array}{ccc}
\hspace*{-1.8cm}
\rho_1(n_1) =\omega_1^{n_1} n_1!
&\hspace*{-1.5cm}\text{and}&
\hspace*{-0.3cm}
\rho_2(n_2,n_1) =\omega_2^{n_2+\kappa_{21}n_1} \Gamma[\gamma_{21} +n_2]\\
\hspace*{-1,8cm} \rho_1(n_1) =\omega_1^{n_1} n_1!
&&
\rho_2(n_2,n_1,n_3) =\omega_2^{n_2+\kappa_{23}n_3 + \kappa_{21} n_1} \Gamma[\gamma_{2} +n_2] \\
\rho_{1}(n_1,n_2) = \omega_1^{n_1+\kappa_{12}n_2} \Gamma[\gamma_{12} +n_1]
&&\hspace*{-0.3cm}  \rho_2(n_2,n_1) = \omega_2^{n_2+\kappa_{21}n_1} \Gamma[\gamma_{21} +n_2]
\\
\rho_{1}(n_1,n_2) = \omega_1^{n_1+\kappa_{12}n_2} \Gamma[\gamma_{12} +n_1]
&&\hspace*{-0.3cm}
\rho_2(n_2,n_3) =\omega_2^{n_2+\kappa_{23}n_3 } \Gamma[\gamma_{23} +n_2]
\\
\rho_{1}(n_1,n_2) = \omega_1^{n_1+\kappa_{12}n_2} \Gamma[\gamma_{12} +n_1]
&&
\rho_2(n_2,n_1,n_3) =\omega_2^{n_2+\kappa_{23}n_3 + \kappa_{21} n_1} \Gamma[\gamma_{2} +n_2] \\
\rho_1(n_1,n_3) =\omega_1^{n_1+\kappa_{13}n_3 } \Gamma[\gamma_{13} +n_1]
&&
\rho_2(n_2,n_1,n_3) =\omega_2^{n_2+\kappa_{23}n_3 + \kappa_{21} n_1} \Gamma[\gamma_{2} +n_2] \\
\;\;
\rho_1(n_1,n_2,n_3) =\omega_1^{n_1+\kappa_{12}n_2 + \kappa_{13} n_3} \Gamma[\gamma_{1} +n_1]
&&
\rho_2(n_2,n_1,n_3) =\omega_2^{n_2+\kappa_{23}n_3 + \kappa_{21} n_1} \Gamma[\gamma_{2} +n_2].
\label{deptsum3DA}
\end{array}
\eea
{\bf Case $(i3)$ $i=1,2$ -} Again the same ideas hold here, either
sums can be decoupled or mutually dependent given a choice $(\rho_1,\rho_3)$.
The following factorials generate independent
sums:
\beq
\begin{array}{ccc}
\hspace*{-1.8cm}
\rho_1(n_1) =\omega_1^{n_1} n_1!
&\text{and}&
\rho_3(n_3,n_2) =\omega_3^{n_3 + \kappa_{32} n_2}
\Gamma[\gamma_{32} + n_3]\\
\rho_1(n_1,n_3) =
\omega_1^{n_1 + \kappa_{13} n_3}
\Gamma[\gamma_{13} + n_1]
&&\quad\!\!
\rho_3(n_3,n_2)=\omega_3^{n_3 + \kappa_{32} n_2}
\Gamma[\gamma_{32} + n_3] .
\end{array}
\label{indeptsum13}
\eeq
Meanwhile, the following one involve dependent sums
\bea
\begin{array}{ccc}
\\
\hspace*{-1.8cm}
\rho_1(n_1) =\omega_1^{n_1} n_1!
&\text{and}&
\rho_3(n_3,n_1,n_2) =
\omega_3^{n_3 +\kappa_{31}n_1 + \kappa_{32}n_2} \Gamma[\gamma_{3} + n_3] \\
\rho_1(n_1,n_3) =\omega_1^{n_1+\kappa_{13}n_3 } \Gamma[\gamma_{13} +n_1]
&&
\rho_3(n_3,n_1,n_2) =
\omega_3^{n_3+\kappa_{31}n_1 + \kappa_{32}n_2} \Gamma[\gamma_{3} + n_3]\\
\rho_{1}(n_1,n_2) = \omega_1^{n_1+\kappa_{12}n_2} \Gamma[\gamma_{12} +n_1]
&&\hspace*{-2.1cm}
\rho_3(n_3) =\omega_3^{n_3} n_3!
\\
\rho_{1}(n_1,n_2) = \omega_1^{n_1+\kappa_{12}n_2} \Gamma[\gamma_{12} +n_1]
&&\hspace*{-0.3cm}
\rho_3(n_3,n_1) =\omega_3^{n_3+\kappa_{31}n_1} \Gamma[\gamma_{31} +n_3]
\\
\rho_{1}(n_1,n_2) = \omega_1^{n_1+\kappa_{12}n_2} \Gamma[\gamma_{12} +n_1]
&&\hspace*{-0.3cm}
\rho_3(n_3,n_2) =\omega_3^{n_3+\kappa_{32}n_2} \Gamma[\gamma_{32} +n_3] \\
\rho_{1}(n_1,n_2) = \omega_1^{n_1+\kappa_{12}n_2} \Gamma[\gamma_{12} +n_1]
&&
\rho_3(n_3,n_1,n_2) =\omega_3^{n_3+\kappa_{31}n_1 + \kappa_{32}n_2} \Gamma[\gamma_{3} +n_3] \\
\;\;\,
\rho_{1}(n_1,n_2,n_3) = \omega_1^{n_1+\kappa_{12}n_2 + \kappa_{13}n_3} \Gamma[\gamma_{1} +n_1]
&&\hspace*{-2.1cm}
\rho_3(n_3) =\omega_3^{n_3} n_3!
\\\;\;\,
\rho_{1}(n_1,n_2,n_3) = \omega_1^{n_1+\kappa_{12}n_2 + \kappa_{13}n_3} \Gamma[\gamma_{1} +n_1]
&&\hspace*{-0.3cm}
\rho_3(n_3,n_1) =\omega_3^{n_3+\kappa_{31}n_1} \Gamma[\gamma_{31} +n_3]
\\\;\;\,
\rho_{1}(n_1,n_2,n_3) = \omega_1^{n_1+\kappa_{12}n_2 + \kappa_{13}n_3} \Gamma[\gamma_{1} +n_1]
&&\hspace*{-0.3cm}
\rho_3(n_3,n_2) =\omega_3^{n_3+\kappa_{32}n_2} \Gamma[\gamma_{32} +n_3] \\
\;\;\,
\rho_{1}(n_1,n_2,n_3) = \omega_1^{n_1+\kappa_{12}n_2 + \kappa_{13}n_3} \Gamma[\gamma_{1} +n_1]
&&\,
\rho_3(n_3,n_1,n_2) =\omega_3^{n_3+\kappa_{31}n_1 + \kappa_{32}n_2} \Gamma[\gamma_{3} +n_3] .
\end{array}
\label{deptsum3DB}
\eea
We are now in position to define VCS classes.

\subsubsection{
 $(\gamma_{13},\gamma_{23})$- and
$(\gamma_{13},\gamma_{32})$-deformed CS and descendants}
\label{ho3D2d2c}

We treat first the so-called VCS classes generated
by independent sums.
According to (\ref{indeptsum}), it proves to be judicious to work
with the last quantities
$\rho_1(n_1,n_3)$ and $\rho_{2}(n_2,n_3)$,
since the previous one can be recovered by some limits.
Hence, we consider
\bea
\rho_1(n_1,n_3) =\omega_1^{n_1+\kappa_{13}n_3 } \Gamma[\gamma_{13} +n_1]\;,
\quad
\rho_2(n_2,n_3) =\omega_2^{n_2+\kappa_{23}n_3}
\Gamma[\gamma_{23} +n_2].
\eea
Introduce two complex variables $z_1=r_1e^{i\theta_1}$ and $z_2=r_2e^{i\theta_2}$,
and the $(\gamma_{13},\gamma_{23})$-deformed class given by the state
\bea
|z_{1}, z_{2}, n_3 \rangle =
\mathcal N(z_{1}, z_{2}, n_3)^{-\frac12}
 \sum_{n_{1},n_{2} = 0}^{\infty}
\frac{
z_{1}^{ n_{1} +\kappa_{13}n_{3} }
z_{2}^{ n_{2} +\kappa_{23}n_{3} }}{
 \left[
\omega_{1}^{ n_{1} +\kappa_{13}n_{3} }
\omega_{2}^{ n_{2} +\kappa_{23}n_{3} }
\Gamma[\gamma_{13} +n_1]
\Gamma[\gamma_{23} +n_2]\right]^{\frac12} }\,
\, |  n_{1}, n_{2},n_{3}\rangle ,
\label{3D2d2c}
\eea
where $n_3$ is the vector index.
The state is normalizable provided
\beq
&&
\mathcal N(z_{1}, z_{2}, n_3)=
\left[\frac{
|z_{1}|^{2\kappa_{13}}
|z_{2}|^{ 2\kappa_{23}}}{
\omega_{1}^{\kappa_{13} }
\omega_{2}^{\kappa_{23}} }\right]^{n_3}
\sum_{n_{1},n_{2} = 0}^{\infty}
\frac{
|z_{1}|^{ 2n_{1} }
|z_{2}|^{ 2n_{2} }}{
 \left[
\omega_{1}^{ n_{1}  }
\omega_{2}^{ n_{2} }
\Gamma[\gamma_{13} +n_1]
\Gamma[\gamma_{23} +n_2]\right]} \crcr
&&
= \left[\frac{
|z_{1}|^{2\kappa_{13}}
|z_{2}|^{ 2\kappa_{23}}}{
\omega_{1}^{\kappa_{13} }
\omega_{2}^{\kappa_{23}} }\right]^{n_3}
\frac{1}{\Gamma[\gamma_{13}]\Gamma[\gamma_{23}]}
\,  _1F_1\left(1,\gamma_{13}; \frac{|z_1|^2}{\omega_1}\right)\,  _1F_1\left(1,\gamma_{23}; \frac{|z_2|^2}{\omega_2}\right)
\eeq
is convergent. This is indeed the case as one easily can check
by a ratio test for each sector $n_1$ and $n_2$ which, in this
case, factorize. Alternatively, one can surely perform a comparison test, as
long as $n_i \geq 1$,
\bea
\frac{1}{\Gamma[\gamma_{13} +n_1]
\Gamma[\gamma_{23} +n_2]} \leq  \frac{1}{n_1!n_2!},
\qquad \gamma_{ij} \geq 1,
\label{compa0}
\eea
and then conclude by Corollary \ref{coro},
since the r.h.s is the term of a
product of exponential series.

The resolution of the identity
requires that
\beq
\int d\mu(z_1,z_2,n_3)\;
|z_{1}, z_{2}, n_3 \rangle
\langle z_{1}, z_{2}, n_3 | = I_{n_3}
= \sum_{n_1,n_2=0}^\infty |n_{1}, n_{2}, n_3 \rangle
\langle n_{1}, n_{2}, n_3 |,
\label{resol3D2d2c}
\eeq
where
\bea
 d\mu(z_1,z_2,n_3) =
\frac{1}{\pi^2}\,\mathcal{N}(|z_1|,|z_2|,n_3) \chi(r_1,r_2,n_3)
\prod_{i=1}^2 r_i dr_i d\theta_i.
\label{mes2deg}
\eea
The resulting moment problem is of the form,
using $u_i=r_i^2$,
\bea
\int du_1 du_2 \;\chi(r_1,r_2,n_3)
\frac{
u_{1}^{ n_{1}+ \kappa_{13}n_3}
u_{2}^{ n_{2}+ \kappa_{23}n_3}}{
\omega_{1}^{ n_{1} +  \kappa_{13} n_3}
\omega_{2}^{ n_{2} +  \kappa_{13} n_3}
}
=\Gamma[\gamma_{13} +n_1]
\Gamma[\gamma_{23} +n_2]
\label{resolu3D2d2c}
\eea
and is solved by $\chi(r_1,r_2,n_3)=\varrho_1(r_1)\varrho_2(r_2)$
where $\varrho_i(r_i)$ are given by (\ref{hodenssmart}).

Remark that (as already discussed in Section \ref{sub:3Dprel})  the exponents of the complex variables
$z_i^{\bullet}$ and frequency $\omega_i^{\bullet}$ have been adjusted in order to exactly match with the argument of the gamma
function $\Gamma[\bullet+1]$ in the moment problem.
Thereby the latter will be solved by a simple exponential as given by  (\ref{hodenssmart}).
All the classes here and after, being designed in this
particular form, possess a resolution of the identity
of this sort. We will not discuss the requirement
of the resolution of the identity
which should be obvious provided that
(a) the phase
 integration in $\theta_i$ yields enough constraints
to project the resolution of the identity on the
real domain, and (b) one ensures
at least that the bounds of integration of the moment problem
are the same as (\ref{hodenssmart}).
This means that all states should be of infinite radius
of convergence in $|z_i|$ under some conditions on
the parameters $\kappa_{jl}$.
(a) will be only discussed when it leads to nontrivial
facts whereas to prove (b) will be our main goal for the remaining part.

Taking the limit $\kappa_{13} \to 0$  in (\ref{3D2d2c}), we get another class of VCS
\bea
|z_{1}, z_{3}, n_3 \rangle =
\mathcal N(z_{1}, z_{2}, n_3)^{-\frac12}
 \sum_{n_{1},n_{2} = 0}^{\infty}
\frac{
z_{1}^{ n_{1}  }
z_{2}^{ n_{2} +\kappa_{23}n_{3} }}{
 \left[
\omega_{1}^{ n_{1} }
\omega_{2}^{ n_{2} +\kappa_{23}n_{3} }
n_1!
\Gamma[\gamma_{23} +n_2]\right]^{\frac12} }\,
\, |  n_{1}, n_{2},n_{3}\rangle ,
\label{3D2d1c}
\eea
and taking both limits $\kappa_{13} \to 0$ and
 $\kappa_{23} \to 0$, one gets the straightforward
extension of the $(1,1)$-class:
\bea
|z_{1}, z_{2},n_3 \rangle =
\mathcal N(z_{1}, z_{2})^{-\frac12}
 \sum_{n_{1},n_{2} = 0}^{\infty}
\frac{
z_{1}^{ n_{1}  }
z_{2}^{ n_{2} }}{
 \left[
\omega_{1}^{ n_{1} }
\omega_{2}^{ n_{2}  }
n_1! n_2!\right]^{\frac12} }\,
\, |  n_{1}, n_{2},n_{3}\rangle = |z_1\rangle \otimes |z_2\rangle \otimes |n_3\rangle,
\label{3D2d0c}
\eea
where in the last expression $|z_i\rangle$ stands for
the CS for the harmonic oscillator in $1D$.
Indeed (\ref{3D2d0c}) can be written, in sloppy symbols,
$|z_{1}, z_{2},n_3 \rangle = \mathcal{N}^{-\frac12} \sum
(\mathcal{N}^{\frac12}  |z_1,n_2\rangle) \otimes |n_3\rangle$,
where $\mathcal{N}^{\frac12}  |z_1,n_2\rangle$
is the non normalized state associated with the $(1,1)$-class
(\ref{ho2d2c}). Furthermore,  (\ref{3D2d1c})
is nothing but the state generated by
$|z_{1}, z_{2},n_3 \rangle = \mathcal{N}^{-\frac12}
 (\mathcal{N}^{\frac12}  |z_1\rangle) \otimes \sum (\mathcal{N}^{\frac12} |z_2,n_3\rangle)$,
where $\mathcal{N}^{\frac12} |z_2,n_3\rangle$ is the non normalized
state associated with (\ref{ho2d1c}) namely the $(\gamma_{23},1)$-VCS class. Both classes (\ref{3D2d1c}) and (\ref{3D2d0c})
are normalized (by a comparison procedure as was performed
in (\ref{compa0}))
and solved by the same densities as for (\ref{3D2d2c}).
Thus, surprisingly, summing an infinite tower of
VCS like was shown before yields again a VCS.
We will denote (\ref{3D2d1c}) and (\ref{3D2d0c})
by $(1,\gamma_{23})$- and $(1,1)$-VCS classes.
Note that taking another limit $\kappa_{23} \to 0$ in (\ref{3D2d2c})
yields then a $(\gamma_{13},1)$-VCS class (with
a similar meaning but on the sector 1,3)
which gives after $\kappa_{31}\to 0$ the same $(1,1)$-class (\ref{3D2d0c}).

Considering now independent sums
provided by Case (13), we have the quantities given
by (\ref{indeptsum13}):
\bea
\rho_1(n_1,n_3) =
\omega_1^{n_1 + \kappa_{13} n_3}
\Gamma[\gamma_{13} + n_1],
\qquad
\rho_3(n_3,n_2)=\omega_3^{n_3 + \kappa_{32} n_2}
\Gamma[\gamma_{32} + n_3]
\eea
which define the $(\gamma_{13},\gamma_{32})$-deformed CS
as
\bea
|z_{1}, z_{3}, n_3 \rangle =
\mathcal N(z_{1}, z_{3}, n_3)^{-\frac12}
 \sum_{n_{1},n_{2} = 0}^{\infty}
\frac{
z_{1}^{ n_{1} +\kappa_{13}n_{3} }
z_{3}^{ n_{3} +\kappa_{32}n_{2} }}{
 \left[
\omega_{1}^{ n_{1} +\kappa_{13}n_{3} }
\omega_{3}^{ n_{3} +\kappa_{32}n_{2}  }
\Gamma[\gamma_{13} +n_1]
\Gamma[\gamma_{32} +n_3]\right]^{\frac12} }\,
\, |  n_{1}, n_{2},n_{3}\rangle .
\label{3D2d2cB}
\eea
The norm of this vector is unity if
\beq
&&
\mathcal N(z_{1}, z_{3}, n_3) =
\left[\frac{
|z_{1}|^{2\kappa_{13} }
|z_{3}|^{ 2  }}{
\omega_{1}^{ \kappa_{13}}
\omega_{3}}\right]^{n_3}
\sum_{n_{1},n_{2} = 0}^{\infty}
\frac{
|z_{1}|^{2 n_{1} }
|z_{3}|^{ 2\kappa_{32}n_{2}}}{
\omega_{1}^{ n_{1} }
\omega_{3}^{\kappa_{32}n_{2}  }
\Gamma[\gamma_{13} +n_1]
\Gamma[\gamma_{32} +n_3] } \crcr
&& = \left[\frac{
|z_{1}|^{2\kappa_{13} }
|z_{3}|^{ 2  }}{
\omega_{1}^{ \kappa_{13}}
\omega_{3}}\right]^{n_3}\frac{1}{\Gamma[\gamma_{13}]}
\fun\left(1; \gamma_{13};
\frac{ |z_{1}|^{2 } }{\omega_1}\right)\sum_{n_{2} = 0}^{\infty}
\frac{
|z_{3}|^{ 2\kappa_{32}n_{2}}}{
\omega_{3}^{\kappa_{32}n_{2}  }
\Gamma[\gamma_{32} +n_3] }.
\label{3D2d2cBnorm}
\eeq
This double series factorizes: the series in $n_1$
is again bounded by an exponential and so
we only have to prove that the series in $n_2$ is convergent.
Using the $\Gamma$-Stirling approximation
\bea
\Gamma[z] \sim \sqrt{\frac{2\pi}{z}}
\left(\frac{z}{e}\right)^{z}
\eea
valid at large $\Re (z)$ and for $ \arg(z) <\pi -\epsilon$
for $\epsilon >0$, the following relation holds
as long as $\kappa_{32} > 0$:
\bea
\lim_{n_2\to \infty}
\frac{\Gamma[1+ \kappa_{32} n_2 +n_3]}{\Gamma[1+ \kappa_{32}(1+ n_2) +n_3]} =
\lim_{n_2\to \infty}
\left[1+ \kappa_{32}(1+ n_2) +n_3\right]^{- \kappa_{32}}=0.
\label{bound}
\eea
This means
that the radius of convergence of the series is infinite again.
Note the important fact that at $\kappa_{32}=0$,
the state is non normalizable, hence the VCS is not defined
at that limit.

The resolution of the identity is again
of the form (\ref{resol3D2d2c})
having an integration measure like (\ref{mes2deg}) including
a factor of $\kappa_{32}$ (canceling a contribution to
a phase integration in $\kappa_{32}\theta_3$) and with moment problem,
similar to (\ref{resolu3D2d2c}) hence solvable by the same
kind of densities. Henceforth, we will not mention
the constant corrections to the measure up to
factors of $\kappa_{i\iche}$ due to unessential phase
integrations which may occur.

We also have here a unique class which can be obtained
at $\kappa_{13}\to 0$, i.e.
\beq
&&
|z_{1}, z_{3}, n_3 \rangle =
\mathcal N(z_{1}, z_{3}, n_3)^{-\frac12}
 \sum_{n_{1},n_{2} = 0}^{\infty}
\frac{
z_{1}^{ n_{1}  }
z_{3}^{ n_{3} +\kappa_{32}n_{2} }}{
 \left[
\omega_{1}^{ n_{1} }
\omega_{3}^{ n_{3} +\kappa_{32}n_{2}  }
n_1!
\Gamma[\gamma_{32} +n_3]\right]^{\frac12} }\,
\, |  n_{1}, n_{2},n_{3}\rangle ,
\label{3D2d2cB1} \\
&&
\mathcal N(z_{1}, z_{3}, n_3) =
 \left[\frac{
|z_{3}|^{ 2  }}{
\omega_{3}}\right]^{n_3}
\exp\left\{
\frac{ |z_{1}|^{2 } }{\omega_1}\right\}
\sum_{n_{2} = 0}^{\infty}
\frac{
|z_{3}|^{ 2\kappa_{32}n_{2}}}{
\omega_{3}^{\kappa_{32}n_{2}  }
\Gamma[\gamma_{32} +n_3] } ,
\eeq
which, as expected, consists in the VCS given by the
other set of generalized factorial in (\ref{indeptsum13})
generating indepedent sums.
Note that $\kappa_{32} \to 0$ is yet forbidden.
We cannot recovered the first class VCS (\ref{3D2d0c})
hence the type of these VCS classes (\ref{3D2d2cB})
and (\ref{3D2d2cB1}) is not of any type yet encountered
so far. This can be easily understood given the fact
that the VCS (\ref{3D2d0c})  is a continuous limit deformation of
another VCS class which can be only recovered at the limit $\kappa_{23}\to 0$ and so at $\kappa_{32}\to \infty$.
We mention also that (\ref{3D2d2cB1}) can be simply viewed
as
$
|z_{1}, z_{3}, n_3 \rangle $ $=\mathcal N^{-\frac12}
(\mathcal N^{\frac12} |z_1 \rangle)  \otimes
\sum  (\mathcal N^{\frac12} |z_{3}, n_3 \rangle)
$
where $ (\mathcal N^{\frac12} |z_{3}, n_3 \rangle)$
is the unormalized fake dual $(1,\gamma_{32})$ (\ref{ho2d3c})
for the sector 2,3.

\subsubsection{
$(\gamma_{i},\gamma_{kl})$-deformed CS and descendants}
\label{ho3D2d9c}
This section deals with the analysis of dependent sums.
We study the most general states and
derive its descendants by continuous limits.

The generalized factorials
\bea
\rho_1(n_1,n_3) =\omega_1^{n_1+\kappa_{13}n_3 } \Gamma[\gamma_{13} +n_1]
\qquad
\rho_2(n_2,n_1,n_3) =
\omega_2^{n_2+\kappa_{23}n_3 + \kappa_{21} n_1} \Gamma[\gamma_{2} +n_2]
\label{genfact3D9c}
\eea
 yield the states
\bea
|z_{1}, z_{2}, n_3 \rangle =
\mathcal N(z_{1}, z_{2}, n_3)^{-\frac12}
 \sum_{n_{1},n_{2} = 0}^{\infty}
\frac{
z_{1}^{ n_{1} +\kappa_{13}n_{3} }
z_{2}^{ n_{2} +\gamma_{2} - 1 }}{
 \left[
\omega_{1}^{ n_{1} +\kappa_{13}n_{3} }
\omega_{2}^{ n_{2} +\gamma_{2} - 1}
\Gamma[\gamma_{13} +n_1]
\Gamma[\gamma_{2} +n_2]\right]^{\frac12} }\,
\, |  n_{1}, n_{2},n_{3}\rangle
\label{3D2d9c}
\eea
with normalization factor
\bea
\mathcal N(z_{1}, z_{2}, n_3) =
 \left[\frac{
|z_{1}|^{2\kappa_{13} }
|z_{2}|^{ 2\kappa_{23}   }}{
\omega_{1}^{ \kappa_{13}}
\omega_2^{\kappa_{23}}}\right]^{n_3}
\sum_{n_1,n_{2} = 0}^{\infty}
\frac{
|z_{1}z_{2}^{\kappa_{21}}|^{2n_{1} }
|z_{2}|^{2 n_{2} } }{
(\omega_1\omega_2^{\kappa_{21}})^{n_{1} }
\omega_{2}^{ n_{2} }
 \Gamma[\gamma_{13} +n_1]
\Gamma[\gamma_{2} +n_2]  } .
\label{normeici}
\eea
In this situation, the sums are dependent and do not factorize.
Therefore we need a criterion to analyze the convergence
of the double series (\ref{normeici}).
Calculating the convergence of row and column series,
one finds that they are indeed of infinite radius
of convergence, $|z_1 z_2^{\kappa_{21}}|>0$
and $|z_2|>0$ which entail also that $|z_1|>0$.
Corollary \ref{coro} is sufficient
to prove the absolute convergence here since, fortunately,
a similar inequality as (\ref{compa0}) can be applied for
$n_i\geq 1$ and $\gamma_{\cdot} \geq 1$.
Thus, the normalization factor converges everywhere
in $\C^2$.

Treating the VCS defined through (\ref{indeptsum})
and (\ref{deptsum3DA}),
Corollary \ref{coro} will be always sufficient in order to prove
the convergence of the double series in all cases provided the fact that the row and column series will be convergent.
Indeed, all VCS normalizations
defined through the generalized factorials
as defined by these cases, can be all bounded by or compared
to the double exponential series with general term
$z^{n_1}_1z^{n_2}_2/(n_1!n_2!)$.
In contradistinction, with classes which will
be defined by (\ref{indeptsum13}) and (\ref{deptsum3DB}),
a greater effort will be required to prove the convergence of the
series.

The following successive limits yield relevant VCS:
taking $\kappa_{13}\to 0$ from (\ref{genfact3D9c})
yields
\bea
\rho_1(n_1) =\omega_1^{n_1} n_1!,
\qquad
\rho_2(n_2,n_1,n_3) =
\omega_2^{n_2+\kappa_{23}n_3 + \kappa_{21} n_1} \Gamma[\gamma_{2} +n_2]
\eea
and, considering again $\kappa_{23} \to 0$, one has
\bea
\rho_1(n_1) =\omega_1^{n_1} n_1 !,
\qquad
\rho_2(n_2,n_1) =
\omega_2^{n_2+ \kappa_{21} n_1} \Gamma[\gamma_{21} +n_2].
\eea
Then it still remains a last limit which is
$\kappa_{21}\to 0$ giving
\bea
\rho_1(n_1) =\omega_1^{n_1} n_1!,
\qquad
\rho_2(n_2) =
\omega_2^{n_2}n_2!
\eea
leading to the $(1,1)$-VCS.
Still from (\ref{genfact3D9c}), but now considering
$\kappa_{21}\to 0$, we get
\bea
\rho_1(n_1,n_3) =\omega_1^{n_1+\kappa_{13}n_3 } \Gamma[\gamma_{13} +n_1],
\qquad
\rho_2(n_2,n_3) =
\omega_2^{n_2+\kappa_{23}n_3 } \Gamma[\gamma_{23} +n_2]
\eea
generating the $(\gamma_{13},\gamma_{23})$-VCS class
(\ref{3D2d2c}) and from these, other limits can be performed.
Finally, by the limit $\kappa_{23} \to 0$ in (\ref{genfact3D9c}), we have
\bea
\rho_1(n_1,n_3) =\omega_1^{n_1+\kappa_{13}n_3 } \Gamma[\gamma_{13} +n_1],
\qquad
\rho_2(n_2,n_1) =
\omega_2^{n_2+\kappa_{21}n_1 } \Gamma[\gamma_{21} +n_2],
\eea
then, we perform a symmetry $(1\leftrightarrow 2)$ in order
to recover another element of the list (\ref{deptsum3DA}):
\bea
\rho_2(n_2,n_3) =\omega_2^{n_2+\kappa_{23}n_3 } \Gamma[\gamma_{23} +n_2],
\qquad
\rho_1(n_1,n_2) =
\omega_1^{n_1+\kappa_{12}n_2 } \Gamma[\gamma_{12} +n_1].
\eea
The number of VCS classes involving dependent sums
and their possible link by deformation
belonging to Case (12) and having (\ref{genfact3D9c})
as ancestor has been exhausted.

Let us now  discuss  Case (13). Starting
by the factorials
\bea
\rho_{1}(n_1,n_2,n_3) = \omega_1^{n_1+\kappa_{12}n_2 + \kappa_{13}n_3} \Gamma[\gamma_{1} +n_1] ,
\qquad
\rho_3(n_3,n_2) =\omega_3^{n_3+\kappa_{32}n_2} \Gamma[\gamma_{32} +n_3],
\label{genfact2dB9c}
\eea
one can build the following states
\bea
|z_{1}, z_{3}, n_3 \rangle =
\mathcal N(z_{1}, z_{3}, n_3)^{-\frac12}
 \sum_{n_{1},n_{2} = 0}^{\infty}
\frac{
z_{1}^{ n_{1} +\gamma_1 - 1 }
z_{3}^{ n_{3}+\kappa_{32}n_2 } }{
 \left[
\omega_1^{n_1+\gamma_1 - 1 }
\omega_3^{n_3+\kappa_{32}n_2 } \Gamma[\gamma_{1} +n_1]
\Gamma[\gamma_{32} +n_3] \right]^{\frac12} }\,
\, |  n_{1}, n_{2}, n_{3} \rangle ,
\label{3D2dB9c}
\eea
with normalization condition
\beq
&&
\mathcal N(z_{1}, z_{3}, n_3) =  \left[\frac{
|z_{1}|^{2\kappa_{13} }
|z_{3}|^{ 2  } }{
\omega_{1}^{ \kappa_{13}}
\omega_3}\right]^{n_3}
\sum_{n_1,n_{2} = 0}^{\infty}
\frac{
|z_{1} |^{ 2n_{1} }
|z_{3}^{\kappa_{32}} z_{1}^{\kappa_{12} } |^{2 n_{2} } }{
\omega_1^{n_{1} }
(\omega_{3}^{\kappa_{32}} \omega_1^{\kappa_{12}})^{n_{2} }
 \Gamma[\gamma_{1} +n_1]
\Gamma[\gamma_{32} +n_3]  } \crcr
&& =
 \left[\frac{
|z_{1}|^{2\kappa_{13} }
|z_{3}|^{ 2  } }{
\omega_{1}^{ \kappa_{13}}
\omega_3}\right]^{n_3}
\sum_{n_{2} = 0}^{\infty}
\frac{1}{\Gamma[\gamma_{1}]}
\frac{
|z_{3}^{\kappa_{32}} z_{1}^{\kappa_{12} } |^{2 n_{2} } }{
(\omega_{3}^{\kappa_{32}} \omega_1^{\kappa_{12}})^{n_{2} }
\Gamma[\gamma_{32} +n_3]  }
\fun\left(1;\gamma_1;\frac{|z_1|^2}{\omega_1}\right).
\label{norm3D2dB9c}
\eeq
The hypothesis of a convergence  theorem can be verified
in this case also. The row series in $n_1$ is convergent
whereas the sum over $n_2$ is only convergent
for $\kappa_{32}>0$ or $\kappa_{12}>0$.
Moreover, using $n_1 \geq 1$,
\bea
\frac{1}{\Gamma[\gamma_{1} +n_1]
\Gamma[\gamma_{32} +n_3]} \leq
\frac{1}{n_1!
\Gamma[\gamma_{32} +n_3]},
\label{gamm323}
\eea
we can infer, following  Corollary \ref{coro}
and the same steps from (\ref{bound}),
that the norm series is convergent for $\kappa_{32}>0$.
A double checking of this statement using Theorem \ref{compa}
is given in Appendix \ref{app:check}.

We can perform the limit $\kappa_{12}\to 0$ in (\ref{genfact2dB9c})
and get
\bea
\rho_{1}(n_1,n_3) = \omega_1^{n_1+ \kappa_{13}n_3} \Gamma[\gamma_{13} +n_1] ,
\qquad
\rho_3(n_3,n_2) =\omega_3^{n_3+\kappa_{32}n_2} \Gamma[\gamma_{32} +n_3]
\eea
which generates $(\gamma_{13},\gamma_{32})$-class (\ref{3D2d2cB});
and then again a unique limit is allowed which is $\kappa_{13}\to 0$
providing the class (\ref{3D2d2cB1}).
Conversely taking first the limit $\kappa_{13}\to 0$
in (\ref{genfact2dB9c}), one gets
\bea
\rho_{1}(n_1,n_2) = \omega_1^{n_1+\kappa_{12}n_2 } \Gamma[\gamma_{12} +n_1] ,
\qquad
\rho_3(n_3,n_2) =\omega_3^{n_3+\kappa_{32}n_2} \Gamma[\gamma_{32} +n_3];
\eea
then  the limit $\kappa_{12}\to 0$ yields
\bea
\rho_{1}(n_1,n_2) = \omega_1^{n_1} n_1!,
\qquad
\rho_3(n_3,n_2) =\omega_3^{n_3+\kappa_{32}n_2} \Gamma[\gamma_{32} +n_3]
\eea
which finally give the same end-point limit (\ref{3D2d2cB1}).

Starting from (\ref{genfact2dB9c}), a third limit  can be performed.
As $\kappa_{32} \to 0$, one has
\bea
\rho_{1}(n_1,n_2,n_3) = \omega_1^{n_1+\kappa_{12}n_2 + \kappa_{13}n_3} \Gamma[\gamma_{1} +n_1] ,
\qquad
\rho_3(n_3,n_2) =\omega_3^{n_3} n_3!
\label{noe}
\eea
and then $\kappa_{13}\to 0$ implies
\bea
\rho_{1}(n_1,n_2) = \omega_1^{n_1+\kappa_{12}n_2} \Gamma[\gamma_{12} +n_1] ,
\qquad
\rho_3(n_3,n_2) =\omega_3^{n_3} n_3! .
\label{noeud}
\eea
All the above states are properly normalized
by the same normalization procedure as that for their ancestor.
 But the state defined by (\ref{noe}) differ drastically from the form of its ancestor (\ref{3D2dB9c}). Its normalizability will not follow
from the same recipe because it involves $\kappa_{32}=0$
prohibited so far. Hence, defining
\beq
&&
|z_{1}, z_{3}, n_3 \rangle =
\mathcal N(z_{1}, z_{3}, n_3)^{-\frac12}
 \sum_{n_{1},n_{2} = 0}^{\infty}
\frac{
z_{1}^{ n_{1} +\gamma_1 - 1 }
z_{3}^{ n_{3} } }{
 \left[
\omega_1^{n_1+\gamma_1 - 1 }
\omega_3^{n_3 } \Gamma[\gamma_{1} +n_1]
n_3!\right]^{\frac12} }\,
\, |  n_{1}, n_{2}, n_{3} \rangle ,
\label{3D2dB9cnoe} \\
&&
\mathcal N(z_{1}, z_{3}, n_3)
 = \left[\frac{|z_3|^2}{\omega_3^{n_3} n_3!}\right]^{n_3}
\sum_{n_{1},n_{2} = 0}^{\infty}
\frac{
|z_{1}|^{ 2(n_{1} +\gamma_1 - 1) }}{
\omega_1^{n_1+\gamma_1 - 1 }
\Gamma[\gamma_{1} +n_1] },
\label{3D2dB9cnoenorm}
\eeq
and $a_{n_1,n_2}$ standing for the general term of the
series (\ref{3D2dB9cnoenorm}) ,
the ratio tests of the row and column series are such that
\bea
\lim_{n_1\to \infty} \frac{a_{n_1+1,n_2}}{a_{n_1,n_2}} = 0 ,
\qquad
\lim_{n_2\to \infty} \frac{a_{n_1,n_2+1}}{a_{n_1,n_2}} =
w_2\lim_{n_2\to \infty}[ \kappa_{12}n_2 + \gamma_{13}  + n_1]^{-\kappa_{12}} \sim
0,
\eea
the last equality holding only for $\kappa_{12}>0$. Hence, row
and column series converge everywhere.

Neither the simple comparison test of Corollary \ref{coro}
nor the ratio comparison given by Theorem \ref{compa}
using some exponentials, can help here for extracting the largest solvable class
(see Appendix \ref{app:check}).
Directly  evaluating the  ratio tests,
one has
\beq
&&
\lim_{n_1,n_2 \to \infty}
\frac{a_{n_1+1,n_2}}{a_{n_1,n_2}}
  =w_1\lim_{n_1,n_2 \to \infty}\frac{1 }{\gamma_{1} +n_1} < 1 ,\\
&&
\lim_{n_1,n_2 \to \infty}
\frac{a_{n_1,n_2+1}}{a_{n_1,n_2}}
\sim w_2\lim_{n_1,n_2 \to \infty}
\frac{1}{
[\kappa_{12}(n_2+1)  + \gamma_{13} +n_1]^{\kappa_{12}} } \leq 1,
\nonumber
\eeq
used in the last inequality  has been  made of the $\Gamma$-Stirling approximation and $\kappa_{12}>1$. The first inequality,
through Theorem \ref{ratio},  allows us to infer the convergence.
Other numerical evidences for that convergence even for
$0<\kappa_{12}\leq 1$ can be found in Appendix \ref{app:check}.

The resolution of the identity of the state (\ref{3D2dB9cnoe})
is not totally straightforward as one may expect.
Given four
integers $(n_1,n_2,n'_1,n'_2)$ defined by twice the
state double series, phase integrations will require
a unique constraint
\bea
(n_1 - n_1') + \kappa_{12}(n_2 - n_2') = 0
\eea
which could possess many solutions. However, the most
interesting would be the one such that
\bea
\kappa_{12} \neq -\frac{n_1 - n_1'}{n_2 - n_2'}
\eea
which may occur for instance for any irrational
value of $\kappa_{12}$. In that situation,
one is led to the unique solutions $n_i=n_i'$,
$i=1,2$. Then the resolution of the identity
can be directly inferred from our ordinary method.
Let us comment that the derivation of the
properties of the states defined by
(\ref{noeud}) is completely similar to that of the VCS (\ref{3D2dB9cnoe}).

Another set of factorials generates a different class
compared to what occurs before. Consider, still in Case (13), the
following
\bea
\rho_{1}(n_1,n_2) = \omega_1^{n_1+\kappa_{12}n_2} \Gamma[\gamma_{12} +n_1] ,
\qquad
\rho_3(n_3,n_1,n_2) =\omega_3^{n_3+\kappa_{31}n_1 + \kappa_{32}n_2} \Gamma[\gamma_{3} +n_3].
\label{genfact3D9c10}
\eea
From these, we define the set of VCS
\bea
|z_{1}, z_{3}, n_3 \rangle =
\mathcal N(z_{1}, z_{3}, n_3)^{-\frac12}
 \sum_{n_{1},n_{2} = 0}^{\infty}
\frac{
z_{1}^{ n_1+\kappa_{12}n_2 }
z_{3}^{ n_{3}+\gamma_{3} -1} }{
 \left[
\omega_1^{n_1 + \kappa_{12}n_2 }
\omega_3^{n_3+\gamma_{3} -1 } \Gamma[\gamma_{12} +n_1]
\Gamma[\gamma_{3} +n_3] \right]^{\frac12} }\,
\, |  n_{1}, n_{2}, n_{3} \rangle
\label{3D2dB10c}
\eea
normalized with
\bea
\mathcal N(z_{1}, z_{3}, n_3) =
\left[
\frac{ |z_{3}|^{2} }{
\omega_{3} }\right]^{ n_3 }
\sum_{ n_1, n_{2} = 0 }^{ \infty }
\frac{
|z_{1} z_{3}^{\kappa_{31}} |^{ 2n_{1} }
|z_{3}^{\kappa_{32}} z_{1}^{\kappa_{12} } |^{2 n_{2} } }{
(\omega_1\omega_{3}^{\kappa_{31}} )^{n_{1} }
(\omega_{3}^{\kappa_{32}} \omega_1^{\kappa_{12}})^{n_{2} }
 \Gamma[\gamma_{12} +n_1]
\Gamma[\gamma_{3} +n_3]  } .
\label{norm123}
\eea
We verify the normalization factor convergence using
similar notations as above.
The row and column series satisfy
\beq
&&
\lim_{n_1\to 0}
\frac{a_{n_1+1,n_2}}{a_{n_1,n_2}} =
w_1\lim_{n_1\to 0}
\frac{
\Gamma[\kappa_{31}n_1+ \gamma_{32}  +n_3] }{
(\gamma_{12} +n_1)
\Gamma[\kappa_{31}(n_1+1)+ \gamma_{32} +n_3]  } =0 ,\crcr
&&
\lim_{n_2\to 0}
\frac{a_{n_1,n_2+1}}{a_{n_1,n_2}} =
w_2\lim_{n_1\to 0}
\frac{\Gamma[\kappa_{12}n_2 +1 +n_1]
\Gamma[ \kappa_{32}n_2 + \gamma_{31}  +n_3] }{
\Gamma[\kappa_{12}(n_2+1) +1+n_1]
\Gamma[\kappa_{32}(n_2+1)+ \gamma_{31}   +n_3]  } =0,
\eeq
where in the second limit, we require either
 $\kappa_{12}>0$ or $\kappa_{32}>0$.
Furthermore, one writes using the monotony of the Gamma function
for positive large arguments
\bea
\frac{1}{\Gamma[ \gamma_{12}  +n_1]\Gamma[\gamma_{3}+n_3]}
\leq \frac{1}{n_1! \Gamma[1+ \kappa_{32}n_2 + \kappa_{31}n_1 + n_3 ]}\leq \frac{1}{n_1! \Gamma[\gamma_{32}  + n_3 ]}
\label{ineq123}
\eea
and therefore the previous analysis for (\ref{gamm323}) holds again and
ensures the convergence of the norm series.
This  result can be differently  checked (see Appendix \ref{app:check}).

At the limit $\kappa_{12}\to 0$, we get from (\ref{genfact3D9c10})
\bea
\rho_{1}(n_1,n_2) = \omega_1^{n_1} n_1! ,
\qquad
\rho_3(n_3,n_1,n_2) =\omega_3^{n_3+\kappa_{31}n_1 + \kappa_{32}n_2} \Gamma[\gamma_{3} +n_3]
\label{genfact13}
\eea
which, again taking $\kappa_{31}\to 0$, leads to the
factorials defining (\ref{3D2d2cB1}).
The convergence of the norm for this case
follows from (\ref{ineq123}) given $\kappa_{32}>0$.

A last VCS class has to be studied. This is the one generated
by
\bea
\rho_{1}(n_1,n_2) = \omega_1^{n_1+\kappa_{12}n_2} \Gamma[\gamma_{12} +n_1],
\qquad
\rho_3(n_3,n_1) =\omega_3^{n_3+\kappa_{31}n_1} \Gamma[\gamma_{31} +n_3]
\label{genfact3D10}
\eea
which entail the class of states of the form
\bea
|z_{1}, z_{3}, n_3 \rangle =
\mathcal N(z_{1}, z_{3}, n_3)^{-\frac12}
 \sum_{n_{1},n_{2} = 0}^{\infty}
\frac{
z_{1}^{ n_1+\kappa_{12}n_2 }
z_{3}^{ n_{3}+\kappa_{31}n_1} }{
 \left[
\omega_1^{n_1 + \kappa_{12}n_2 }
\omega_3^{n_3+\kappa_{31}n_1} \Gamma[\gamma_{12} +n_1]
\Gamma[\gamma_{31} +n_3] \right]^{\frac12} }\,
\, |  n_{1}, n_{2}, n_{3} \rangle .
\label{3D2dB11c}
\eea
A normalization condition can be formulated as
\beq
&&
\mathcal N(z_{1}, z_{3}, n_3) =
\left[
\frac{ |z_{3}|^{2} }{
\omega_{3} }\right]^{ n_3 }
\sum_{ n_1, n_{2} = 0 }^{ \infty }
\frac{
|z_{1} z_{3}^{\kappa_{31}} |^{ 2n_{1} }
| z_{1}^{\kappa_{12} } |^{2 n_{2} } }{
(\omega_1\omega_{3}^{\kappa_{31}} )^{n_{1} }
\omega_1^{\kappa_{12}n_{2} }
 \Gamma[\gamma_{12} +n_1]
\Gamma[\gamma_{31} +n_3]  } .
\eeq
The latter series converges on $\C^2$ for $\kappa_{12}>0$
since
\bea
\frac{1}{ \Gamma[\gamma_{12} +n_1]
\Gamma[\gamma_{31} +n_3]  }
\leq \frac{1}{ \Gamma[\gamma_{12} +n_1] n_3!  }
\eea
and the r.h.s term is nothing but a part of (\ref{noeud})
which has been already studied.
Thus, for $\kappa_{31}\to 0$,
(\ref{3D2dB11c}) tends to the VCS defined by
(\ref{noeud}).

\subsubsection{
$(\gamma_{1},\gamma_{2})$- and $(\gamma_{1},\gamma_{3})$-
deformed CS and descendants}
\label{ho3D2d3c}

We pursue the analysis on dependent sums (\ref{deptsum3DA}).
Focusing on
\bea
\rho_1(n_1,n_2,n_3) =\omega_1^{n_1+\kappa_{12}n_2 + \kappa_{13} n_3} \Gamma[\gamma_{1} +n_1],
\quad
\rho_2(n_2,n_1,n_3) =\omega_2^{n_2+\kappa_{23}n_3 + \kappa_{21} n_1} \Gamma[\gamma_{2} +n_2],
\label{fact3D2d4cA}
\eea
we construct the set of states with two deformation parameters
$(\gamma_{1},\gamma_{2})$ as follows:
\bea
|z_{1}, z_{2}, n_3 \rangle =
\mathcal N(z_{1}, z_{2}, n_3)^{-\frac12}
 \sum_{n_{1},n_{2} = 0}^{\infty}
\frac{
z_{1}^{ n_{1} +\gamma_1 - 1 }
z_{2}^{ n_{2} +\gamma_2 - 1 }}{
 \left[
\omega_1^{n_1+\gamma_1 - 1}
\omega_2^{n_2+\gamma_2 - 1} \Gamma[\gamma_{1} +n_1]
\Gamma[\gamma_{2} +n_2]\right]^{\frac12} }\,
\, |  n_{1}, n_{2},n_{3}\rangle
\label{3D2d4cA}
\eea
with the normalization factor
\beq
&&
\mathcal N(z_{1}, z_{2}, n_3) =
 \sum_{n_{1},n_{2} = 0}^{\infty}
\frac{
|z_{1}|^{2( n_{1} +\gamma_1 - 1) }
|z_{2}|^{2( n_{2} +\gamma_2 - 1 )} }{
 \left[
\omega_1^{n_1+\gamma_1 - 1}
\omega_2^{n_2+\gamma_2 - 1} \Gamma[\gamma_{1} +n_1]
\Gamma[\gamma_{2} +n_2]\right] } \crcr
&&
 =  \left[\frac{
|z_{1}|^{2\kappa_{13} }
|z_{2}|^{ 2\kappa_{23}   }}{
\omega_{1}^{ \kappa_{13}}
\omega_2^{\kappa_{23}}}\right]^{n_3}
\sum_{n_1,n_{2} = 0}^{\infty}
\frac{
|z_{1}z_{2}^{\kappa_{21}}|^{2n_{1} }
|z_{2}z_{1}^{\kappa_{12}}|^{2 n_{2} } }{
(\omega_1\omega_2^{\kappa_{21}})^{n_{1} }
(\omega_{2}\omega_1^{\kappa_{12}})^{ n_{2} }
 \Gamma[\gamma_{1} +n_1]
\Gamma[\gamma_{2} +n_2]  } .
\label{normgam12}
\eeq
Even though the series does not factorize,
by simple comparison $\Gamma[\gamma_i + n_i]\geq  n_i!$ for
$\gamma_i \geq 1$ and for $n_i\geq 1$,
it is direct to prove that (\ref{normgam12})
is convergent for all $|z_{1}z_{2}^{\kappa_{21}}|>0$,
and all $|z_{2}z_{1}^{\kappa_{12}}|>0$, and therefore
absolutely convergent for all $z_i\in \C$ $i=1,2$.

Addressing the resolution of the identity of these states
in the form (\ref{resol3D2d2c}), we find
that the preliminary phase integration in $(\theta_1,\theta_2)$ gives a kind of consistency condition
\bea
n_1 - n_1'  + \kappa_{12}(n_2 - n_2') =0 \quad \Leftrightarrow
\quad
n_2 - n_2'  + \kappa_{21}(n_1 - n_1') =0
\eea
which is indeed trivially satisfied, and we are led to the moment problem solved by our ordinary technique.

Let us study the solvable classes induced by limit procedures
from (\ref{3D2d4cA}).
We can perform the limit  $\kappa_{13}\to 0$, and find the
corresponding to the generalized factorials
\bea
\rho_1(n_1,n_2) =\omega_1^{n_1+\kappa_{12}n_2} \Gamma[\gamma_{12} +n_1],
\quad
\rho_2(n_2,n_1,n_3) =\omega_2^{n_2+\kappa_{23}n_3 + \kappa_{21} n_1} \Gamma[\gamma_{2} +n_2].
\label{fact3D2d4c1}
\eea
From these quantities, take the limit $\kappa_{23}\to 0$ and get
\bea
\rho_1(n_1,n_2) =\omega_1^{n_1+\kappa_{12}n_2} \Gamma[\gamma_{12} +n_1],
\quad
\rho_2(n_2,n_1) =\omega_2^{n_2+ \kappa_{21} n_1} \Gamma[\gamma_{21} +n_2]
\label{fact3D2d4c2}
\eea
defining another VCS class and then the procedure stops.
The class  (\ref{fact3D2d4c2})  corresponds to an extension
of the $(\gamma_{12},\gamma_{21})$-doubly deformed
VCS (\ref{ho2d4c}) where, in addition, a sum is carried out on the second index $n_2$. Thus (\ref{ho2d4c}) are generators of the class
defined by (\ref{fact3D2d4c2}).

Note that we could have done first the second limit
$\kappa_{23}\to 0$ yielding a different intermediate step
which defines nothing but a symmetric (and so not a new)
class of VCS determined by the symmetric generalized factorials
of (\ref{fact3D2d4c1}) under $(1\leftrightarrow 2)$.
From this state, then perform $\kappa_{13}\to 0$ giving the same final
VCS class determined by (\ref{fact3D2d4c2}).
All these VCS limits have a convergent normalization factor
since the above comparison criterion does not
depend on the $\kappa$ parameters.

We have another set of generalized factorials
in Case (13) (\ref{deptsum3DB}) yet
defining another set of $(\gamma_1,\gamma_3)$-deformation
of VCS that can be reported here also. These are given
by
\bea
\rho_{1}(n_1,n_2,n_3) = \omega_1^{n_1+\kappa_{12}n_2 + \kappa_{13}} \Gamma[\gamma_{1} +n_1] ,
\quad
\rho_3(n_3,n_1,n_2) =\omega_3^{n_3+\kappa_{31}n_1 + \kappa_{32}n_2} \Gamma[\gamma_{3} +n_3]
\eea
from which we define the set of states
\bea
|z_{1}, z_{3}, n_3 \rangle =
\mathcal N(z_{1}, z_{3}, n_3)^{-\frac12}
 \sum_{n_{1},n_{2} = 0}^{\infty}
\frac{
z_{1}^{ n_{1} +\gamma_1 - 1 }
z_{3}^{ n_{3} +\gamma_3 - 1 }}{
 \left[
\omega_1^{n_1+\gamma_1 - 1}
\omega_3^{n_3+\gamma_3 - 1} \Gamma[\gamma_{1} +n_1]
\Gamma[\gamma_{3} +n_3]\right]^{\frac12} }\,
\, |  n_{1}, n_{2},n_{3}\rangle .
\label{3D2d4cB}
\eea
The normalization factor can be computed as
\beq
&&
\mathcal N(z_{1}, z_{2}, n_3) =
 \sum_{n_{1},n_{2} = 0}^{\infty}
\frac{
|z_{1}|^{2( n_{1} +\gamma_1 - 1) }
|z_{3}|^{2( n_{3} +\gamma_3 - 1 )} }{
 \left[
\omega_1^{n_1+\gamma_1 - 1}
\omega_3^{n_3+\gamma_3 - 1} \Gamma[\gamma_{1} +n_1]
\Gamma[\gamma_{3} +n_3]\right] } \crcr
&&
 =  \left[\frac{
|z_{1}|^{2\kappa_{13} }
|z_{3}|^{ 2  } }{
\omega_{1}^{ \kappa_{13}}
\omega_3}\right]^{n_3}
\sum_{n_1,n_{2} = 0}^{\infty}
\frac{
|z_{1} z_{3}^{ \kappa_{31} } |^{ 2n_{1} }
|z_{3}^{\kappa_{32}} z_{1}^{\kappa_{12} } |^{2 n_{2} } }{
(\omega_1 \omega_3^{\kappa_{31}})^{n_{1} }
(\omega_{3}^{\kappa_{32}} \omega_1^{\kappa_{12}})^{n_{2} }
 \Gamma[\gamma_{1} +n_1]
\Gamma[\gamma_{3} +n_3]  } .
\eeq
Corollary \ref{coro} can be applied here noting that
at large arguments
\bea
 \frac{1}{\Gamma[\gamma_{1} +n_1]
\Gamma[\gamma_{3} +n_3]  } \leq
\frac{1}{n_1! \;\Gamma[\gamma_{3} +n_3]  }
\eea
and the r.h.s. is part of the series defined by
(\ref{genfact13}) which has been already determined to be convergent
everywhere using (\ref{ineq123}) and $\kappa_{32}>0$.

The resolution of the identity passes  again through
a well defined $(\theta_1,\kappa_{31}\theta_3)$ phase integrations yielding an obvious statement:
\bea
n_1 - n_1'  + \kappa_{12}(n_2 - n_2') =0 \quad \Leftrightarrow
\quad
 n_1 - n_1'  + \kappa_{13}\kappa_{32}( n_2 - n_2') =0
\eea
and the resolution of the moment problem
can be carried out through the same
routine with solution given by (\ref{hodenssmart}).

The limit $\kappa_{12}\to 0$ yields
\bea
\rho_{1}(n_1,n_3) = \omega_1^{n_1+ \kappa_{13}n_3} \Gamma[\gamma_{13} +n_1] ,
\quad
\rho_3(n_3,n_1,n_2) =\omega_3^{n_3+\kappa_{31}n_1 + \kappa_{32}n_2} \Gamma[\gamma_{3} +n_3] .
\eea
Note that we cannot go further since any other limits
would be undefined: $\kappa_{31}\to 0$ implies $\kappa_{13}\to
\infty$ (or vice-versa) and $\kappa_{32}\to 0$ leads to
an infinite series summing over an integrand free of the index summation $n_2$.
Nevertheless, from the beginning, we can take the
limit $ \kappa_{32}\to 0$ such that the quantities become
\bea
\rho_{1}(n_1,n_2,n_3) = \omega_1^{n_1+\kappa_{12}n_2 + \kappa_{13}} \Gamma[\gamma_{1} +n_1] ,
\quad
\rho_3(n_3,n_1,n_2) =\omega_3^{n_3+\kappa_{31}n_1} \Gamma[\gamma_{31} +n_3]
\eea
which, for similar reasons, are an end-point of finding a
well defined limit.

\begin{figure}[t]
\centering{
\includegraphics[width=160mm]{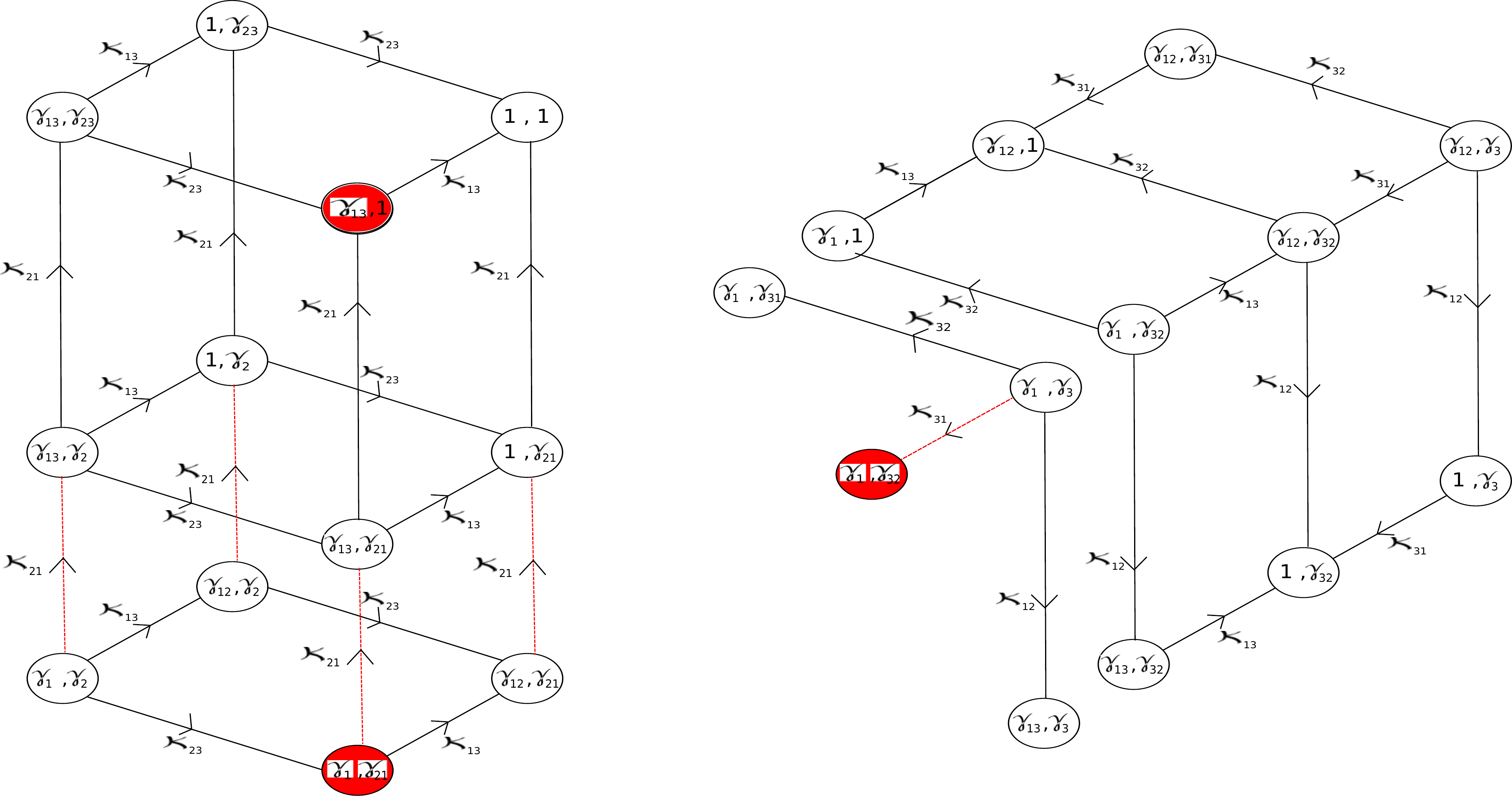}
\put(-400,-15){Case (12) VCS}
\put(-150,-15){Case (13) VCS} }
\caption{\footnotesize $3D$ Harmonic oscillator solvable
VCS classes with two degrees of freedom organized through the
relation ``ancestor and descendant''. Classes
highlighted (in red) have been already listed:
$(1,\gamma_{23})\equiv(\gamma_{13},1)$,
and
$(\gamma_{1},\gamma_{21})\equiv(\gamma_{12},\gamma_{2})$,
by symmetry $(1\leftrightarrow2)$. Red arrows are undefined limits.}
\label{fig:cube}
\end{figure}

Figure (\ref{fig:cube}) gives a summary of
different classes and their descendants.

\subsection{Some VCS with three degrees of freedom}

We sketch here the construction of VCS
with three degrees of freedom.

Consider the generalized factorials
\beq
&&
\rho_{1}(n_1,n_2,n_3) = \omega_1^{n_1+\kappa_{12}n_2 + \kappa_{13}} \Gamma[\gamma_{1} +n_1] ,
\quad
\rho_2(n_2,n_1,n_3) =\omega_2^{n_2+\kappa_{21}n_1 + \kappa_{23}n_3} \Gamma[\gamma_{2} +n_2] , \crcr
&&
\rho_3(n_3,n_1,n_2) =\omega_3^{n_3+\kappa_{31}n_1 + \kappa_{32}n_2} \Gamma[\gamma_{3} +n_3]
\eeq
from which we define the set of states maximally
 $(\gamma_1,\gamma_2,\gamma_3)$-deformed
given by
\beq
&&
|z_{1}, z_{2},z_{3}, n_3 \rangle =
\mathcal N(z_{1}, z_{2}, z_{3}, n_3)^{-\frac12}  \times \cr\cr
&&
 \sum_{n_{1},n_{2} = 0}^{\infty}
\frac{
z_{1}^{ n_{1} +\gamma_1 - 1 }
z_{2}^{ n_{2} +\gamma_2 - 1 }
z_{3}^{ n_{3} +\gamma_3 - 1 }}{
 \left[
\omega_1^{n_1+\gamma_1 - 1}
\omega_2^{n_2+\gamma_2 - 1}
\omega_3^{n_3+\gamma_3 - 1}
\Gamma[\gamma_{1} +n_1]
\Gamma[\gamma_{2} +n_2]
\Gamma[\gamma_{3} +n_3]\right]^{\frac12} }\,
\, |  n_{1}, n_{2},n_{3}\rangle
\label{3D3dmax}
\eeq
with normalization factor
\beq
&&
\mathcal N(z_{1}, z_{2}, n_3) =
   \left[\frac{
|z_{1}|^{2\kappa_{13} }
|z_{2}|^{2\kappa_{23} }
|z_{3}|^{ 2  } }{
\omega_{1}^{ \kappa_{13}}
\omega_{2}^{ \kappa_{23}}
\omega_3}\right]^{n_3} \crcr
&&
\sum_{n_1,n_{2} = 0}^{\infty}
\frac{
|z_{1} z_{3}^{ \kappa_{31} } z_{2}^{\kappa_{21}}|^{ 2n_{1} }
|z_{2}z_{3}^{\kappa_{32}} z_{1}^{\kappa_{12} } |^{2 n_{2} } }{
(\omega_1 \omega_2^{\kappa_{21}}\omega_3^{\kappa_{31}})^{n_{1} }
( \omega_2\omega_{3}^{\kappa_{32}} \omega_1^{\kappa_{12}})^{n_{2} }
\Gamma[\gamma_{2} +n_2]
 \Gamma[\gamma_{1} +n_1]
\Gamma[\gamma_{3} +n_3]  }
\eeq
converging because bounded by the double exponential
series up to some factor function of $n_3$. One can show that the class (\ref{3D3dmax})
is a VCS class which are integrated to unity according
to our formalism.

Finally, the straightforward generalization of the $(1,1)$ VCS (\ref{ho2d1c}) and (\ref{3D2d0c}) becomes a $(1,1,1)$ VCS made with
simple factorials in the three sector as follows:
\beq
&&
|z_{1}, z_{2},z_{3}, n_3 \rangle =
\mathcal N(z_{1}, z_{2}, z_{3})^{-\frac12}
 \sum_{n_{1},n_{2} = 0}^{\infty}
\frac{
z_{1}^{ n_{1}  }
z_{2}^{ n_{2}  }
z_{3}^{ n_{3} }  }{
 \left[
\omega_1^{n_1} \omega_2^{n_2}
\omega_3^{n_3}
n_1! n_2!n_3!\, \right]^{\frac12} }
 \, |  n_{1}, n_{2},n_{3}\rangle,
\label{3D3dmin} \\
&&
\mathcal N(z_{1}, z_{2}, z_{3},n_3) =
\left(\frac{|z_3|^2}{\omega_3}\right)^{n_3}
\exp \left\{\frac12( |z_1|^2 + |z_2|^2)\right\}.
\eeq
This is just a state proportional to $|z_1\rangle \otimes |z_2\rangle \otimes |n_3\rangle$.
The classes (\ref{3D3dmax}) and (\ref{3D3dmin}) are
the one which can be simply inferred for the
harmonic oscillator and
always remain solvable in any dimension.
Between these states, corresponding to the maximally deformed
and not deformed one, a number of intermediate states
with a less number of deformation  parameters
occurs, but their solvability is not  guaranteed.

\section{Conclusion}
\label{concl}

This work has been devoted to an extension of solvable classes
of VCS for the harmonic oscillator in $2D$ and $3D$.
By a combinatorics involving  different possible
partitions of the energy of the system
and by assigning to each partition a couple of variables
and thereby building the corresponding set of VCS,
we improve some partial results known in the literature \cite{gazeau-novaes,thirulo,thirulo-saad}.
We focus on two basic requirements that the VCS ought to satisfy: a normalizability condition and a resolution of unity on the Hilbert space.
The first requirement has  mainly involved
recent techniques providing criteria for
double series convergence, and an optimization of  the computations:
all VCS states of this kind are normalizable.
The second was performed in a row by noting an interesting
fact appearing at the very basic level: the exponent
of the continuous variable can be always cast in a way in order
to conspire with  generalized factorial
for solving the Stieljes moment problem
in a particular simple way.
Moreover, that resolution of the identity is proved
to be valid for a wide range of the frequency parameters
of the VCS, making these latter maximal in the space of these parameters.
Interestingly, this work has found new connections
with the (non)unicity of the measure integrating
the VCS to unity \cite{nuss,lassere}.
Only the aforementioned basic axioms have been proved, nevertheless the so-called Gazeau-Klauder
properties could be certainly implemented from our results.
In addition, a primary classification scheme has been investigated
according to the number of degrees of freedom
and complexity of the $\gamma_{\cdot}$ parameters.
Another way to understand these classes is through
a deformed theory: VCS classes are consistent
frequency dependent deformations of one another built out
of  ordinary CS living in a subspace of a higher rank
tensor Hilbert space. The links are realized in
the deformation parameter space.

An attempt of classification of these classes
of VCS has been provided, though, one has to acknowledge,
deserves to be definitely refined and precise.
This work is under current investigations \cite{hounk}.
Furthermore, we have mainly focused the existence proof
of extended classes of VCS however, and notably,
one should investigate the statistical properties of these classes
of states. It is known that \emph{canonical} VCS of the $(1,1)$-class
form satisfy the ordinary properties
of CS, i.e. that they are \emph{intelligent}.
This same question has to be addressed for the remaining VCS classes
here. Yet more extensions of previous results
could be investigated also. For instance,
it will be interesting
to make a sense of statistical properties of these VCS according
to  Gazeau-Novaes formalism \cite{gazeau-novaes}
using Berezin-Lieb inequalities  for multidimensional CS.

\section*{Acknowledgments}

J.B.G. would like to thank M. Zeltser and  J.B. Lassere
for helpful discussions.
This work is partially supported by the ICTP through the
OEA-ICMPA-Prj-15. The ICMPA is in partnership with
the Daniel Iagolnitzer Foundation (DIF), France.
Research at Perimeter Institute is supported by the Government of Canada through Industry
Canada and by the Province of Ontario through the Ministry of Research and Innovation.

\appendix

\renewcommand{\theequation}{\Alph{section}.\arabic{equation}}
\setcounter{equation}{0}

\section{Solving  generalized moment problems}
\label{app:solving}
We consider the generalized moment problem as given
in (\ref{genmom})
 \beq
\int du_1 du_2\; \chi(u_1,u_2,n_2)\;
\left[\frac{ u_1^{\alpha_1} u_2^{\beta_2 \kappa_2 } }
{ \omega_1^{\alpha'_1} \omega_2^{ \beta'_2 \kappa_2} }
\right]^{n_1}
\left[
\frac{ u_1^{ \beta_1 \kappa_1 } }{
\omega_1^{ \beta'_1 \kappa_1} }
\right]^{n_2}
\left[
\frac{ u_2^{ \alpha_2 } }{ \omega_2^{\alpha'_2 } }
\right]^{n_2}= R_1(n_1,n_2) R_2(n_2).
\label{genmomapp}
\eeq
A density solution of (\ref{genmomapp}) is not
uniquely defined. We have however
a single constraint on $\chi$: it should not depend on $n_1$.
Hence, we are led to the following choices:

(a) All dependencies in $n_2$ can be simplified
using the fact that $\chi$ may depend on $n_2$
and all variables $u_1,u_2$. Then
introduce, by hand, the correct dependence in $n_2$
which could generate  the generalized factorial
$ R_1(n_1,n_2) R_2(n_2)$. For instance,
assume that the density $\chi$  is of the form
\beq
\chi(u_1,u_2,n_2)=
\chi'(u_1,u_2,n_2)
 \left[
\frac{ u_1^{ \beta_1 \kappa_1 } }{
\omega_1^{ \beta'_1 \kappa_1} }
\right]^{-n_2}
\left[
\frac{ u_2^{ \alpha_2 } }{ \omega_2^{\alpha'_2 } }
\right]^{-n_2}   r_2^{n_2}\; e^{-r_2^{n_2}}.
\label{densimpl}
\eeq
The remaining density
$\chi'(u_1,u_2,n_2)$ has the unique purpose to
integrate the combined new variable
$\tilde{u}_1 = u_1^{\alpha_1} u_2^{\beta_2 \kappa_2 }$
in order to get at first $R_1(n_1,n_2)$;
the quantity $R_2(n_2)$ can be recovered
without ambiguity and then the problem (\ref{genmomapp}) will be solved. However, considering this option
boils down to cancel  $z_2^{n_2}/\omega^{n_2}_2$
at the very beginning in the VCS and thereby
to redefine them as an  one degree of freedom VCS with new variable $z_1 z_2^{\kappa}$. Therefore, using blindly this option may lead to already known VCS.
Moreover, implementing this option, one
may need to introduce the same kind of terms
that have been already simplified, then
this method might be not very efficient.

(b) All dependencies in $n_2$, are not simplified and
one tries to carry out a strict change of variables
in order to solve a decoupled moment problem.
This option does have an advantage: it is the one relevant
when discussing VCS in general and when, in particular,
the second index $n_2$ is summed. In that situation, the density $\chi$ should depend on the variables $u_1$ and $u_2$ but
not on $n_2$. For instance, keeping in mind (\ref{genmomapp}), one considers the following change of
variables
\bea
u_1 \to \tilde{u}_1 =  u_1^{\alpha_1} u_2^{\beta_2 \kappa_2 }   \qquad \text{and} \qquad
u_2 \to \tilde{u}_2  =  u_2^{\alpha_2} u_1^{\beta_1 \kappa_1 }
\eea
with Jacobian
\bea
J = (\alpha_1 \alpha_2 - \beta_1 \beta_2) u_1^{\alpha_1 + \beta_1 \kappa_1 -1}
u_2^{\alpha_2 + \beta_2 \kappa_2 -1}
\label{cov}
\eea
which could vanish without further assumption on the parameters.
For instance the case $\alpha_i =\beta_i=1$
which is relevant for our discussion
is not allowed. Consequently,
this option has a disadvantage:
it is dependent on the change of variable
(which could appear singular)
and so restricts the kind of solvable
moment problems.

(c) A third option is to use a mixed formalism:
one can choose to simplify or not the dependence in
$n_2$ but always in  such a way
that the change of variable (\ref{cov})
will appear non singular.
The resolution of
the moment problem with its generalized factorials
strongly depends on the kind of integrand.
Simplifications have to be chosen appropriately.
In a specific instance, consider that
the moment integral yields $n_1!n_2!$.
Using the dependence on $n_2$ of $\chi$,
remove all dependencies as $u_1^{\beta_1 \kappa_1 n_2}$ so that
we end up with a triangular change of variables:
\bea
u_1 \to \tilde{u}_1 =  u_1^{\alpha_1} u_2^{\beta_2 \kappa_2 }   \qquad \text{and} \qquad
u_2 \to \tilde{u}_2  =  u_2^{\alpha_2}
\label{tricov}
\eea
which is always invertible for $\alpha_i\neq 0$, $i=1,2$.
However,  if the product of generalized factorials is of the
form $\Gamma[\gamma_1 + n_1] n_2!$,
simplifying $u_1^{\beta_1 \kappa_1 n_2}$
will be harmful since only a power $X_1^{n_1 +\beta_1 \kappa_1 n_2}X_2^{n_2}$ with $X_{1,2}$ some variables,
could produce such a result. Hence reintroducing
by hand the missing term is the only way out.
In that above situation, then do not perform a simplification and still
there is a change of variables which is regular
(we will give more precision afterwards).
It should be emphasized also that
this method cannot be reported for
higher order degrees of freedom or summing
over $n_2$, but preserves the
two degrees of freedom in the VCS (the issue
of  option (a) is cured) and offers
always a solution of the moment problem
(hence the disadvantage of option (b) is avoided).

In the following, guided by the order of efficiency in finding
solution of the moment problem though providing non listed VCS, we will use the third option.

We introduce the following terminology:
\emph{to recombine} a term $x$
is to perform a change of variable in another
variable, say $y$ and $y\to yx$, in order to simplify $x$.
The main steps for solving any moment
problem (\ref{genmomapp}) for Subsection \ref{sub:ho2d} (VCS with
two degrees of freedom summing over $n_1$
built with generalized factorial $R_1(n_1,n_2) R_2(n_2,n_1)$)
using option (c) are the following:

(0)  Always consider a variable as
$r_i^{(\bullet)}/\omega_i^{(\bullet')}$,
i.e. a ratio between a variable and its  frequency
before integrating it. Even though, the frequencies $\omega_i$ could be regarded just as dressing factors,
these provide actual continuous deformation parameters
giving a sense of the VCS classification.

(1)  - If $R_i(n_i,n_{\iche})=\Gamma[\gamma_i + n_i]$,
for all $i=1,2$, then do not simplify or recombine \emph{a priori}
any of the variables.

- Given a couple $i,\iche \in \{1,2\}$, $i\neq\iche$,
if  $R_i(n_i,n_{\iche})=\Gamma[\gamma_i + n_i]$
and $R_{\iche}(n_{\iche})=n_{\iche} !$,
then recombine ($\iche=2$) or simplify ($\iche=1$)
only $u_{\iche}$.

- If $R_{i}(n_i) = n_i!$ for $i=1,2$, then recombine
$u_2$ and simplify $u_1$.

(2) Use the ansatz $\chi(u_1,u_2,n_2) = \varrho_1(u_1,u_2,n_2)\varrho_2(u_2)$
for solving the moment problem
where $\varrho_1(u_1,u_2,n_2)$
will be used to integrate the single
variable $u_1$ (and simplifying the maximum of
factors) and $\varrho_2(u_2)$
for the second variable $u_2$.

(3) As a convention,
all extra Jacobian factors coming from
the change of variables $u_1 \to \tilde{u}_1$
(resp. $u_2 \to \tilde{u}_2$)
 should be reabsorbed by $\varrho_1$
(resp. by $\varrho_2$).

This program will not give of course a direct solution
of the generalized moment problem
but, at least, it provides an unique way for dealing
with the combinatorics of extra factors generated by
the variables $u_1$ and $u_2$.

We now apply this program
to the problem (\ref{genmomapp}).
First, we need to specify the generalized factorials.
Consider $R_1(n_1)=n_1!$ and $R_2(n_2)=n_2!$
describing a first class of VCS of the kind (\ref{ho2d2c}). Then,
by step (1), we simplify the extra factor in $u_1^{\beta_1\kappa_1 n_2}$, by considering
the density
\bea
\chi(u_1,u_2,n_2)=
\varrho_1'(u_1,u_2,n_2)
 \left[
\frac{ u_1^{ \beta_1 \kappa_1 } }{
\omega_1^{ \beta'_1 \kappa_1} }
\right]^{-n_2}
  \varrho_2 (u_2) ,
\eea
and substituting this in the problem and recombining $u_2^{\beta_2\kappa_2 n_1}$,  we get
\beq
\int du_1 du_2\; \varrho_1'(u_1,u_2,n_2)\;\varrho_2(u_2)
\left[\frac{ u_1^{\alpha_1} u_2^{\beta_2 \kappa_2 } }
{ \omega_1^{\alpha'_1} \omega_2^{ \beta'_2 \kappa_2} }
\right]^{n_1}
\left[
\frac{ u_2^{ \alpha_2 } }{ \omega_2^{\alpha'_2 } }
\right]^{n_2}= R_1(n_1,n_2) R_2(n_2).
\label{genmom2app}
\eeq
Perform a change of variables
$
u_1 \to \tilde{u}_1 =  u_1^{\alpha_1} u_2^{\beta_2 \kappa_2 }$ and
$
u_2 \to \tilde{u}_2  =  u_2^{\alpha_2}
$
with minor Jacobians given by
$[\alpha_1 u^{\alpha_1-1}_1u_2^{ \beta_2 \kappa_2 } ]$
and $ [ \alpha_2 u_2^{\alpha_2 -1}  ]$, respectively.
Use $\varrho'_1$ and $\varrho_2$ to absorb these
terms, respectively.  The following densities given in radial variables solve the problem (\ref{genmomapp})
\beq
&&
\varrho_1(r_1,r_2,n_2) =
 \alpha_1
\frac{r^{2(\alpha_1-1)}_1r_2^{2 \beta_2 \kappa_2 }
}{\omega^{\alpha'_1}_1  \omega_2^{\beta'_2 \kappa_2 }}
\left[
\frac{ \omega_1^{\beta'_1  \kappa_1} }
{r_1^{ 2\beta_1\kappa_1 }}
\right]^{n_2}
e^{ - \frac{ r^{2\alpha_1}_1r_2^{ 2\beta_2 \kappa_2 }}{ \omega^{\alpha'_1}_1  \omega_2^{\beta'_2 \kappa_2 } }}, \quad
\varrho_2(r_2)=\alpha_2 \frac{1}{\omega^{\alpha'_2}_2}
r_2^{2(\alpha_2-1)} e^{-\frac{r^{2\alpha_2}_2}{\omega^{\alpha'_2}_2}},\crcr
&&
\chi(r_1,r_2,n_2) = \alpha_1 \alpha_2
\frac{r^{2(\alpha_1-1)}_1r_2^{2 (\alpha_2 +\beta_2 \kappa_2 -1)}
}{\omega^{\alpha'_1}_1
\omega_2^{ \alpha'_2 +\beta'_2 \kappa_2}}
\left[
\frac{ \omega_1^{\beta'_1  \kappa_1} }
{r_1^{ 2\beta_1\kappa_1 }}
\right]^{n_2}
e^{ - \frac{ r^{2\alpha_1}_1r_2^{ 2\beta_2 \kappa_2 }}{ \omega^{\alpha'_1}_1  \omega_2^{\beta'_2 \kappa_2 } }
-\frac{r^{2\alpha_2}_2}{\omega^{\alpha'_2}_2}}.
\label{soludeforapp}
\eeq
The calculations are  more involved using instead
$R_1(n_1,n_2)= \Gamma[\gamma_1+ n_1]$ and
$R_2(n_2)=n_2!$ which are the data for second class
VCS (\ref{ho2d1c}).
By step (1), we should not simplify $u_1^{\beta_1 \kappa_1 n_2}$ and only recombine $u_2^{\beta_2 \kappa_2 n_1}$.
Coming back to (\ref{genmom2app}), we have
\beq
\int du_1 du_2\; \chi_1(u_1,u_2,n_2)
\left[\frac{ u_1^{\alpha_1} u_2^{\beta_2 \kappa_2 } }
{ \omega_1^{\alpha'_1} \omega_2^{ \beta'_2 \kappa_2} }
\right]^{n_1 + \kappa_1 n_2}
\left[\frac{ u_1^{(\beta_1  -\alpha_1)\kappa_1 } \omega_2^{ \beta'_2 } }
{ \omega_1^{(\beta'_1  -\alpha'_1)\kappa_1}
u_2^{\beta_2 }} \right]^{n_2}
\left[
\frac{ u_2^{ \alpha_2 } }{ \omega_2^{\alpha'_2 } }
\right]^{n_2}
\label{genmom3app}
\eeq
Clearly, in order to avoid problems with Jacobians in a double
change of variables, we should simplify
the ratio appearing as  $u_1^{(\beta_1  -\alpha_1)\kappa_1}/ \omega_1^{(\beta'_1  -\alpha'_1)\kappa_1} $ using $\chi$.
However, we have still two choices
for the variable $u_2$: either to recombine
all dependencies in $u_2$ in one
variable
$u_2^{\alpha_2 - \beta_2}/
\omega_2^{\alpha_2 -\beta_2}$,
or to simplify again $u_2^{ - \beta_2}/
\omega_2^{ -\beta_2}$ using $\chi$.
The first choice will lead us to a
Jacobian with factor $\alpha_2 - \beta_2$,
which could vanish and so it is not
the best option for seeking general solutions. We will therefore simplify
the factor $u_2^{ - \beta_2}/
\omega_2^{ -\beta_2}$, set
\bea
\chi(u_1,u_2,n_2) = \varrho_1'(u_1,u_2,n_2)\left[\frac{ u_1^{(\beta_1  -\alpha_1)\kappa_1 } \omega_2^{ \beta'_2 } }
{ \omega_1^{(\beta'_1  -\alpha'_1)\kappa_1}
u_2^{\beta_2 }} \right]^{-n_2} \varrho_2(u_2)
\eea
and come to a similar problem as obtained previously
for which one gets the solutions
\beq
&&
\varrho_1(r_1,r_2,n_2) = \alpha_1
\frac{r_1^{2(\alpha_1-1)} r_2^{ 2\beta_2 \kappa_2 }
}{\omega^{\alpha'_1}_1  \omega_2^{\beta'_2 \kappa_2 }}
\left[\frac{\omega_1^{(\beta'_1 - \alpha'_1) \kappa_1}
r_2^{ 2 \beta_2  }}
{r_1^{ 2(\beta_1 - \alpha_1 )\kappa_1 }
\omega_2^{ \beta'_2 }}
\right]^{n_2}
e^{ - \frac{ r^{2\alpha_1}_1r_2^{ 2\beta_2 \kappa_2 }}{ \omega^{\alpha'_1}_1  \omega_2^{\beta'_2 \kappa_2 } }},\crcr
&&
\varrho_2(r_2)=\alpha_2 \frac{1}{\omega^{\alpha'_2}_2}
r_2^{2(\alpha_2-1)} e^{-\frac{r^{2\alpha_2}_2}{\omega^{\alpha'_2}_2}}.
\label{soludefor2app}
\eeq

Next, according to the same formalism, we solve
the moment problem of the states (\ref{ho2d3csub})
determined by
$R_2(n_2,n_1)=\Gamma[\gamma_2 + n_2]$
and $R_1(n_1)=n_1!$.
Then $u_2$ should not be recombined or simplified
without more considerations whereas
$R_1(n_1)=n_1!$ implies that $u_1^{\beta_1 \kappa_1 n_2}$ should be
simplified.
Using the same routine, we write the moment problem
associated with these states:
\beq
&&
\int du_1 du_2\; \chi_1(u_1,u_2,n_2)
\left[
\frac{ u_2^{\alpha_2 }  }
{ \omega_2^{\alpha'_2 }  }
\right]^{n_2 + \kappa_2 n_1}
\left[\frac{ u_1^{ \alpha_1 }u_2^{ (\beta_2-\alpha_2)\kappa_2} }
{\omega_1^{\alpha'_1 }\omega_2^{ (\beta'_2-\alpha'_2)\kappa_2 } }
\right]^{n_1}
\left[
\frac{ u_1^{  \beta_1}  }
{\omega_1^{\beta'_1 } }
\right]^{ \kappa_1n_2}.
\label{genmom4}
\eeq
Changing variables as
$\tilde{u}_2 = u_2^{\alpha_2}$ and
$\tilde{u}_1 = u_1^{ \alpha_1 }u_2^{ (\beta_2-\alpha_2)\kappa_2}$,
one can solve the problem by setting
 $\chi(r_1,r_2,n_2)  = \varrho_1(r_1,r_2,n_2) \varrho_{2}(r_2)$
such that
\beq
&&
\varrho_1(r_1,r_2,n_2) = \alpha_1
\frac{r_1^{2(\alpha_1-1)} r_2^{ 2(\beta_2-\alpha_2)\kappa_2 }
}{\omega^{\alpha'_1}_1  \omega_2^{(\beta'_2-\alpha'_2) \kappa_2 }}
\left[\frac{\omega_1^{\beta'_1}}
{r_1^{ 2\beta_1 }}
\right]^{ \kappa_1n_2}
e^{ - \frac{ r^{2\alpha_1}_1r_2^{ 2(\beta_2-\alpha_2) \kappa_2 }}{ \omega^{\alpha'_1}_1  \omega_2^{(\beta'_2-\alpha'_2) \kappa_2 } }},\crcr
&&
\varrho_2(r_2)=\alpha_2 \frac{1}{\omega^{\alpha'_2}_2}
r_2^{2(\alpha_2-1)} e^{-\frac{r^{2\alpha_2}_2}{\omega^{\alpha'_2}_2}}.
\eeq

\section{Checking the convergence of norm series }
\label{app:check}

We provide here other arguments justifying the
norm series convergence by using different criteria than the one used in the main text. These additional tests remain instructive
for they offer the correct way
to manipulate and optimize the convergence criteria
afforded by the different theorems. This will enable us to achieve
less constraints on the  parameters $\kappa$ and thus
to get larger VCS classes.

\noindent{\bf Checking Eq.(\ref{norm3D2dB9c})}
As a first double checking, let us come back on
the convergence of the norm series (\ref{norm3D2dB9c}).
We can evaluate comparison test ratios
of Theorem \ref{compa}, by defining $w_i>0$, $i=1,2$,  $b_{n_1,n_2} $ the general term of the exponential series $e^{w_1 +w_2 }$ and
\bea
b_{n_1,n_2} = \frac{w_1^{n_1}w_2^{n_2}}{n_1! n_2!} >0,
\qquad
a_{n_1 ,n_2} = \frac{w_1^{n_1}w_2^{n_2} }{
\Gamma[\gamma_{1} +n_1]
\Gamma[\gamma_{32} +n_3] }>0.
\eea
We have
\beq
&&
\frac{a_{n_1+1,n_2}b_{n_1,n_2}}{a_{n_1,n_2}b_{n_1+1,n_2}}
 =
\frac{ 1+n_1 }{\gamma_{1} +n_1 } \leq 1 ,\crcr
&&
\frac{a_{n_1,n_2+1}b_{n_1,n_2}}{a_{n_1,n_2}b_{n_1,n_2+1}}
 =
\frac{(n_2+1)\Gamma[\kappa_{12} n_2 +\gamma_{13} +n_1]
\Gamma[\kappa_{32}n_2 +1+n_3]}{
\Gamma[\kappa_{12}(n_2+1)  + \gamma_{13} +n_1]
\Gamma[\kappa_{32}(n_2+1) +1+n_3] }.
\eeq
Using the $\Gamma$-Stirling asymptote at large arguments,
one gets
\beq
&&
\frac{a_{n_1,n_2+1}b_{n_1,n_2}}{a_{n_1,n_2}b_{n_1,n_2+1}}
\sim \crcr
&&
e^{\kappa_{12} +\kappa_{32}}  (n_2+1)\frac{
 (\frac{\kappa_{12} n_2 +\gamma_{13} +n_1}{1})^{\kappa_{12} n_2 +\gamma_{13} +n_1}
  (\frac{\kappa_{32}n_2 +1+n_3}{1})^{\kappa_{32}n_2 +1+n_3} }{
 (\frac{\kappa_{12}(n_2+1)  + \gamma_{13} +n_1}{1})^{\kappa_{12}(n_2+1)  + \gamma_{13} +n_1}
(\frac{\kappa_{32}(n_2+1) +1+n_3}{1})^{\kappa_{32}(n_2+1) +1+n_3}} \crcr
&&
\sim
\frac{ (n_2+1)}{
[\kappa_{12}(n_2+1) +\gamma_{13}+n_1]^{\kappa_{12}}
[\kappa_{32}(n_2+1) +1+n_3]^{\kappa_{32}}
}.
\eeq
The latter is bounded by $1$ at large $n_1$ and $n_2$
for $\kappa_{32}>0$ (condition induced by $n_1\to \infty$)
and $\kappa_{12}+ \kappa_{32}>1$ (condition induced
by $n_2 \to\infty$).  Thus, by Theorem \ref{compa} providing
a sufficient condition, by insisting to compare
the exponential series with the present norm series,
we could miss some possible solutions.

We can optimize the test, by comparing
now  $a_{n_1,n_2}$ with
\bea
b'_{n_1,n_2} = \frac{w_1^{n_1}w_2^{n_2}}{n_1! \Gamma[\gamma_{32} + n_3]}
\eea
which is the term of a previous norm series which proves to be
convergent for $\kappa_{32}>0$. One finds the ratios
\beq
&&
\frac{a_{n_1+1,n_2}b'_{n_1,n_2}}{a_{n_1,n_2}b'_{n_1+1,n_2}}
 =
\frac{1+ n_1 }{ \gamma_{1}+n_1 } \leq 1, \crcr
&&
\frac{a_{n_1,n_2+1}b_{n_1,n_2}}{a_{n_1,n_2}b_{n_1,n_2+1}}
 =
\frac{\Gamma[\kappa_{12} n_2 +\gamma_{13} +n_1]
\Gamma[\kappa_{32}n_2 +1+n_3]\Gamma[1+ \kappa_{32}(n_2+1)+n_3]}{
\Gamma[\kappa_{12}(n_2+1)  + \gamma_{13} +n_1]
\Gamma[\kappa_{32}(n_2+1) +1+n_3]\Gamma[\kappa_{32}n_2 +1+n_3] } \crcr
&&
 =
\frac{\Gamma[\kappa_{12} n_2 +\gamma_{13} +n_1]}{
\Gamma[\kappa_{12}(n_2+1)  + \gamma_{13} +n_1]}
\sim \frac{1}{(\kappa_{12}(n_2+1)  + \gamma_{13} +n_1)^{\kappa_{12}}} \sim 0 \leq 1
\label{ratio1}
\eeq
which hold at large $n_i$ using the $\Gamma$-Stirling approximation
and in the last, we only require that $\kappa_{12}>0$
turning out to be less stringent than the above condition
$\kappa_{12}+ \kappa_{32}>1$ (we will come back
on the second inequality in the next verification).
Theorem \ref{compa} can be applied now.

Let us inspect the limit case $\kappa_{12}=0$
by unfolding the same calculations: one ends up with
two ratios (\ref{ratio1}), at large $n_i$ limits,
bounded by $1$. Then Theorem \ref{compa} is again
applied and the case $\kappa_{12}=0$
is not to exclude and will lead to  relevant VCS.

\begin{figure}[t]
\centering{
\includegraphics[width=50mm]{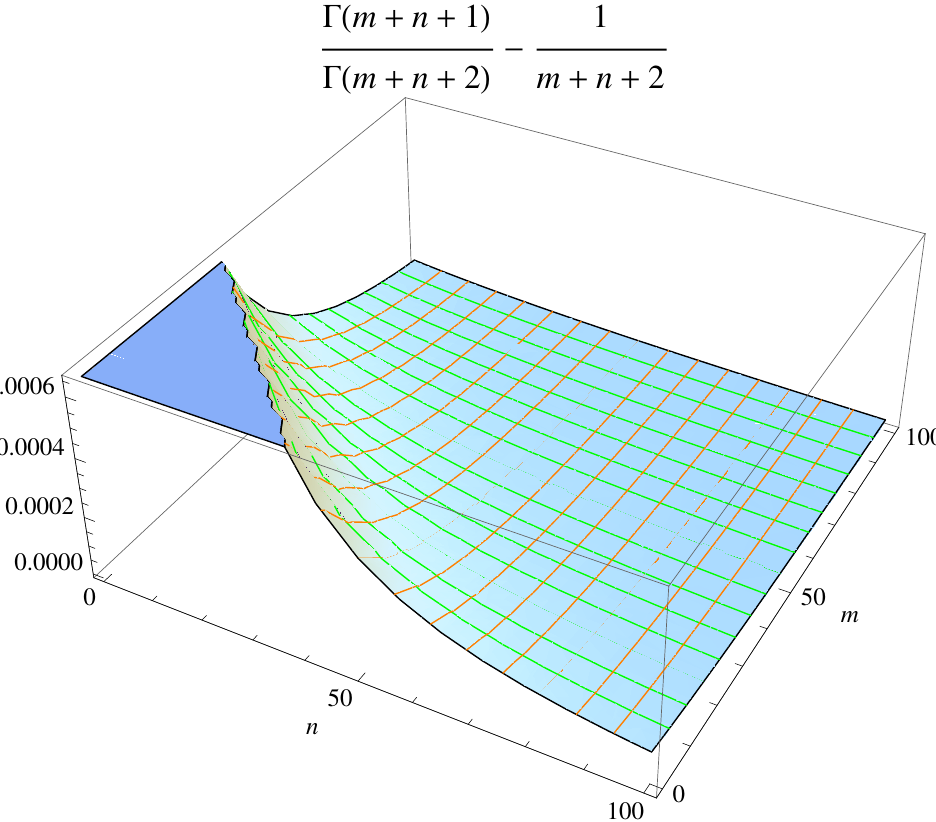}
\includegraphics[width=50mm]{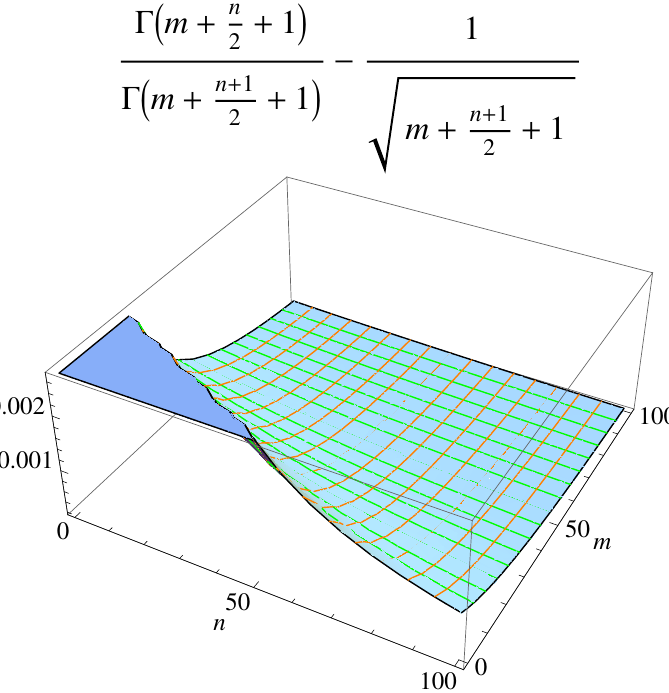}
\includegraphics[width=50mm]{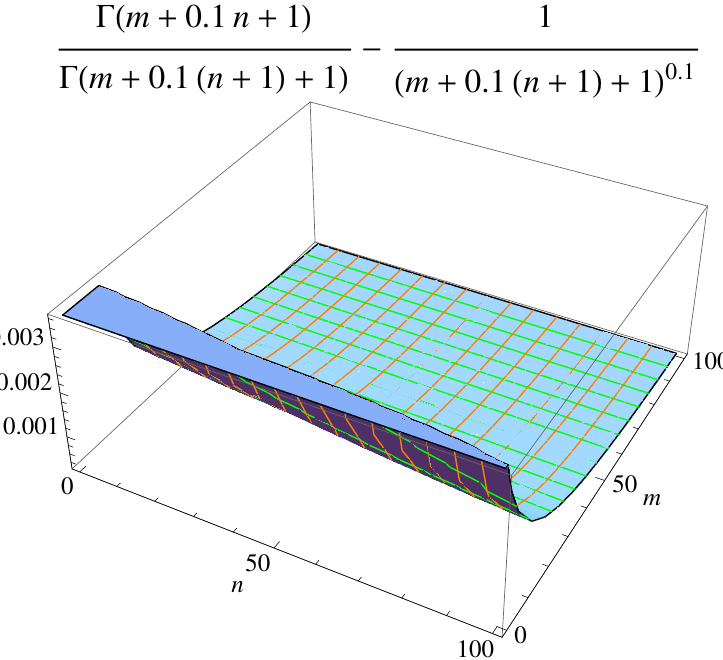}
\includegraphics[width=50mm]{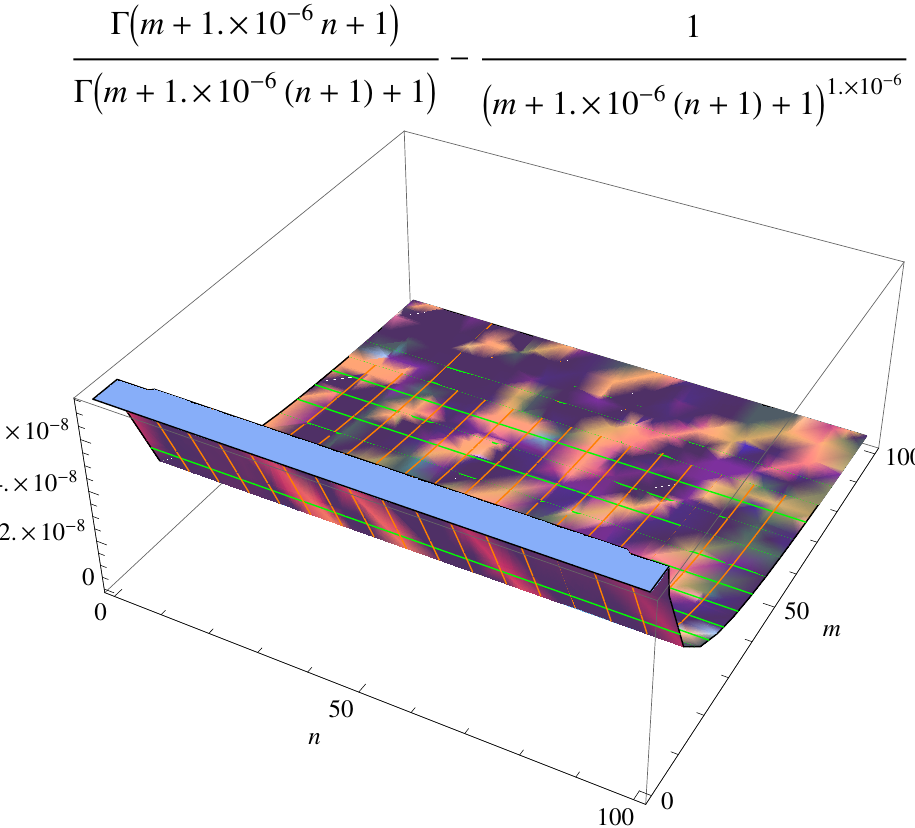}
}
\caption{\footnotesize{The difference
$ \frac{ \Gamma[\gamma_{13} + \kappa_{12} n_2 + n_1 ]}
{\Gamma[\gamma_{13} + \kappa_{23}(n_2+1) +n_1]}
 -[\gamma_{13} + \kappa_{23}(n_2+1) +n_1]^{-\kappa_{12}} $ at large $n_1=m$ and $n_2=n$ for different parameters
$\kappa_{12}\in \{ 1,1/2,1/10,1/10^6 \}$.}}
\label{fig:numdegam}
\end{figure}

\noindent{\bf Checking Eq.(\ref{3D2dB9cnoenorm})}
A second verification is in order for the norm
(\ref{3D2dB9cnoenorm}).
Let us evaluate the ratio-comparison with double exponential
term $b_{n_1,n_2}$
\beq
&&
\frac{a_{n_1+1,n_2}b_{n_1,n_2}}{a_{n_1,n_2}b_{n_1+1,n_2}}
  =\frac{ 1+n_1 }{\gamma_{1} +n_1}  \leq 1, \\
&&
\frac{a_{n_1,n_2+1}b_{n_1,n_2}}{a_{n_1,n_2}b_{n_1,n_2+1}}
 =
\frac{(n_2+1)\Gamma[\kappa_{12} n_2 +\gamma_{13} +n_1]}{
\Gamma[\kappa_{12}(n_2+1)  + \gamma_{13} +n_1] }
\sim
\frac{(n_2+1)}{
[\kappa_{12}(n_2+1)  + \gamma_{13} +n_1]^{\kappa_{12}} } \leq 1,
\nonumber
\eeq
in the last inequality used has been  made of the $\Gamma$-Stirling approximation and $\kappa_{12}>1$.
In fact, $\kappa_{12}>0$ is enough for proving the
convergence of the series (again comparison with exponential is a
too strong requirement). We have to prove
\beq
&&
\lim_{n_1,n_2 \to \infty} \frac{a_{n_1+1,n_2}}{a_{n_1,n_2}}
  =w_1\lim_{n_1,n_2 \to \infty} \frac{ 1 }{\gamma_{13} + \kappa_{23}n_2 +n_1} <1 ,\crcr
&&
\lim_{n_1,n_2 \to \infty} \frac{a_{n_1,n_2+1}}{a_{n_1,n_2}}
  =w_2\lim_{n_1,n_2 \to \infty} \frac{ \Gamma[\gamma_{13} + \kappa_{12} n_2 + n_1 ] }{\Gamma[\gamma_{13} + \kappa_{23}(n_2+1) +n_1]} <1,
\eeq
and use Theorem \ref{ratio}. The bound of the first ratio is obvious and this should be
sufficient to end the proof. However, let us check the second
ratio bound, because also it is involved in the proof of (\ref{ratio1}).
The second test ratio can be shown using again our favorite
$\Gamma$ approximation and
\bea
\lim_{n_1,n_2 \to \infty}
\left[ \frac{ \Gamma[\gamma_{13} + \kappa_{12} n_2 + n_1 ] }{\Gamma[\gamma_{13} + \kappa_{23}(n_2+1) +n_1]}
 -\frac{1}{[\gamma_{13} + \kappa_{23}(n_2+1) +n_1]^{\kappa_{23}}} \right] \sim 0
\eea
which holds for all $\kappa_{12}>0$.
The main problem being  at low $\kappa_{12}\sim 0$ but $\kappa_{12}\neq 0$, some numerics show that this is indeed the case (see Figure \ref{fig:numdegam}).

\noindent{\bf Checking Eq.(\ref{norm123})}
We perform the checking of the convergence
of the series (\ref{norm123}), writting
\bea
a_{n_1,n_2} = \frac{w_1^{n_1}w_2^{n_2}}{\Gamma[ \gamma_{12}  +n_1]\Gamma[\gamma_{3}+n_3]} \qquad
b_{n_1,n_2} =\frac{w_1^{n_1}w_2^{n_2}}{n_1! \Gamma[1+ \kappa_{32}n_2 + \kappa_{31}n_1 + n_3 ]}.
\label{ab2}
\eea
First, we check that the intermediate term
yields an everywhere convergent series:
\beq
&&
\lim_{n_1\to \infty} \frac{b_{n_1+1,n_2}}{b_{n_1,n_2}}=
w_1\lim_{n_1\to \infty}
\frac{\Gamma[1+ \kappa_{32}n_2 + \kappa_{31}n_1 + n_3 ]}{(n_1+1) \Gamma[1+ \kappa_{32}n_2 + \kappa_{31}(n_1+1) + n_3 ]}
= 0 ,\crcr
&&
\lim_{n_2\to \infty} \frac{b_{n_1,n_2+1}}{b_{n_1,n_2}}=w_2
\lim_{n_2\to \infty}
\frac{\Gamma[1+ \kappa_{32}n_2 + \kappa_{31}n_1 + n_3 ]}{\Gamma[1+ \kappa_{32}(n_2+1) + \kappa_{31}n_1 + n_3 ]}
= 0, \qquad \kappa_{32} >0,\cr\cr
&&
\lim_{n_1,n_2\to \infty} \frac{b_{n_1+1,n_2}}{b_{n_1,n_2}}
\sim
w_1\lim_{n_1,n_2\to \infty}
\frac{1}{(n_1+1) [1+ \kappa_{32}n_2 + \kappa_{31}(n_1+1) + n_3 ]^{\kappa_{31}}}
\sim 0 <1 ,\crcr
&&
\lim_{n_1,n_2\to \infty} \frac{b_{n_1,n_2+1}}{b_{n_1,n_2}}
\sim w_2
\lim_{n_1,n_2\to \infty}
\frac{1}{[1+ \kappa_{32}(n_2+1) + \kappa_{31}n_1 + n_3 ]^{\kappa_{32}}}
\sim 0 <1.
\eeq
The first inequality is valid for all $\kappa$'s
whereas the second is only valid for $\kappa_{32}>0$
which is consistent with the initial constraint on the column
series.
Theorem \ref{ratio} ensures that the series defined by $b_{n_1,n_2}$
is convergent everywhere.

Now, we can verify the hypothesis of Theorem \ref{compa}
using (\ref{ab2}).
At large $n_1$ and $n_2$, the following holds
\beq
&&
\frac{a_{n_1+1,n_2}b_{n_1,n_2}}{a_{n_1,n_2}b_{n_1+1,n_2}}
 =
\frac{(n_1+1)}{(\gamma_{12} +n_1)}\leq 1, \crcr
&&
\frac{a_{n_1,n_2+1}b_{n_1,n_2}}{a_{n_1,n_2}b_{n_1,n_2+1}}
 =
\frac{ \Gamma[\kappa_{12}n_2 + 1+n_1]
\Gamma[\kappa_{32}n_2 + \gamma_{31} +n_3]
\Gamma[\kappa_{32}(n_2+1) + \gamma_{31} +n_3] }{
\Gamma[\kappa_{12}(n_2+1) + 1+n_1]
\Gamma[\kappa_{32}(n_2 +1)+ \gamma_{31} +n_3]
\Gamma[\kappa_{32}n_2 + \gamma_{31} +n_3]  }  \crcr
&&
= \frac{ \Gamma[\kappa_{12}n_2 + 1+n_1] }{
\Gamma[\kappa_{12}(n_2+1) + 1+n_1] }
\sim \frac{1}{[\kappa_{12}(n_2+1) + 1+n_1]^{\kappa_{12}}}
\sim 0 \leq 1,
\eeq
valid for $\kappa_{12}>0$. The proof for the case $\kappa_{12}=0$
is the same as the convergence for the series defined
by $b_{n_1,n_2}$ (\ref{ab2}). Hence, the norm series (\ref{norm123})
converges everywhere.

\end{document}